\documentclass[9pt,a4paper,oneside,twocolumn]{extarticle}

\usepackage[margin=0.65in]{geometry}

\date{}

\usepackage[T1]{fontenc}
\usepackage{tgtermes}

\newcommand\blfootnote[1]{%
	\begingroup
	\renewcommand\thefootnote{}\footnote{#1}%
	\addtocounter{footnote}{-1}%
	\endgroup
}

\usepackage[numbers,sort&compress]{natbib}

\usepackage{amssymb}
\usepackage{amsmath}
\DeclareMathOperator*{\argmax}{arg\,max}

\usepackage[british]{babel}

\usepackage{algorithm}
\usepackage{algpseudocode}
\usepackage[beginLComment=//~,endLComment=]{algpseudocodex}
\algnewcommand{\LineComment}[2]{\vspace{0.5em} \Statex \hspace{#1em} \textcolor{gray}{\textit{// #2}}}

\usepackage{adjustbox}
\usepackage{array}
\newcolumntype{R}[2]{%
    >{\adjustbox{angle=#1,lap=\width-(#2)}\bgroup}%
    l%
    <{\egroup}%
}

\usepackage{pdflscape}
\usepackage{lscape}

\PassOptionsToPackage{hyphens}{url}\usepackage{hyperref}
\definecolor{aquamarine}{RGB}{11, 153, 193}
\hypersetup{
	colorlinks=true,   
	linkcolor=aquamarine,
	citecolor=aquamarine,
	urlcolor=aquamarine
}

\usepackage{graphicx}

\usepackage{authblk}

\usepackage[dvipsnames]{xcolor}
\usepackage[table]{xcolor}
\definecolor{DSBlue}{RGB}{52, 101, 164}
\definecolor{DSYellow}{RGB}{230, 233, 5}
\definecolor{DSGreen}{RGB}{22, 130, 83}
\definecolor{DSGrey}{RGB}{128, 128, 128}

\usepackage[decisionutilitycolor]{influence-diagrams}

\usepackage{longtable}
\usepackage{booktabs}
\usepackage{arydshln}

\usepackage{multirow}

\usepackage{booktabs}
\usepackage{amssymb}
\usepackage{pifont}
\usepackage{tikz}
\usetikzlibrary{tikzmark, arrows.meta, calc}

\usepackage{fontawesome5}
\usepackage{xcolor}
\newcommand{\srcInf}{\textcolor{darkgray}{\faIndustry}}
\newcommand{\srcAca}{\textcolor{darkgray}{\faFlask}}
\newcommand{\srcGAI}{\textcolor{darkgray}{\faRobot}}
\usepackage{wasysym}
\newcommand{\fcircle}{\textcolor{black}{\CIRCLE}}
\newcommand{\hlcircle}{\textcolor{black}{\LEFTcircle}}
\newcommand{\ecircle}{\textcolor{black}{\Circle}}

\newcommand{\amark}{\textcolor{black}{$\bigstar$}}

\usepackage[section]{placeins}
\usepackage{caption}
\usepackage{subcaption}

\usepackage{titlesec}
\titleformat{\section}{\large\bfseries}{\thesection}{1em}{}
\titleformat{\subsection}{\normalsize\bfseries}{\thesubsection}{1em}{}
\titleformat{\subsubsection}{\normalsize\itshape}{\thesubsubsection}{1em}{}
\usepackage{caption}
\titlespacing*{\section}{0pt}{2ex plus 0.5ex minus 0.2ex}{0.8ex plus 0.2ex}

\usepackage{enumitem}
\setlist{itemsep=3pt, parsep=0pt, topsep=5pt}

\newcommand{\orcid}[1]{\href{https://orcid.org/#1}{\textcolor[HTML]{A6CE39}{\faOrcid}}}

\begin{document}

\let\WriteBookmarks\relax
\def\floatpagepagefraction{1}
\def\textpagefraction{.001}

\title{\LARGE \textbf{A Knowledge-Based Multi-Agent Framework for Security Control Recommendation}}

\author[1,2,*]{Carolina Fern\'{a}ndez-Mart\'{i}nez \orcid{0000-0003-1865-7177}}
\author[1]{Shuaib Siddiqui \orcid{0000-0002-8257-4969}}
\author[2]{Vanesa Daza \orcid{0000-0003-0583-7929}}
\affil[1]{Cybersecurity \& Blockchain Research Group, i2CAT Foundation, Barcelona, 08039, Spain}
\affil[2]{Department of Information and Communication Technologies, Universitat Pompeu Fabra, Barcelona, 08018, Spain}


\twocolumn[
\begin{@twocolumnfalse}
	\maketitle
	\vspace{-2ex}
	\noindent\hrulefill\par\vspace{0.75ex}
	 \noindent\large\textbf{Abstract}\par\vspace{0.75ex}
	 \normalsize
		\noindent Hardening IT on-premises environments can be a daunting task for teams without access to adequate cybersecurity expertise. In this regard, Decision Support Systems (DSS) with embedded expert knowledge can assist users by guiding them with security recommendations to meet their objectives. This work proposes a Security DSS that recommends security control sub-families given minimal user requirements indicating coverage of different security dimensions. It leverages a curated, unified dataset from both well-known Information Security (InfoSec) and academic sources. This DSS is defined as a non-zero-sum, simultaneous game that is grounded in a Multi-Agent Influence Diagram (MAID) model and explores the decision space over 7 security dimensions or agents, using no-regret online learning to ultimately find the security control sub-families that best fit the requirements while incurring minimal under- and over-provisioning of security resources. This work was validated in terms of performance and accuracy, among others, for varying dataset sizes. It shows exceptional satisfaction coverage results of 99\% when using as little as $\sim$65\% of the SW-implementable security controls, running in 1.2-35.7 seconds; and more moderate coverage results of 73\%-77\% when using $\sim$29\% of the controls, resolving in 0.8-13.8 seconds.

	\vspace{3ex}\par\noindent

	\vspace{0.75ex}\par\noindent\hrulefill
	\vspace{4ex} 
\end{@twocolumnfalse}
]

\blfootnote{
	\textcopyright~2026. This manuscript version is made available under the CC-BY-NC-SA 4.0 license. The final published version of record is available in Knowledge-Based Systems at \url{https://doi.org/10.1016/j.knosys.2026.116558}.
	\\
	\indent *Corresponding author
}


\section{Introduction}
\label{sec:introduction}

System administrators and operators find themselves required to survey, design and secure environments constructed from an ever-expanding ecosystem of architectures, specifications, frameworks and tools.
The public and hybrid cloud paradigm has handed a varying degree of these responsibilities over to hyperscalers, depending on the number and type of businesses' services an IT company may host there.
However, the adoption of the private cloud approach has grown over the last few years \cite{Lenschow_Kalia_Sandler_El-Assal_2024}.
Whilst using hybrid, multi-cloud environments is the norm, the investment in private cloud infrastructure is increasing at twice the rate of public clouds \cite{Cloud_Usage_and_Management_Trends_2025}.
Some of the adduced reasons for this shift include increasingly stringent regulatory compliance, translating into data sovereignty and isolated workloads (e.g. AI) on premises.

These hybrid and private cloud approaches require organisations to retain in-house knowledge regarding general and security-related configurations.
Modern tooling used in both local and hybrid setups becomes more complex as it encompasses more moving parts that must be continuously managed (e.g. tools for granular and stricter identity management, authentication and authorisation) \cite{Syed_Shah_Shaghaghi_Anwar_Baig_Doss_2022}.
At the same time, the scarcity of qualified cybersecurity experts \cite{Ahmed_Hossain_Fazio_Lezzi_Islam_2024} makes it more difficult to properly assess the significance and impact of adopting security-related tooling, not to mention the interpretation of the security control frameworks and threat modelling.
While methodologies like DevSecOps aim to involve all actors to reduce the threat surface early in the software development lifecycle, setting up and hardening operational environments with third-party tools is more fragmented and prone to error.

Because of that, teams without proper security expertise face a three-fold problem: (i) deciding which security controls to survey and prioritise first from a large, abstract pool; (ii) selecting subsets of controls that best mitigate their organisation's requirements and processes; and (iii) optimising this selection to minimise both the vulnerable attack surface and the invested time, effort and budget.
In some occasions, this task is complex even for cybersecurity experts, as the optimal control selection requires reaching a trade-off between under- and over-provisioning from vast catalogues \cite{Rahman_Rahman_Williams_2025} with varying contributions, which makes a comprehensive survey even more complex.
Specifically, the vulnerability surface of the organisation must be maintained or reduced, whilst ensuring minimal resources are committed to it.
A proper balance ultimately minimises costs, and is thus also reviewed from the economic side \cite{a15060211}.

The above complexity can be tamed by introducing Security Decision Support Systems (DSS).
Such systems guide operators during the initial surveying stage, bridging the knowledge gap by internally evaluating the security control frameworks according to formalised requirements.
By providing tailored, optimised recommendations, a Security DSS allows operators to harden their environment more efficiently by maximising compliance and minimising invested time and effort.

\subsection{Contributions}

This work proposes a Security DSS that embodies expert knowledge as a Knowledge Graph (KG) and a Game Theory framework that recommends minimal security control sub-families to best meet the requirements over the security dimensions.
These dimensions represent the system's data protection objectives, encompassing the CIA triad (confidentiality, integrity, availability) and others (authenticity, accountability, non-repudiation, privacy) \cite{Cawthra_Ekstrom_Lusty_Sexton_Sweetnam_2020,Cherdantseva_Hilton_2013}.
The game evaluates the agents' utility functions while exploring the decision space to maximise required security coverage and minimise under- and over-provisioning. Finally, a minimal set of best-suited control sub-families is identified and recommended.
The following make this possible:
\begin{enumerate}
    \item A curated and unified dataset that maps the Security and Privacy Control Families (SPCFs) and Sub-Families (SPCSFs) from NIST SP 800-53 rev5 \cite{NIST_800_53_r5} to 38 columns with data of interest: their contribution to cover the security dimensions, to mitigate threat vectors from the STRIDE-LM model; as well as detailed and aggregated scores from other works measuring effectiveness on mitigating attacker's techniques and scores from expert surveys, among others described in Section \ref{sec:implementation_dataset:structure}.
    \item A Multi-Agent Influence Diagram (MAID), as an extension to classical Bayesian Network (BN) \cite{Jensen_Nielsen_2007} models. These encode the coverage per security dimension and the total score from the dataset above as a KG with embedded probabilistic influences across random variables within MAID nodes.
    \item A multi-agent, non-zero-sum, simultaneous game-theoretic framework using no-regret online learning algorithms to heavily reduce the computational complexity of exploring the decision space. It evaluates the agents' (security dimensions) utilities from the MAID's KG and maximises the overall payoff, returning the joint profile that best meets the requirements. This is used as upper boundary by the recommender stage to maximise accuracy while minimising over-provisioning.
\end{enumerate}

The remainder of this paper introduces first the related work in Section \ref{sec:related-work}. The main sections detail the contents and generation process of the curated security dataset in Section \ref{sec:implementation_dataset} and the design and implementation of the MAID-based security recommender that leverages such dataset in Section \ref{sec:implementation_model}. 
Section \ref{sec:evaluation} compiles the evaluation methodology, metrics and tests of the security recommender and its baseline.
Section \ref{sec:discussion-conclusion} concludes with an overview of the system along with its assumptions and identified limitations for both the security dataset and recommender; as well as the concluding remarks and future work directions.


\section{Related work}
\label{sec:related-work}

Related academic literature is documented for each contribution from this work: (i) assessing and modelling the impact of the security controls available to users; and (ii) modelling the security decision support problem to find suitable recommendations.

\subsection{Security control's importance evaluation}
\label{sec:related-work:sec-control-importance}

For a security decision process, a knowledge base must exist that (i) categorises the different security mechanisms or controls to enact; and that (ii) scores these appropriately to the task at hand, i.e. depending on specific criteria.
Three academic works were identified that perform an evaluation of security mechanisms, whether using its own taxonomy or basing on existing ones, such as the list of SPCFs and SPCSFs provided by NIST SP 800-53 rev5 standard \cite{NIST_800_53_r5}.
This list contains (i) 20 families (SPCFs); (ii) 300 active intermediate controls, or sub-families (SPCSFs); and (iii) 714 active control enhancements.
Whilst other standardised security and privacy controls catalogues exist, such as ISO/IEC 27002:2022 \cite{ISO_IEC_27002} or MITRE D3FEND \cite{Kaloroumakis_Smith_2020}, NIST SP 800-53 is chosen for its higher relevance in previous literature, its periodic revisions and its open availability both from reports \cite{NIST_800_53_r5} and repositories \cite{NIST_OSCAL_800_53_r5_repo} (whose common format allows for easy programmatic analysis).\\
To identify the most impacting security controls, \cite{Liu_Shore_Yeoh_Jiang_Zeadally_2025} performed a three-round interview with cybersecurity management layer experts.
This work provides its own taxonomy for the security controls and provides scores and a mapping towards NIST CSF 2.0, used to improve organisation's cybersecurity plans.
In \cite{Hermann_Schneider_Tony_Yardim_Peldszus_Berger_Scandariato_Sasse_Naiakshina_2025}, a simplified taxonomy of functional security dimensions was produced with five categories of security controls. The authors analysed which controls could be implemented by software and provided code examples for popular security frameworks.
The NIST SP 800-53 SPCSFs had its effectiveness assessed in \cite{Rahman_Rahman_Williams_2025}, based on their ability to mitigate 188 adversarial techniques from the MITRE ATT\&CK framework.
This study considered 7 metrics per SPCSF, where Technique Coverage (TEC) is the prominent one.
They filtered the SPCSFs by relevance (107 with $TEC > 0$), they applied k-means clustering to separate the top 21 critical SPCSFs that perform the best (i.e. more risk-efficient and versatile) for organisations to prioritise.

\subsection{Probabilistic security and risk decision support}
\label{sec:related-work:sec-decision}

Both probabilistic and game-theoretic approaches are used in recommender and decision support scenarios; starting from the more probability-oriented classification and suggestion and incorporating also the motivation of different agents that are to converge. The latter typically consider Nash Equilibrium \cite{Nash_1950} as a stable state reached when agents cannot improve their outcome by acting unilaterally --- yet others exist, such as Stackelberg, subgame perfect, epsilon or correlated equilibria.

Probabilistic models, such as those based on the Bayes' Theorem, allow updating the probability of a hypothesis, given its likelihood and the conditioned probability of some evidence (known data) given that hypothesis.
Bayes' theorem is at the core of BNs \cite{Jensen_Nielsen_2007,Koller_Friedman_2010}, which allow modelling and updating events (as random variables, depicted as nodes) based on known information or beliefs (evidence) and use these to infer the events' unknown joint probabilities (hypothesis) conditioned on selected subsets of evidence. BNs can leverage this for recommender systems.
In \cite{Huang_Poskitt_Shar_2025}, an adaptive DSS for cybersecurity incident mitigation leverages the Directed Acyclic Graph (DAG) within a BN as well as a custom, domain-specific model data for better computational efficiency.
It applies Bayesian confidence calibration on well-known vulnerability and exploit scoring systems to compute the exposure probability, applying multi-objective optimisation for three pillars (attack likelihood, impact severity and system availability) to recommend mitigating controls.
\cite{Zhang_Malacaria_2021} proposes a cybersecurity DSS that converts the online optimisation to Mixed-Integer Conic Programming and runs over probabilistic attack graphs using Bayesian Stackelberg Games. They model the defender (leader) and attacker (follower) and account for the uncertainty on the latter's capabilities; ultimately selecting a subset of security controls from 24 defined from their university, so to counteract potential or ongoing attacks.

Game Theory approaches can also be used to model systems that consider the behaviour of agents under either collaborative or competing incentives and propose the most strategically beneficial behaviour.
Some authors combine these with Influence Diagrams to provide more descriptive and versatile frameworks, and at the same time, ensure fairness across involved agents and overall stability.
A MAID is used in \cite{Vonk_Kononova_Back_Sweijs_2025} to model an attacker-defender environment, modelling four categories of custom counter-hybrid measures that range from the technical to the policy level; then calculating the subgame perfect equilibrium that balances each agent's payoff and ultimately analysing results per counter-hybrid measure.
In \cite{Leveille_Jaskolka_2024}, a game-theoretic DSS evaluates the effectiveness of strategies used by an analyst to protect according to the attacker's objective. It uses a 2-player (security analyst and attacker), zero-sum, one-shot algorithm, considering attacker profiles and filtering on budget effectiveness. The end-user is left to provide the security controls' catalogue and expert data evaluation.
Another game-theoretic approach to cybersecurity decision is provided in \cite{Chen_Han_Jajodia_Lindelauf_Subrahmanian_Xiong_2020}, where a 2-player game models two competing countries and guides them on the optimal strategy regarding disclosure of vulnerabilities. In \cite{Bashir_Shamszaman_Song_Han_2026}, attacker and defender populations are modelled as a 2-player asymmetric, non-cooperative, non-zero sum game where each can either act or not. Based on Evolutionary Game Theory, this helps investigating the dynamics of the players' interactions, including the costs, benefits and defence probability and providing suggestions for the organisations' defence.
Adversarial Risk Analysis (ARA) is a related approach that allows decision making with strategic opponents and uncertain outcome; modelling different types of (non collaborative) agents.
Authors from \cite{Camacho_Couce-Vieira_Arroyo_Insua_2025} propose a risk analysis framework for AI-based systems, modelling both the defender's and attacker's expected utilities with multiple attributes.
They consider, among others, the portfolio costs and organisational risk attitude for the defender; and follow a Bayesian probabilistic approach with Monte Carlo sampling on the attacker to model its uncertainty.
With this, potential portfolios of (generic) cost-effective security controls are suggested.
On the ground of forensic investigations against strategic adversaries, \cite{Nisioti_Loukas_Rass_Panaousis_2021} proposes a two-player (investigator and attacker), non-zero-sum game that finds the Nash Equilibrium through the Lemke–Howson algorithm \cite{Lemke_Howson}.
Utility functions model benefits and costs per actor and attempt to maximise each actor's payoff, considering the other actor's actions: uncovering the attack (investigator) or hide traces (attacker).
A multi-stage cyber attack is modelled as multiple Bayesian games of incomplete information that use a graph of attack actions.
Bayesian probabilities also model here the uncertainty regarding the type of attacker, and the ultimate goal is to generate an optimal strategy for the investigator that also balances the attacker's type and the potential benefit and cost per action.

\subsection{Gap analysis}

When assessing the security controls' contribution for a security decision scenario, the current literature shows significant fragmentation between taxonomies \cite{Hermann_Schneider_Tony_Yardim_Peldszus_Berger_Scandariato_Sasse_Naiakshina_2025} that capture relations and the SW-implementable assessment, expert managers' surveys \cite{Liu_Shore_Yeoh_Jiang_Zeadally_2025} focusing on cybersecurity management and maturity assessment and datasets providing several technical metrics per control \cite{Rahman_Rahman_Williams_2025} and mappings to threats, yet not exhaustively computed for all the security catalogue in use.
This work combines relevant attributes (such as score and SW-implementability) from the previously indicated academic datasets into a unified one. It also integrates other attributes from standardised catalogues of security and privacy controls and also mapping controls to their impact on threats, following well-known threat modelling. This considerably enriches the resulting dataset, providing enough granularity (per SPCSF) to facilitate the construction of KGs.

Besides this, applying these full datasets as KG in game-theoretic models to obtain exact, optimal results (or equilibria) introduces considerable scalability issues. Since finding exact equilibria can be Polynomial Parity Arguments on Directed graphs (PPAD)-complete \cite{Daskalakis_2009}, such computational complexity restricts analysed implementations to 2-player games with small decision domains \cite{Zhang_Malacaria_2021,Vonk_Kononova_Back_Sweijs_2025,Leveille_Jaskolka_2024,Chen_Han_Jajodia_Lindelauf_Subrahmanian_Xiong_2020,Bashir_Shamszaman_Song_Han_2026,Camacho_Couce-Vieira_Arroyo_Insua_2025,Nisioti_Loukas_Rass_Panaousis_2021}.
Without simplifications like zero-sum games \cite{Leveille_Jaskolka_2024}, obtaining exact optimal solutions using Nash, Stackelberg or subgame perfect equilibria becomes directly intractable as the decision space grows, as influenced from e.g. number of agents or decision domains \cite{Zhang_Malacaria_2021,Vonk_Kononova_Back_Sweijs_2025,Leveille_Jaskolka_2024,Nisioti_Loukas_Rass_Panaousis_2021}.
To bypass exact calculations, other approaches rely on iterative Monte Carlo simulations \cite{Camacho_Couce-Vieira_Arroyo_Insua_2025}, which sample random utilities to estimate the attacker's actions but can incur heavy computational overhead for large control sets; as well as on Evolutionary Game Theory \cite{Bashir_Shamszaman_Song_Han_2026}, which handles probabilities of interactions at population-level and thus can lose significant granularity.

Overall, the literature does not adequately cover the combination addressed in this work. Existing works either evaluate security controls from specific perspectives (e.g. expert judgement, implementability or threat mitigation) or model cybersecurity decisions through probabilistic and game-theoretic approaches that typically focus on smaller control sets, two-player (e.g. attacker/defender) setups, or cost-driven objectives.
What remains insufficiently addressed is a knowledge-based framework that integrates heterogeneous control-level evidence into a unified representation, expresses requirements in terms of security dimensions and delivers granular objectives while supporting scalable multi-agent recommendation of Security and Privacy Control Sub-Families.
Therefore, this work models a non-zero-sum, simultaneous game with granular interactions across 7 agents. By relating each security dimension to an agent, the game can account for the differentiated interactions (e.g. cooperative or competitive) across each type of agent.
By avoiding exact calculations and applying instead no-regret online learning algorithms, this approach bypasses the PPAD-complete scalability issue and offers a scalable approach that converges faster on approximate optimal joint profiles from the explored decision space and refines these into optimal solutions.
Consequently, the resulting DSS runs in up to 35.7 seconds. The KG is recreated from the latest available dataset for every execution; thus maximising operations on an up-to-date dataset.


\section{Security dataset}
\label{sec:implementation_dataset}

\begin{figure}[htbp]
    \centering
    \includegraphics[width=0.95\columnwidth]{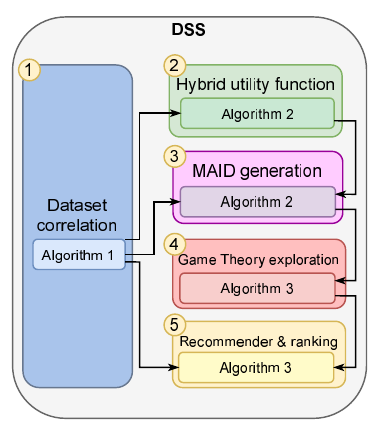}
    \caption{Security DSS system overview}
    \label{fig:implementation_dataset:dss-system-overview}
\end{figure}

A security dataset is first constructed to provide relevant information for the security controls.
This dataset acts as the cornerstone of the proposed Security DSS, as its information feeds the different decisions and information modelling present in the DSS architectural building blocks.
Fig. \ref{fig:implementation_dataset:dss-system-overview} illustrates the relations (dependencies) across these blocks, their mapping to the three described algorithms and their order of execution as part of the Security DSS.

The structure and contents of the dataset are explained below, along with its generation process and the analysis of the resulting data.

\subsection{Structure}
\label{sec:implementation_dataset:structure}

This dataset relates security controls (SPCSFs) from well-established frameworks in the cybersecurity community to both existing expert- and ML-generated scores provided by academic works. It also performs a simple quantification process on the SPCSF's impact on mitigating threats and on covering or addressing the seven security dimensions, as taken from well-known open Information Security (InfoSec) sources. With these scores and information, three partial and one global scores are generated per SPCSF.

To extract security controls highly relevant to the community, the NIST SP 800-53 rev5 \cite{NIST_800_53_r5} catalogue was chosen. Beyond its $\sim$20-year life and its adoption in both engineering and academic communities, it is accessible in machine-friendly formats that facilitate automated processing, e.g. OSCAL \cite{NIST_OSCAL_800_53_r5_repo}. It is structured in 20 control families, with each family containing multiple controls, here called SPCSFs; and each control potentially having multiple control enhancements. These SPCSFs families comprise specific security topics, from access control (AC) to Supply Risk Management (SR). It contains 324 SPCSFs (300 of these not being withdrawn). These conform the rows of the proposed dataset $DS_{8}$.

\begin{table*}[htbp]
    \centering
    \caption{Provenance and relation across groups of features/columns per dataset}
    \label{tab:implementation-dataset:structure}
    \resizebox{1\textwidth}{!}{
    \begin{tabular}{@{} lc ccccccc cc @{}}
        \toprule
        & & \multicolumn{7}{c}{\textbf{Source attributes}} & \multicolumn{2}{c}{\textbf{Derived scores}} \\
        \cmidrule(lr){3-9} \cmidrule(l){10-11}
        \textbf{Dataset} & \textbf{Source} & \textbf{NIST (2+2)} & \textbf{SW-impl. (3)} & \textbf{Sec. dim. cov. (8)} & \textbf{STRIDE-LM cov. (7)} & \textbf{Expert scores (5)} & \textbf{Taxonomy (1)} & \textbf{ATT\&CK \& impact (6)} & \textbf{Partial (3)} & \textbf{Total (1)} \\
        \midrule
        $DS_{1}$ & \srcInf & \fcircle~\ecircle & \hlcircle~\ecircle~\ecircle & \ecircle & \ecircle & \ecircle & \ecircle & \ecircle & \ecircle & \ecircle \\
        $DS_{2*}$ & \srcAca & \hlcircle~\ecircle & \ecircle~\ecircle~\ecircle & \ecircle & \ecircle & \tikzmarknode{n2}{\fcircle} & \ecircle & \ecircle & \ecircle & \ecircle \\
        $DS_{3}$ & \srcInf & \fcircle~\ecircle & \ecircle~\ecircle~\ecircle & \ecircle & \ecircle & \ecircle & \ecircle & \ecircle & \ecircle & \ecircle \\
        $DS_{4}$ & \srcGAI & \fcircle~\ecircle & \ecircle~\fcircle~\ecircle & \fcircle & \ecircle & \ecircle & \ecircle & \ecircle & \ecircle & \ecircle \\
        $DS_{5}$ & \srcInf & \fcircle~\ecircle & \ecircle~\ecircle~\ecircle & \ecircle & \tikzmarknode{n5}{\fcircle} & \ecircle & \ecircle & \ecircle & \ecircle & \ecircle \\
        $DS_{6}$ & \srcAca & \fcircle~\ecircle & \ecircle~\ecircle~\fcircle & \ecircle & \ecircle & \ecircle & \fcircle & \ecircle & \ecircle & \ecircle \\
        $DS_{7}$ & \srcAca & \fcircle~\fcircle & \ecircle~\ecircle~\ecircle & \ecircle & \ecircle & \ecircle & \ecircle & \tikzmarknode{n7}{\fcircle} & \ecircle & \ecircle \\
        $DS_{8}$ & \amark & \ecircle~\ecircle & \ecircle~\ecircle~\ecircle & \ecircle & \ecircle & \ecircle & \ecircle & \ecircle & \tikzmarknode{tgt_p}{\amark} & \tikzmarknode{tgt_t}{\amark} \\
        \bottomrule
        \addlinespace
        \multicolumn{9}{@{}l}{\small \textbf{Source (dataset):} \srcInf~InfoSec \quad \srcAca~Academy \quad \srcGAI~Generative AI (GenAI) \quad \amark~Derived
        \quad --- \quad
        \textbf{Source (column):} \fcircle~Native \quad \hlcircle~Inferred \quad \ecircle~Absent \quad \amark~Derived}
    \end{tabular}
    \begin{tikzpicture}[
        remember picture, 
        overlay,
        >=Stealth,
        fusion/.style={->, thick, darkgray!80!black, rounded corners=4pt}
    ]
        \draw[fusion] ($(n2.east) + (3pt, 0)$) -- ++(19pt,0) |- ($(tgt_p.west) + (-3pt, 4pt)$);
        \draw[fusion] ($(n5.east) + (3pt, 0)$) -- ++(12pt,0) |- ($(tgt_p.west) + (-3pt, 0)$);
        \draw[fusion] ($(n7.east) + (3pt, 0)$) -- ++(12pt,0) |- ($(tgt_p.west) + (-3pt, -4pt)$);
        \draw[fusion] ($(tgt_p.east) + (3pt, 0)$) -- ($(tgt_t.west) + (-3pt, 0)$);
    \end{tikzpicture}
    }
\end{table*}

Table \ref{tab:implementation-dataset:structure} groups the 38 columns available in $DS_{8}$, indicating the number of features or columns per group, the originating source ($DS_{i}$) and arrows indicating which were used to derive aggregated partial and total scores.
All sources used to construct the dataset $DS_{8}$ are identified as $DS_{i}, i \in [1,7]$ and summarised in Table \ref{tab:implementation-dataset:nomenclature}.
The data transformations used to generate the dataset are detailed in Algorithm \ref{alg:implementation-dataset:dataset}.
Overall, the generated dataset $DS_{8}$ contains 300 SPCSFs and 38 columns with extracted and generated data properly correlated.

\subsubsection{Source datasets}

As indicated in Table \ref{tab:implementation-dataset:structure}, three types of datasets were used: from InfoSec communities, from academic papers and GenAI-interpreted.

Three academic works, introduced in Section \ref{sec:related-work:sec-control-importance}, were used in this contribution to deliver averaged scores for SPCSFs.
These are (i) expert-based scores per SPCSF in $DS_{2*}$ \cite{Liu_Shore_Yeoh_Jiang_Zeadally_2025}; (ii) taxonomy and SW-implementability criteria in $DS_{6}$ \cite{Hermann_Schneider_Tony_Yardim_Peldszus_Berger_Scandariato_Sasse_Naiakshina_2025}; and (iii) the analysed mitigation impact, considering the MITRE ATT\&CK framework in $DS_{7}$ \cite{Rahman_Rahman_Williams_2025}.
InfoSec sources from NIST were also used; both from (i) the OSCAL catalogue ($DS_{1}$) \cite{NIST_OSCAL_800_53_r5_repo} and from (ii) the CyberSecurity Framework (CSF) ($DS_{3}$); as well as from the CSFtools site \cite{CSF_Tools_NIST_Special_Publication_800_53_Revision_5_2022} for both (iii) the impact of the NIST 800-53 rev5 SPCSFs on mitigating Microsoft's STRIDE-LM threat vectors ($DS_{5}$).
GenAI was used to analyse the coverage of each SPCSF towards the security dimensions ($DS_{4}$), based on the text-based description from the InfoSec source (i).

As introduced above, two well-known threat frameworks are used in $DS_{8}$. One of them is the MITRE ATT\&CK, considered from $DS_{7}$ and involving a taxonomy of adversarial tactics composed of techniques, as well as detection strategies and mitigations per technique. The other one is Microsoft's STRIDE-LM, providing seven threat vectors that can be leveraged by an attacker to compromise an environment: spoofing (S), tampering (T), repudiation (R), information disclosure (I), denial of service (D), elevation of privilege (E) and lateral movement (LM).

\subsubsection{Dataset attributes}

The columns provided in $DS_{8}$ contain mostly attributes correlated from the other $DS$ sources. The grouped columns from Table \ref{tab:implementation-dataset:structure} are explained below.

The two NIST columns provide the identifier and name per SPCSF for most $DS$ (except for $DS_{2*}$), which are used for direct (or indirect) correlation across datasets. Two other columns from $DS_{7}$ provide the NIST baseline and priority from both NIST SP 800-53 rev5 and rev4, respectively. The priority was removed in r5, yet is deemed of interest here.
The interpretation of the implementability through SW of a SPCSF is provided in three columns that are either inferred from attributes from $DS_{1}$ or directly extracted from $DS_{4}$ and $DS_{6}$.
The contribution (coverage) of each SPCSF to the seven considered security dimensions is provided in $DS_{4}$ as real values between 0 and 1, with the constraint that all of them sum up to 1 for normalisation. This comes from interpreting the SPCSF descriptions from OSCAL $DS_{1}$ and CSFtools site \cite{CSF_Tools_NIST_Special_Publication_800_53_Revision_5_2022}, which was left to a GenAI model. For human inspection, an extra column was provided with its justification of the interpretation.
Another measure of impact is the contribution of each control enhancement per SPCSF to cover the seven threat vectors from the STRIDE-LM framework. These columns are included in $DS_{5}$ and contain a list of comma-separated SPCSF's control enhancements.
A custom taxonomy column coming from $DS_{6}$ is added to potentially help establishing categories of SPCSFs depending on their kind of actions and benefit.
Six columns are used to denote the impact of SPCSFs as mitigating mechanisms, considering the MITRE ATT\&CK framework. These are fetched from $DS_{7}$ and encode varying metrics, such as the number of techniques or mean number of tactics covered.
Finally, scores are derived from previous attributes. Three partial scores are calculated (i) from the scores given by cybersecurity experts($DS_{2*}$), by averaging these; (ii) from STRIDE-LM ($DS_{5}$), the mitigation efficacy is computed using a simple ratio based on the number of controls enhancements involved in mitigation from all available in each SPCSF; and (iii) from MITRE ATT\&CK ($DS_{7}$), the mitigation efficacy averages its metrics after performing on-the-fly normalisation. The total score is obtained by averaging these three partial scores.

The model detailed in Section \ref{sec:implementation_model} uses the NIST identifiers, the SW-implementability criteria, the security dimension coverage and the total score --- obtained from the aggregated values in the partial scores described in Table \ref{tab:implementation-dataset:structure}.

%
%
\begin{algorithm}
\caption{Dataset generation}
\begin{algorithmic}[1]

\Require $DS_{i} \neq \emptyset ~ \forall i \in [1, 7]$
\Ensure $DS_{8}\neq \emptyset$

\LineComment{-3.5}{Extract Security Families from $DS_{1}$ \cite{NIST_OSCAL_800_53_r5_repo}}
\For{each $g$ in $DS_{1}.catalog.group$} \label{alg:implementation-dataset:dataset:line:s1}
    \State $g_{a} \gets g.controls ~s.t.~ g.status \notin withdrawn$ \label{alg:implementation-dataset:dataset:line:s2}
    \State $nsf[g] \gets |g_{a}|$ \label{alg:implementation-dataset:dataset:line:s3}
    \For{each $d_{1i}$ in $g_{a}$} \label{alg:implementation-dataset:dataset:line:s4}
        \State $DS_{8}[d_{1i}].impl_1 \gets d_{1i}.status \notin withdrawn ~ \bigcap ~ d_{1i}.implementation \in system$ \label{alg:implementation-dataset:dataset:line:s5}
    \EndFor
\EndFor

\LineComment{-1.5}{Merge $DS_{2A}$ and $DS_{2B}$ as $DS_{2}$ \cite{Liu_Shore_Yeoh_Jiang_Zeadally_2025}}
\State $DS_{2} \gets merge(DS_{2A}.process\_id = DS_{2B}.process\_id)$ \label{alg:implementation-dataset:dataset:line:s6}
\LineComment{-1.5}{Merge $DS_{4}$ and $DS_{5}$}
\State $DS_{8}.llm\_* \gets merge(DS_{8}.id = DS_{4}.id)$ \label{alg:implementation-dataset:dataset:line:s7}
\State $DS_{8}.stridelm\_* \gets \text{merge(}DS_{8}\text{.id = }DS_{5}.id\text{)}$ \label{alg:implementation-dataset:dataset:line:s8}
\LineComment{-1.5}{Merge $DS_{6}$ \cite{Hermann_Schneider_Tony_Yardim_Peldszus_Berger_Scandariato_Sasse_Naiakshina_2025}}
\State $DS_{8}.impl_{2} \gets \text{merge(}DS_{8}\text{.id = }DS_{6}.id\text{})$ \label{alg:implementation-dataset:dataset:line:s9}

\LineComment{-1.5}{Incorporate scores from $DS_{2}$ by relating to $DS_{3}$}
\For{each $d_{8i}$ in $DS_{8}$} \label{alg:implementation-dataset:dataset:line:s10}
    \For{each $d_{3j}$ in $DS_{3}$} \label{alg:implementation-dataset:dataset:line:s11}
        \If{$d_{8i}\text{.id = }d_{3j}\text{.reference\_document}$} \label{alg:implementation-dataset:dataset:line:12}
            \State $fam_{d3} \gets \text{family(}d_{3j}\text{.focal\_document})$ \label{alg:implementation-dataset:dataset:line:s13}
                \For{each $d_{2k}$ in $\mathrm{filter\_fam(}DS_{2}\mathrm{, fam_{d3})}$} \label{alg:implementation-dataset:dataset:line:s14}
                    \State $d_{8i}\mathrm{.stats} \gets \mathrm{stat\_metrics(}d_{2k}\mathrm{)}$ \label{alg:implementation-dataset:dataset:line:s15}
                \EndFor
        \EndIf
    \EndFor
\EndFor

\LineComment{-1.5}{Compute score from $DS_{2}$, $DS_{5}$ and $DS_{7}$ \cite{Rahman_Rahman_Williams_2025}}
\For{each $d_{8i}$ in $DS_{8}$} \label{alg:implementation-dataset:dataset:line:s16}
    \State $d_{8i}.score_{DS_2} \gets avg(mms(d_{8i}\text{.stats}[s_{j}]\text{))} ~ \forall s \in \vec{S}$ \label{alg:implementation-dataset:dataset:line:s17}
    \State $av_{7c} \gets avg(mms(DS_{7}.cm)) $ \label{alg:implementation-dataset:dataset:line:s18}
    \State $av_{7n} \gets \frac{avg(DS_{7}.nm)}{DS_{7}.nm_{max}} ~ \forall n \in \vec{DS_{7}.nm}$ \label{alg:implementation-dataset:dataset:line:s19}
    \State $d_{8i}.score_{DS_7} \gets avg(av_{7c}, ~av_{7n})$ \label{alg:implementation-dataset:dataset:line:s20}
    \State $d_{8i}.score_{DS_5} \gets \frac{|sct({c})|}{nsf[g]+1} \forall c \subset DS_{5} \mid g \in SPCSF$ \label{alg:implementation-dataset:dataset:line:s21}
    \State $av_{8i} \gets avg(d_{8i}.score_{DS_2}, d_{8i}.score_{DS_7}$,\\ $d_{8i}.score_{DS_5})$ \label{alg:implementation-dataset:dataset:line:s22}
    \State $d_{8i}.score \gets p_{i} \cdot av_{8i} \mid p_{i} =
      \begin{cases}
        1 & if \bigcup\limits_{k=1}^{2} DS_{8}.impl_{k} \ge 1 \\
        0 & \text{otherwise}
      \end{cases}$ \label{alg:implementation-dataset:dataset:line:s23}
\EndFor

\end{algorithmic}
\label{alg:implementation-dataset:dataset}
\end{algorithm}

\setlength{\tabcolsep}{5pt}
\begin{table}[h!]
	\caption{Notation for Algorithm \ref{alg:implementation-dataset:dataset}}
	\scalebox{0.85}{
		\begin{tabular}{cll}
			\cline{1-3}
			\textbf{Term} & \multicolumn{2}{l}{\textbf{Description}} \\
			\cline{1-3}
			\noalign{\vspace{0.2em}}
			\multicolumn{1}{l}{\textbf{Inputs}} & \multicolumn{2}{l}{} \\
			$DS_{i}$ & \multicolumn{2}{l}{Existing dataset or ground truth} \\
			($i \in$ & $DS{_1}$ & 800-53r5 (OSCAL) catalogue \cite{NIST_OSCAL_800_53_r5_repo} \\
			$[1,7]$) & $DS_{2*}$ & Taxonomy \& CSF 2.0 scores map \\
			& \multicolumn{2}{l}{\quad\quad\quad containing $DS_{2A}$ and $DS_{2B}$ \cite{Liu_Shore_Yeoh_Jiang_Zeadally_2025}}\\
			& $DS_{3}$ & CSF 2.0 to 800-53r5 map \cite{nist_olir_datasets} \\
			& $DS_{4}$ & Gemini 2.5 Pro to security dim. map \\
			& $DS_{5}$ & 800-53r5 to STRIDE-LM map \cite{CSF_Tools_NIST_Special_Publication_800_53_Revision_5_2022} \\
			& $DS_{6}$ & 800-53r5 to implementable map \cite{Hermann_Schneider_Tony_Yardim_Peldszus_Berger_Scandariato_Sasse_Naiakshina_2025} \\
			& $DS_{7}$ & 800-53r3 to scores map \cite{Rahman_Rahman_Williams_2025} \\
			\noalign{\vspace{0.4em}}
			\multicolumn{1}{l}{\textbf{Output}} & \multicolumn{2}{l}{} \\
			$DS_{8}$ & \multicolumn{2}{l}{Generated dataset from research \& InfoSec} \\
			\noalign{\vspace{0.4em}}
			\multicolumn{1}{l}{\textbf{Internals}} & \multicolumn{2}{l}{} \\
			$\vec{S}$ & \multicolumn{2}{l}{Statistical columns \cite{Liu_Shore_Yeoh_Jiang_Zeadally_2025}} \\
			$f: mms(\vec{S})$ & \multicolumn{2}{l}{Min-max scale / normalisation from Eq. \eqref{eq:implementation-dataset:generation:min-max-scaling}} \\
			$cm$ & \multicolumn{2}{l}{$DS_{7}$'s custom metrics (except TAC)} \\
			$nm$ & \multicolumn{2}{l}{$DS_{7}$'s NIST metrics (baseline, priority)} \\
			$f: sct(c)$ & \multicolumn{2}{l}{STRIDE-LM threats covered per $c \in$ SPCSF} \\
			$nsf[g]$ & \multicolumn{2}{l}{No. of active SPCSF within each SPCF} \\
			& \multicolumn{2}{l}{$|\mathrm{SPCSF}| \subset g_{a} ~\forall ~g_{a}~ \in \mathrm{SPCF}$} \\
			$p_{i}$ & \multicolumn{2}{l}{SW-implementability of a SPCSF} \\
			\noalign{\vspace{0.2em}}
			\cline{1-3}
		\end{tabular}
	}
	\label{tab:implementation-dataset:nomenclature}
\end{table}

\subsection{Generation process}
\label{sec:implementation_dataset:generation_process}

First, the NIST OSCAL catalogue \cite{NIST_OSCAL_800_53_r5_repo} is processed to identify (i) active SPCSFs per SPCF group ($g_{a}$) and (ii) SW-implementable criteria per SPCSF, determined by OSCAL labels and depending on whether the SPCSF is not withdrawn from previous releases and it is labelled as either "implementation-level" or "system" -- but not as "organization", which are deemed more strategic. These are covered in Algorithm \ref{alg:implementation-dataset:dataset}, lines \ref{alg:implementation-dataset:dataset:line:s1}-\ref{alg:implementation-dataset:dataset:line:s5}.

Now, the data for each SPCSF from the covered works is correlated and aggregated.
First, the two datasets from $DS_{2*}$ ($DS_{2A}$, $DS_{2B}$) \cite{Liu_Shore_Yeoh_Jiang_Zeadally_2025} are merged, using the common \textit{"process\_id"} column. These hold (i) the security ontology resulting from the interviews as one dataset; and (ii) the score for each category in the security ontology as another one. This is covered in Algorithm \ref{alg:implementation-dataset:dataset}, line \ref{alg:implementation-dataset:dataset:line:s6}. It is worth noting that the "merge" function is implemented as a left join, which takes as ground truth all items at the first (left) parameter.

Then, the $DS_{4}$ mapping is merged into the final dataset, considering the full list of SPCSFs (Algorithm \ref{alg:implementation-dataset:dataset}, line \ref{alg:implementation-dataset:dataset:line:s7}).
$DS_{4}$ was manually compiled upon querying the Gemini 2.5 Pro model on the CSFtools \cite{CSF_Tools_NIST_Special_Publication_800_53_Revision_5_2022} reference to identify the contribution (a real number between 0 and 1) of each SPCSF towards the fulfilment of the security dimensions: confidentiality (C), integrity (I), availability (A), authenticity (Au), accountability (Ac), non-repudiation (Nr) and privacy (Pr).
While relying on expert knowledge per SPCSF yields a more accurate operational ground truth, this LLM-generated data serves well the primary objective of validating the appropriateness of the framework across varying values. 
In operational environments, cybersecurity experts should replace these values upon a manual, detailed approach; e.g. inspired by guidance from authoritative sources such as CNSSI 1253 \cite{CNSSI_1253_Security_Categorization_and_Control_Selection_for_National_Security_Systems_2022} or from programmatic, statistic analysis of descriptions from the OSCAL catalogue \cite{NIST_OSCAL_800_53_r5_repo}.

After this, the final dataset is extended with a mapping ($DS_{5}$) that stores the contribution of each SPCSF (and internal control enhancements) towards the mitigation of each STRIDE-LM threat vectors. This mapping was obtained manually per SPCSF in CSFtools \cite{CSF_Tools_NIST_Special_Publication_800_53_Revision_5_2022} and merged in Algorithm \ref{alg:implementation-dataset:dataset}, line \ref{alg:implementation-dataset:dataset:line:s8}.
A second criterion to determine the SW-implementation, as provided by $DS_{6}$ \cite{Hermann_Schneider_Tony_Yardim_Peldszus_Berger_Scandariato_Sasse_Naiakshina_2025}, is also integrated here (Algorithm \ref{alg:implementation-dataset:dataset}, line \ref{alg:implementation-dataset:dataset:line:s9}).

At this point, the previously generated $DS_{2}$ is merged into the final dataset. To do so, the $DS_{3}$ mapping is used for correlating keys; as it relates controls from NIST SP 800-53 r5 and those from NIST CyberSecurity Framework (CSF) 2.0 \cite{nist_olir_datasets}.
In particular, the ID of each SPCSF in the final dataset is compared against each item from the "reference\_document" column in $DS_{3}$. If these match, the SPCSF's family (i.e. its SPCF) is retrieved. The available statistical metrics from $DS_{2}$ (i.e. 36 processes from its custom taxonomy) are incorporated by using both (i) the internally provided mapping between this custom taxonomy to NIST CSF 2.0 (column "mapping\_nist\_csf", post-processed in the "filter\_fam" method) and (ii) the $DS_{3}$ mapping described above ("focal\_document" column). These statistical metrics consider the mean, median, mode, Standard Deviation (SD) and Inter-Quartile Range (IQR). This is covered in Algorithm \ref{alg:implementation-dataset:dataset}, lines \ref{alg:implementation-dataset:dataset:line:s10}-\ref{alg:implementation-dataset:dataset:line:s15}.

\begin{table}[t]
	\caption{Top 20 SW-implementable SPCSFs \& contribution from works}
	\centering
	\scalebox{1}{
		\begin{tabular}{>{\bfseries}lcccc}
			\toprule
			\textbf{} & \multicolumn{1}{c}{} & \multicolumn{3}{c}{\textbf{Partial score generated per source}} \\
			\cmidrule(l){3-5}
			\small{\textbf{SPCSF}} & \small{\textbf{Total}} & \cellcolor{DSBlue!50} \small{\textbf{\#2 \cite{Rahman_Rahman_Williams_2025}} ($DS_{7}$)} & \cellcolor{DSYellow!50} \small{\textbf{\#3 \cite{Liu_Shore_Yeoh_Jiang_Zeadally_2025}} ($DS_{2*}$)} & \cellcolor{DSGreen!50} \small{\textbf{\#4} ($DS_{5}$)} \\
			\midrule
			\textbf{{\cellcolor{DSGrey!50}} SI-3} &  0.87633 & {\cellcolor{DSBlue!50}} 0.73287 & {\cellcolor{DSYellow!50}} 0.89614 & {\cellcolor{DSGreen!50}} 1.00000 \\
			\textbf{{\cellcolor{DSGrey!50}} CM-7} &  0.86460 & {\cellcolor{DSBlue!50}} 0.72341 & {\cellcolor{DSYellow!50}} 0.87041 & {\cellcolor{DSGreen!50}} 1.00000 \\
			\textbf{{\cellcolor{DSBlue!50}} SI-4} &  0.85704 & {\cellcolor{DSBlue!50}} 0.85878 &  0.79568 &  0.91667 \\
			\textbf{{\cellcolor{DSBlue!50}} AC-3} &  0.84544 & {\cellcolor{DSBlue!50}} 0.71865 &  0.81766 &  1.00000 \\
			\textbf{{\cellcolor{DSGrey!50}} SC-7} &  0.81730 & {\cellcolor{DSBlue!50}} 0.60155 &  0.85035 & {\cellcolor{DSGreen!50}} 1.00000 \\
			\textbf{{\cellcolor{DSBlue!50}} SI-7} &  0.80637 & {\cellcolor{DSBlue!50}} 0.56957 &  0.84953 &  1.00000 \\
			\textbf{{\cellcolor{DSBlue!50}} AC-4} &  0.77137 & {\cellcolor{DSBlue!50}} 0.57407 &  0.80457 &  0.93548 \\
			\textbf{{\cellcolor{DSBlue!50}} AC-2} &  0.76486 & {\cellcolor{DSBlue!50}} 0.59733 &  0.77417 &  0.92308 \\
			\textbf{{\cellcolor{DSGrey!50}} SI-10} &  0.73619 & {\cellcolor{DSBlue!50}} 0.45530 & {\cellcolor{DSYellow!50}} 0.89614 &  0.85714 \\
			\textbf{{\cellcolor{DSGrey!50}} AC-17} &  0.72794 & {\cellcolor{DSBlue!50}} 0.39921 &  0.78461 & {\cellcolor{DSGreen!50}} 1.00000 \\
			\textbf{{\cellcolor{DSYellow!50}} SC-28} &  0.70501 &  0.21889 & {\cellcolor{DSYellow!50}} 0.89614 &  1.00000 \\
			\textbf{{\cellcolor{DSGrey!50}} SC-4} &  0.70336 &  0.22396 & {\cellcolor{DSYellow!50}} 0.88612 & {\cellcolor{DSGreen!50}} 1.00000 \\
			\textbf{{\cellcolor{DSGrey!50}} SC-39} &  0.70105 &  0.22103 & {\cellcolor{DSYellow!50}} 0.88211 & {\cellcolor{DSGreen!50}} 1.00000 \\
			\textbf{{\cellcolor{DSGrey!50}} SC-8} &  0.69072 &  0.17602 & {\cellcolor{DSYellow!50}} 0.89614 & {\cellcolor{DSGreen!50}} 1.00000 \\
			\textbf{{\cellcolor{DSYellow!50}} SC-12} &  0.68915 &  0.17133 & {\cellcolor{DSYellow!50}} 0.89614 &  1.00000 \\
			\textbf{{\cellcolor{DSGrey!50}} SC-16} &  0.68896 &  0.17073 & {\cellcolor{DSYellow!50}} 0.89614 & {\cellcolor{DSGreen!50}} 1.00000 \\
			\textbf{{\cellcolor{DSGrey!50}} IA-4} &  0.68591 & {\cellcolor{DSBlue!50}} 0.27311 &  0.78461 & {\cellcolor{DSGreen!50}} 1.00000 \\
			\textbf{{\cellcolor{DSYellow!50}} SC-32} &  0.68562 &  0.16071 & {\cellcolor{DSYellow!50}} 0.89614 &  1.00000 \\
			\textbf{{\cellcolor{DSYellow!50}} SC-11} &  0.68562 &  0.16071 & {\cellcolor{DSYellow!50}} 0.89614 &  1.00000 \\
			\textbf{{\cellcolor{DSBlue!50}} IA-2} &  0.66727 & {\cellcolor{DSBlue!50}} 0.55055 &  0.78461 &  0.66667 \\
			\bottomrule
		\end{tabular}
	}
	\label{tab:implementation-dataset:composition:top-20-contribution-impl}
\end{table}


\begin{figure}[h]
	\centering
	\begin{minipage}[t]{0.72\columnwidth}
		\begin{subfigure}{\columnwidth}
			\includegraphics[width=\columnwidth]{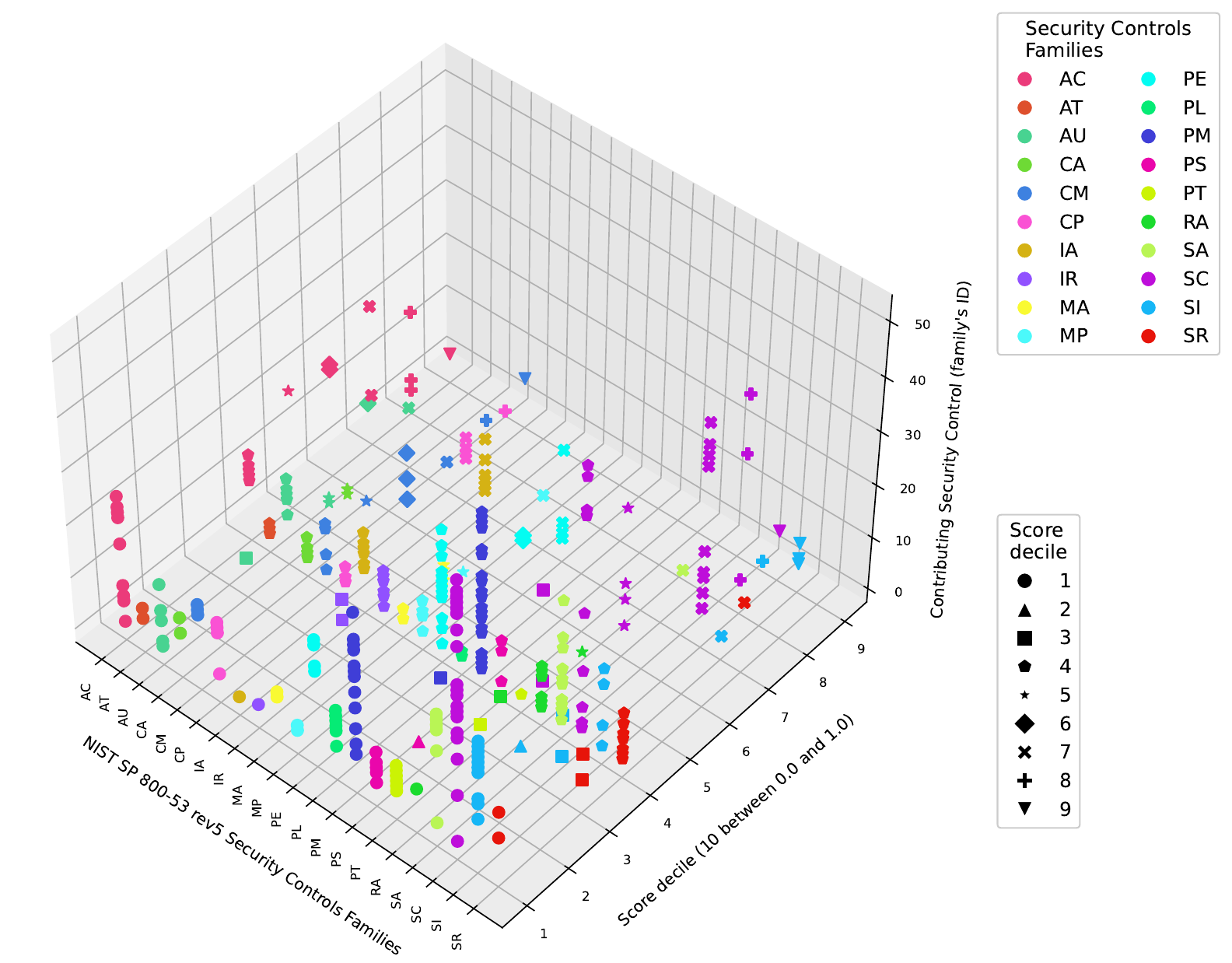} 
			\caption{All SPCFs (global score)}
			\label{fig:implementation-dataset:dataset-score-all}
		\end{subfigure}
		\medskip
		\begin{subfigure}{\columnwidth}
			\includegraphics[width=\columnwidth]{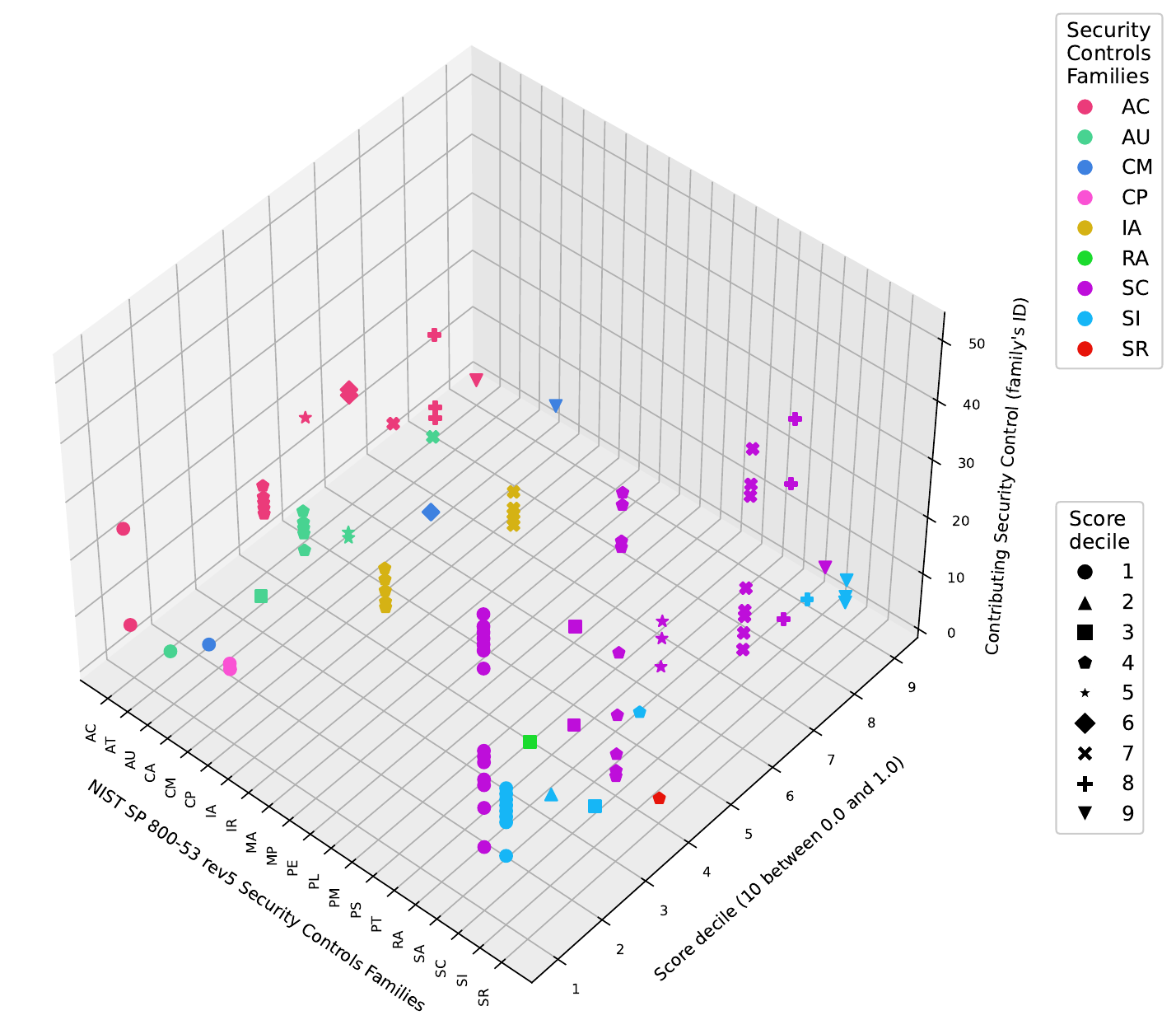}
			\caption{SW-implementable SPCFs (global score)}
			\label{fig:implementation-dataset:dataset-score-implementable}
		\end{subfigure}
	\end{minipage}
	\caption{Security controls contributing to each global score's decile}
	\label{fig:implementation-dataset:dataset-score-decils}
\end{figure}

\begin{equation}
\label{eq:implementation-dataset:generation:min-max-scaling}
mms(\vec{S}) = \frac{s-s_{min}}{s_{max}-s_{min}} ~ \forall ~ s \in \vec{S}
\end{equation}

The last block in the algorithm performs a series of calculations to (i) provide a single score from the considered academic works $DS_{2}$ \cite{Liu_Shore_Yeoh_Jiang_Zeadally_2025} and $DS_{7}$ \cite{Rahman_Rahman_Williams_2025}; (ii) generate a score for the STRIDE-LM contribution; and (iii) average all 3 scores, conditioned to any of the SW-implementation criteria ($impl_1$ or $impl_2$) being true.
First, each statistical set of metrics from $DS_2$ is first normalised applying the $mms$ function for min-max scaling, as per Eq. \eqref{eq:implementation-dataset:generation:min-max-scaling}, to ensure that each value $s$ inside meets $0 \le s \le 1$. Then, the value of each statistical metric for any given SPCSF is taken and aggregated into the intermediate $d_{8i}.score_{DS_2}$. This takes place in Algorithm \ref{alg:implementation-dataset:dataset}, line \ref{alg:implementation-dataset:dataset:line:s17}.

%
\begin{table}[t]
\caption{No. of SW-implementable (total) SPCSFs covering security dimensions}
\setlength{\tabcolsep}{1.5pt}
\centering
\scalebox{0.85}{
\begin{tabular}{lccccccc}
\toprule
\textbf{} & \multicolumn{3}{c}{\textbf{CIA triad}} & \multicolumn{4}{c}{\textbf{Extra Dimensions}} \\
\cmidrule(r){2-4} \cmidrule(l){5-8}
\small{\textbf{SPCF}} & \small{\textbf{C}} & \small{\textbf{I}} & \small{\textbf{A}} & \small{\textbf{Au}} & \small{\textbf{Ac}} & \small{\textbf{Nr}} & \small{\textbf{Pr}} \\
\cmidrule(r){1-1} \cmidrule(r){2-4} \cmidrule(l){5-8}
\midrule
\textbf{AC} & 14 (22) & 14 (20) & 2 (3) & 2 (2) & 3 (4) & 0 (0) & 2 (6) \\
\textbf{AT} & 0 (3) & 0 (3) & 0 (3) & 0 (0) & 0 (3) & 0 (0) & 0 (3) \\
\textbf{AU} & 1 (2) & 1 (1) & 2 (2) & 1 (1) & 6 (11) & 2 (2) & 0 (0) \\
\textbf{CA} & 0 (6) & 0 (6) & 0 (3) & 0 (0) & 0 (5) & 0 (0) & 0 (0) \\
\textbf{CM} & 1 (6) & 3 (14) & 1 (4) & 1 (1) & 0 (5) & 0 (0) & 0 (1) \\
\textbf{CP} & 1 (1) & 2 (4) & 2 (12) & 0 (0) & 0 (1) & 0 (0) & 0 (0) \\
\textbf{IA} & 4 (4) & 4 (4) & 0 (0) & 10 (13) & 3 (4) & 0 (0) & 0 (0) \\
\textbf{IR} & 0 (4) & 0 (3) & 0 (6) & 0 (0) & 0 (3) & 0 (0) & 0 (1) \\
\textbf{MA} & 0 (6) & 0 (6) & 0 (3) & 0 (1) & 0 (0) & 0 (0) & 0 (0) \\
\textbf{MP} & 0 (8) & 0 (5) & 0 (1) & 0 (0) & 0 (2) & 0 (0) & 0 (0) \\
\textbf{PE} & 0 (10) & 0 (6) & 0 (11) & 0 (0) & 0 (4) & 0 (0) & 0 (0) \\
\textbf{PL} & 0 (7) & 0 (7) & 0 (6) & 0 (0) & 0 (6) & 0 (0) & 0 (1) \\
\textbf{PM} & 0 (12) & 0 (12) & 0 (15) & 0 (0) & 0 (18) & 0 (0) & 0 (12) \\
\textbf{PS} & 0 (6) & 0 (7) & 0 (1) & 0 (0) & 0 (5) & 0 (0) & 0 (0) \\
\textbf{PT} & 0 (1) & 0 (0) & 0 (0) & 0 (0) & 0 (3) & 0 (0) & 0 (8) \\
\textbf{RA} & 1 (7) & 1 (7) & 0 (6) & 0 (0) & 0 (3) & 0 (0) & 0 (2) \\
\textbf{SA} & 0 (8) & 0 (16) & 0 (16) & 0 (0) & 0 (1) & 0 (0) & 0 (2) \\
\textbf{SC} & 26 (31) & 29 (33) & 8 (13) & 7 (7) & 2 (4) & 0 (0) & 3 (3) \\
\textbf{SI} & 11 (17) & 12 (17) & 5 (12) & 0 (0) & 2 (3) & 0 (0) & 4 (5) \\
\textbf{SR} & 1 (8) & 1 (10) & 1 (8) & 0 (2) & 0 (4) & 0 (0) & 0 (0) \\
\\[-0.8em] \hdashline \\[-0.8em]
\textbf{Total} & 60 (169) & 67 (181) & 21 (125) & 21 (27) & 16 (89) & 2 (2) & 9 (44) \\
\bottomrule
\end{tabular}
}
\label{tab:implementation-dataset:mapping:ciatunp}
\end{table}

%
%
\begin{figure}[h]
\centering
\begin{minipage}[t]{0.75\columnwidth}
    \begin{subfigure}{\columnwidth}
        \includegraphics[width=\columnwidth]{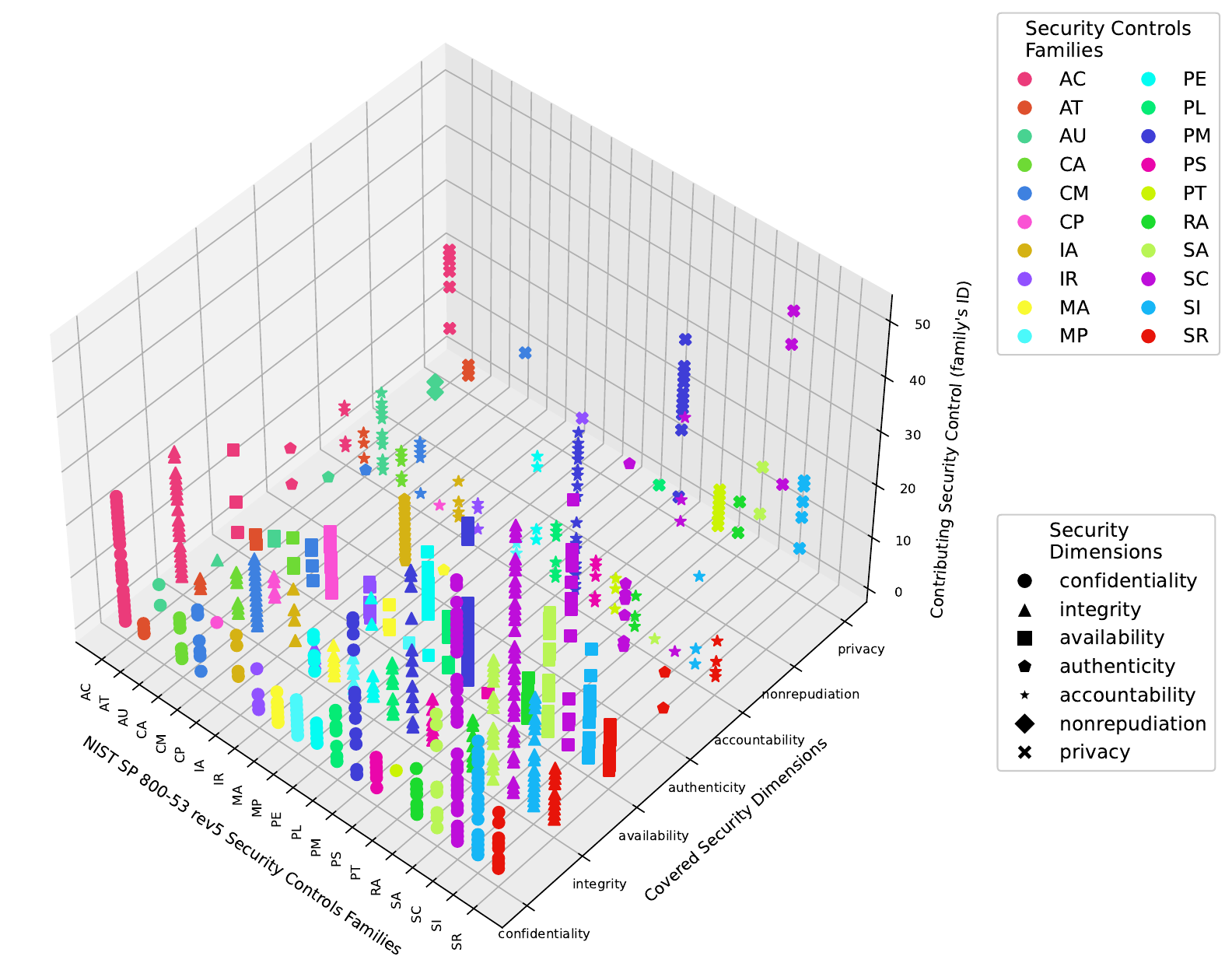}
        \caption{All SPCFs (security dimensions)}
        \label{fig:implementation-dataset:dataset-secfeats-all}
    \end{subfigure}
    \medskip
    \begin{subfigure}{\columnwidth}
        \includegraphics[width=\columnwidth]{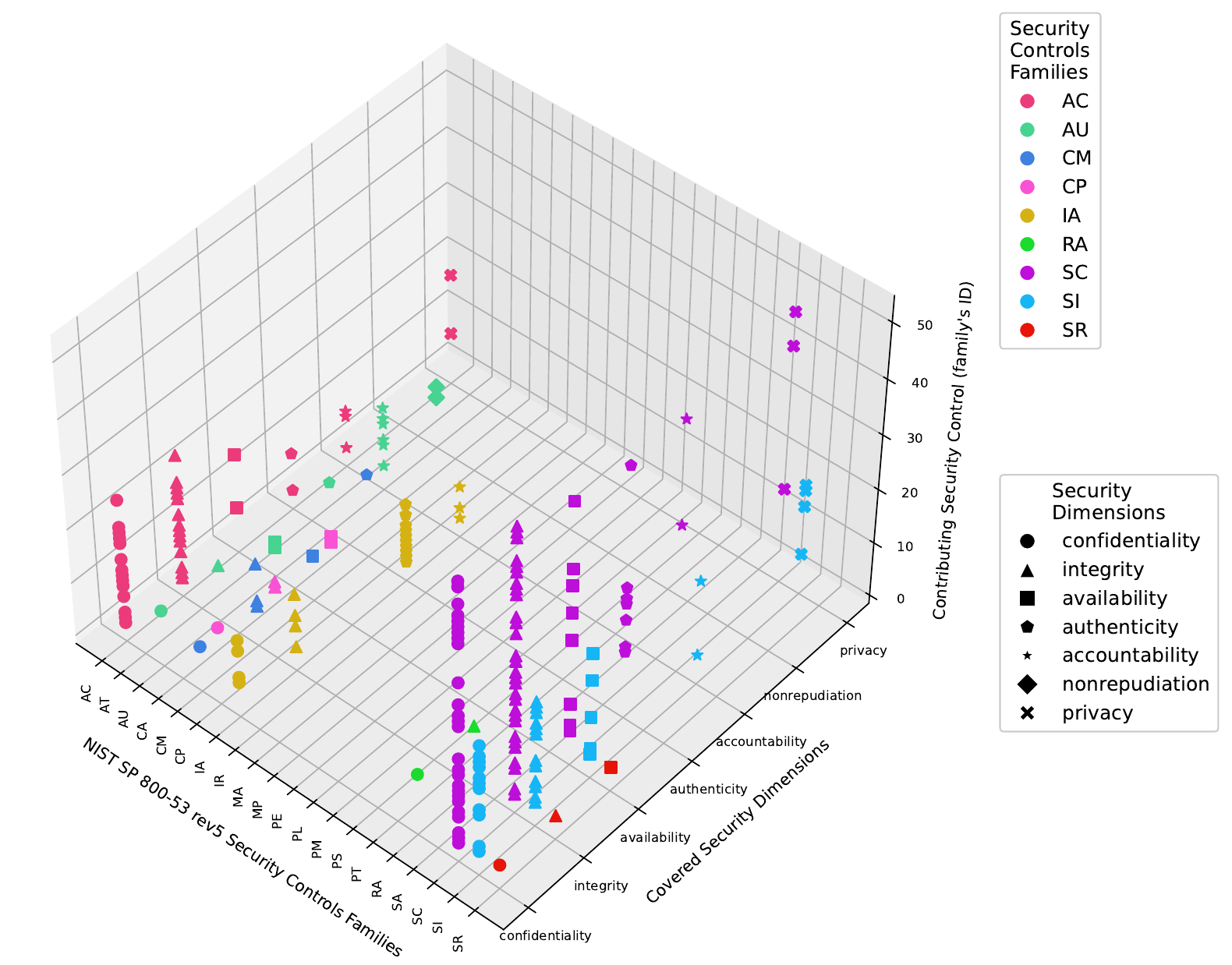}
        \caption{SW-implementable SPCFs (security dimensions)}
        \label{fig:implementation-dataset:dataset-secfeats-implementable}
    \end{subfigure}
\end{minipage}
\caption{No. of SPCFs contributing to security dimensions}
\label{fig:implementation-dataset:dataset-secfeats}
\end{figure}

%
%
\begin{table*}[t!]
  \caption{Summary with the number and statistical values for total and SW-implementable SPCFs}
  \setlength{\tabcolsep}{4pt}
  \centering
    \resizebox{\textwidth}{!}{
    \begin{tabular}{lccccccccccc}
    \toprule
    \textbf{} & \multicolumn{2}{c}{\textbf{Scope}} & \multicolumn{3}{c}{\textbf{Contribution}} & \multicolumn{5}{c}{\textbf{Score statistics}} \\
    \cmidrule(r){2-3} \cmidrule(l){4-6} \cmidrule(l){7-12}
    \small{\textbf{SPCF}} & \small{\textbf{Impl}} & \small{\textbf{Total}} & \small{\textbf{\#2 \cite{Rahman_Rahman_Williams_2025}} ($DS_{7}$)} & \small{\textbf{\#3 \cite{Liu_Shore_Yeoh_Jiang_Zeadally_2025}} ($DS_{2*}$)} & \small{\textbf{\#4} ($DS_{5}$)} & \small{\textbf{Min}} & \small{\textbf{Mean}} & \small{\textbf{Mode}} & \small{\textbf{Max}} & \small{\textbf{SD}} & \small{\textbf{IQR}} \\
    \cmidrule(r){1-1} \cmidrule(l){2-3} \cmidrule(l){4-6} \cmidrule(l){7-12}
    \textbf{AC} & 15 & 23 & 15 (23) & 11 (16) & 11 (13) & 0.06174 & 0.48610 & 0.06174 & 0.84544 & 0.23577 & 0.34124 \\
    \textbf{AT} & 0 & 5 & 0 (5) & 0 (0) & 0 (0) & 0 & 0 & 0 & 0 & 0 & 0 \\
    \textbf{AU} & 10 & 15 & 10 (15) & 5 (10) & 6 (9) & 0.07143 & 0.36807 & 0.07143 & 0.66181 & 0.14086 & 0.08928 \\
    \textbf{CA} & 0 & 8 & 0 (8) & 0 (6) & 0 (0) & 0 (0) & 0 & 0 & 0 & 0 & 0 \\
    \textbf{CM} & 3 & 14 & 3 (14) & 2 (11) & 1 (4) & 0.05357 & 0.48267 & 0.05357 & 0.86460 & 0.33278 & 0.40552 \\
    \textbf{CP} & 2 & 12 & 2 (12) & 0 (8) & 0 (6) & 0.05357 & 0.05357 & 0.05357 & 0.05357 & 0 & 0 \\
    \textbf{IA} & 10 & 13 & 10 (13) & 10 (13) & 5 (6) & 0.30915 & 0.49091 & 0.30915 & 0.68591 & 0.16686 & 0.33852 \\
    \textbf{IR} & 0 & 9 & 0 (9) & 0 (8) & 0 (0) & 0 & 0 & 0 & 0 & 0 & 0 \\
    \textbf{MA} & 0 & 7 & 0 (7) & 0 (1) & 0 (4) & 0 & 0 & 0 & 0 & 0 & 0 \\
    \textbf{MP} & 0 & 8 & 0 (8) & 0 (2) & 0 (5) & 0 & 0 & 0 & 0 & 0 & 0 \\
    \textbf{PE} & 0 & 22 & 0 (22) & 0 (18) & 0 (7) & 0 & 0 & 0 & 0 & 0 & 0 \\
    \textbf{PL} & 0 & 8 & 0 (8) & 0 (2) & 0 (0) & 0 & 0 & 0 & 0 & 0 & 0 \\
    \textbf{PM} & 0 & 32 & 0 (32) & 0 (20) & 0 (0) & 0 & 0 & 0 & 0 & 0 & 0 \\
    \textbf{PS} & 0 & 9 & 0 (9) & 0 (3) & 0 (1) & 0 & 0 & 0 & 0 & 0 & 0 \\
    \textbf{PT} & 0 & 8 & 0 (8) & 0 (1) & 0 (1) & 0 & 0 & 0 & 0 & 0 & 0 \\
    \textbf{RA} & 1 & 9 & 1 (9) & 1 (8) & 0 (0) & 0.22588 & 0.22588 & 0.22588 & 0.22588 & 0 & 0 \\
    \textbf{SA} & 0 & 17 & 0 (17) & 0 (11) & 0 (2) & 0 & 0 & 0 & 0 & 0 & 0 \\
    \textbf{SC} & 40 & 47 & 40 (47) & 19 (21) & 19 (24) & 0.03436 & 0.35038 & 0.05357 & 0.81730 & 0.25662 & 0.57374 \\
    \textbf{SI} & 15 & 22 & 15 (22) & 5 (9) & 5 (7) & 0.04447 & 0.29980 & 0.05357 & 0.87633 & 0.32869 & 0.51410 \\
    \textbf{SR} & 1 & 12 & 1 (12) & 1 (8) & 0 (3) & 0.35899 & 0.35899 & 0.35899 & 0.35899 & 0 & 0 \\
    \bottomrule
    \end{tabular}
    } 
  \label{tab:implementation-dataset:composition:summary-stats}
\end{table*}

Then, values from $DS_{7}$ are incorporated.
It is worth noting that, before this process can start, the publicly available dataset for $DS_{7}$ \cite{Rahman_Rahman_Williams_2025} seemed to require minor modifications in the publicly provided scripts in order to generate the full dataset. To that end, both the "main" and "utils" files were modified to (i) properly compute the severity centrality from the given patterns; (ii) use the latest file names; and (iii) extract all values from the clustering method.
With that covered, all but one effectiveness custom metrics ($DS_{7}.cm$) provided by $DS_{7}$ \cite{Rahman_Rahman_Williams_2025} are used; which makes it 6 out of the original 7 metrics. Such metric is the Tactic Coverage (TAC), which provides granular scalars per SPCF that are already considered in the already used mean TAC (MTAC) metric. These values are handled similarly to the $DS_{2}$ ones: first normalised using the $mms$ function, then averaged (Algorithm \ref{alg:implementation-dataset:dataset}, line \ref{alg:implementation-dataset:dataset:line:s18}).
Besides the custom metrics, this dataset incorporates the NIST-defined baseline and priority metrics ($DS_{7}.nm$) per SPCF; which are used to prioritise the implementation of a given control depending on the baseline to cover (i.e. lower must be implemented first) and priority, both of which range from 0 to 4. Each of these values per SPCSF is averaged and normalised simply considering its maximum value (Algorithm \ref{alg:implementation-dataset:dataset}, line \ref{alg:implementation-dataset:dataset:line:s19}).

Then, both custom ($DS_{7}.cm$) and NIST ($DS_{7}.nm$) metrics are averaged in Algorithm \ref{alg:implementation-dataset:dataset}, line \ref{alg:implementation-dataset:dataset:line:s20}. It is worth noting that both the CR ($cm$) and the NIST priority ($nm$) metrics follow essentially different approaches than other metrics, since smaller values denote higher importance there; and must thus be transformed to convey a similar approach whilst still keeping their significance.\\
The third and final score relates to the contribution towards the mitigation of the STRIDE-LM threat vectors, previously introduced from $DS_{5}$. In Algorithm \ref{alg:implementation-dataset:dataset}, line \ref{alg:implementation-dataset:dataset:line:s21}; each SPCSF is iterated to count how many of its main (e.g. AC-9) or its control enhancements (e.g. AC-9(3)) contribute towards mitigating the STRIDE-LM's threat vectors. That number is divided by the total number of control enhancements available per SPCSF, which is programmatically obtained from the OSCAL catalogue $DS_{1}$ -- and adding one more to count also the SPCSF itself. This provides a measure of the effectiveness of a given SPCSF into mitigating such threats.\\
Finally, the three scores above are averaged and then the final score per SPCSF is calculated, conditioned on whether either one or the other implementation criteria ($p_{i}$, considering both $impl_{1}$, $impl_{2}$) is met.
The generated dataset, correlated from the previous source, is available at $DS_{8}$.

\subsection{Composition analysis}
\label{sec:implementation_dataset:composition}

Table \ref{tab:implementation-dataset:composition:top-20-contribution-impl} enumerates the top 20 SPCSFs that are SW-implementable, ordered from highest to lowest and according to the total score calculated in this work (Algorithm \ref{alg:implementation-dataset:dataset}, line \ref{alg:implementation-dataset:dataset:line:s23}).
It shows each SPCSF, its total score and the degree of contribution from each of the contemplated works and datasets.
Additionally, a colour code is used to represent if any given SPCSF is also present in the top 20 ranked controls of each considered dataset by showing its colour: blue for \#2 \cite{Rahman_Rahman_Williams_2025}, yellow for \#3 \cite{Liu_Shore_Yeoh_Jiang_Zeadally_2025} and green for \#4; as well as gray for contributions from 2 or more datasets.
Those without colour indicate that these are still contributed from these other datasets, yet further down in their ranked list.
There, the highest contributing SPCFs are SC, AC and SI with an average of 9.25, 4.25 and 2.5 contributed SPCSF items, respectively.
Some SPCF categories (AT, CA, IR, MA, MP, PE, PL, PM, PS, PT, SA) are contributed only by non-SW-implementable SPCSF items that relate to awareness, assessment, authorisation, monitoring, incident response and management, among others.
The rest of the categories (AC, AU, CM, CP, IA, RA, SC, SI, SR) are considered instead; contributing to access control, accountability, configuration management and protection efforts, among others.
More information can be found in the dataset's ancillary files, such as the top 20 total (i.e. SW-implementable or not) SPCSF --also in Fig. \ref{fig:implementation-dataset:dataset-score-all}-- and the number of contributing SPCSF controls per SPCF and their statistical metrics and spread.
Fig. \ref{fig:implementation-dataset:dataset-score-implementable} complements this by placing each SPCSF into each of the 10 deciles (in the Y-axis), depending on its total calculated score.

Besides the score distribution, Table \ref{tab:implementation-dataset:mapping:ciatunp} and Fig. \ref{fig:implementation-dataset:dataset-secfeats} show the number and IDs of the SPCSFs contributing to the security dimensions, considering both the CIA triad with confidentiality (C), integrity (I) and availability (A); along with authenticity (Au), accountability (Ac), non-repudiation (Nr) and privacy (Pr).
Fig. \ref{fig:implementation-dataset:dataset-secfeats} renders the contribution of all 20 (Fig. \ref{fig:implementation-dataset:dataset-secfeats-all}) and the 9 SW-implementable (Fig. \ref{fig:implementation-dataset:dataset-secfeats-implementable}) SPCFs.

Finally, Table \ref{tab:implementation-dataset:composition:summary-stats} summarises some relevant data.
From the total 300 SPCSFs, 97 are considered SW-implementable.
Regarding the contribution of the considered works towards these SW-implementable SPCSFs, \#2 \cite{Rahman_Rahman_Williams_2025} is the most comprehensive, as its measurements yield $> 0$ for 96 out of these 97; whereas \#3 \cite{Liu_Shore_Yeoh_Jiang_Zeadally_2025} provides 53 and \#4 considers 47, as per the "Contribution" columns of Table \ref{tab:implementation-dataset:composition:summary-stats}.
The minimum, average, mode, maximum, SD and IQR statistical metrics are provided right after for the SW-implementable SPCSF items belonging to each SPCF category.
As observed in Fig. \ref{fig:implementation-dataset:dataset-score-implementable}, the SC, AC and SI SPCFs are the most prevalent in the highest deciles (roughly around 6 to 9) for the top 20 SPCSFs from Table \ref{tab:implementation-dataset:composition:top-20-contribution-impl}.

\subsubsection{Key findings}

From the analysis of the dataset's content, the following considerations are highlighted:
\begin{itemize}
    \item The highest scored SW-implementable SPCSFs lie in the SI, SC, IA, AC and CM SPCFs (deciles 6 to 9); with the majority coming from SC. Unsurprisingly, these cover data (e.g. configuration) integrity, protect system communications and control access to data for confidentiality. This is consistent with $DS_{7}$ \cite{Rahman_Rahman_Williams_2025}, whose top 10 controls (per $TEC$ metric) consist only of these SPCSFs and which covers also 85\% of the whole dataset's upper $TEC$.
    \item Considering the SW-implementable SPCSFs, the best covered security dimensions are confidentiality and integrity, followed by availability and authenticity; whereas the least covered ones are privacy, accountability and non-repudiation. The SW-implementable SPCFs with higher overall coverage are SC, AC, IA and SI.
    \item When considering total SPCFs, some SPCFs cover well a specific security dimension that cannot be covered by contained SW-implementable SPCSFs (e.g. the total number of SPCSFs in PE and PM covers availability well), but the implementable SPCSFs that can best cover it belong to SC and SI instead --- yet with a high scoring variability. This highlights the direct impact of non-SW-implementable SPCSFs (on planning, physical and personal data protection) to ensure data availability and privacy. On the other hand, these non-SW-implementable SPCFs usually score low in the aggregated $DS_{7}$ \cite{Rahman_Rahman_Williams_2025} data and often also in the final score.
    \item The IA and SC SPCFs (identification and authentication \& system and information integrity) contribute towards authenticity (with scores ranging between 0.8-0.5 and 0.5-0.1, respectively); no matter whether considering total or SW-implementable SPCSFs.
\end{itemize}


\section{Security recommender}
\label{sec:implementation_model}

The knowledge-based multi-agent framework, explained below, leverages the previous dataset $DS_{8}$ to encode as KG in the MAID model $\mathcal{M}$ that recommends SPCSFs.

\subsection{Background}
\label{sec:implementation_model:background}

MAIDs \cite{Koller_Milch_2003} extend from Influence Diagrams (IDs) \cite{Howard_Matheson_2005}, and IDs extend themselves from BNs \cite{Pearl_1986}.
On the one hand, BNs $B = (G, \mathbb{P})$ are represented with (i) a Directed Acyclic Graph (DAG) $G = (V, E)$ encoding random variables (events) $X_{i}$ as vertices and the influence across these as directed edges; and (ii) Conditional Probability Distributions (CPD), encoding the probability $\mathbb{P}$ of an event $X_{i}$ occurring based on that of its parent nodes $Pa(X_{i})$, i.e. $\mathbb{P}(X_{i}|Pa(X_{i}))$.
On the other hand, IDs consider (i) the nodes $V$ in BNs, representing $X_{i}$, as chance nodes $C$, and also include two new types of nodes: (ii) decision variables, for the agents to select as part of their strategy; and (iii) utility variables, to represent the agent's preferences. Thus, the resulting graph could be formalised as $G = (C, D, U)$ and can be used to more easily analyse the probabilistic dependencies across $X_{i}$ ($C$) and $D$.
A MAID $\mathcal{M} = (G, \theta)$ \cite{Hammond_Fox_Everitt_Abate_Wooldridge_2021,Everitt_Ortega_Barnes_Legg_2022} consists of (i) a MAID $G = (\mathcal{A}, V, E)$, where $A$ is the set of agents and $(V, E)$ is a graph defined by nodes $V = (C, D, U)$ to represent chance, decision and utility nodes, respectively, and edges $E = V \times V$; along with (ii) a parameterisation $\theta$ that contains, per node in $V$, a finite domain $\text{Dom}(X_{i}) \in \mathbb{R} ~\forall~ X_{i} \in V$; a real domain $\text{Dom}(X_{i}) \in \mathbb{R} ~\forall~ X_{i} \in U$; and CPDs for chance $C$ and utility $U$ nodes.
$\text{Dom(}X_{i}\text{)}$ will be defined, from now on, as $\mathbb{D}_{X_{i}}$.
The parents of a decision node $Pa(D)$ ($C$ or other $D$) contain the information set available (observable) for $D$ to decide; utility nodes $U$ usually have no children and are probabilistically determined by their parents $Pa(U)$ ($C$ or $D$) \cite{Koller_Milch_2003}; and its DAG and properties offer advantages regarding complexity and explainability regarding the traditional Extensive Form Game representations \cite{Hammond_Fox_Everitt_Abate_Wooldridge_2021}.

\subsection{Problem and model definition}
\label{sec:implementation_model:problem-model-definition}

This work intends to recommend minimal SPCSFs to fulfil the requirements indicating the degree of coverage of the 7 considered security dimensions.
To that end, a MAID $\mathcal{M}$ models the security dimensions as agents $A$, encodes SPCSFs' coverage as chance nodes $C$, uses SPCSFs' domain in the decision nodes $D$ and formally defines the utility function for agents $A$ as the utility nodes $U$.
Chance nodes internally encode each SPCSF's SW-implementability and score from $DS_{8}$; whereas utility nodes embody the utility formulation for each agent, formalising its behaviour from its relation to other agents (which can tend from collaboration to competition) and the contribution received from each SPCSF, among others. 
Edges between these represent how much a SPCSF contributes to cover an agent.
Fig. \ref{fig:implementation-model:maid-example} depicts a simple MAID representing C, D and U nodes as circles, squares and diamonds; respectively. Dotted edges denote the information links or information available to an agent/decision (e.g. $Pa(D) \rightarrow D$), whereas solid edges indicate the dependency over other nodes into the calculation of the utility or payoff of the agent.

\begin{figure}[hbtp]
    \centering
    \begin{influence-diagram}
      \node (help) [draw=none] {};
      \node (P1) [above = of help, decision, player1] {$D_1$};
      \node (U1) [right = of help, utility, player1] {$U_1$};
      \node (P2) [below = of help, decision, player2] {$D_2$};  
      \node (U2) [left = of help, utility, player2] {$U_2$};
      \node (SFi) at (U2|-P1) {$SF_i$};
      \edge {SFi,P1,P2} {U2};
      \node (SFj) at (U1|-P1) {$SF_j$};
      \edge {SFj,P1,P2} {U1};
      \edge {P1,P2} {U1};
      \edge[information] {SFi} {P1};
      \edge[information] {P1} {P2};
      \node (PN) [right = of SFj, decision, player4] {$D_N$};
      \node (UN) [below = of PN, utility, player4] {$U_N$};
      \edge[information] {SFj} {P1};
      \edge[information] {SFj} {PN};
      \edge {PN} {UN};
    \end{influence-diagram}

\caption{MAID sample with C, D and U nodes}
\label{fig:implementation-model:maid-example}
\end{figure}

%
%
\begin{figure*}
    \centering
    \includegraphics[width=0.95\textwidth]{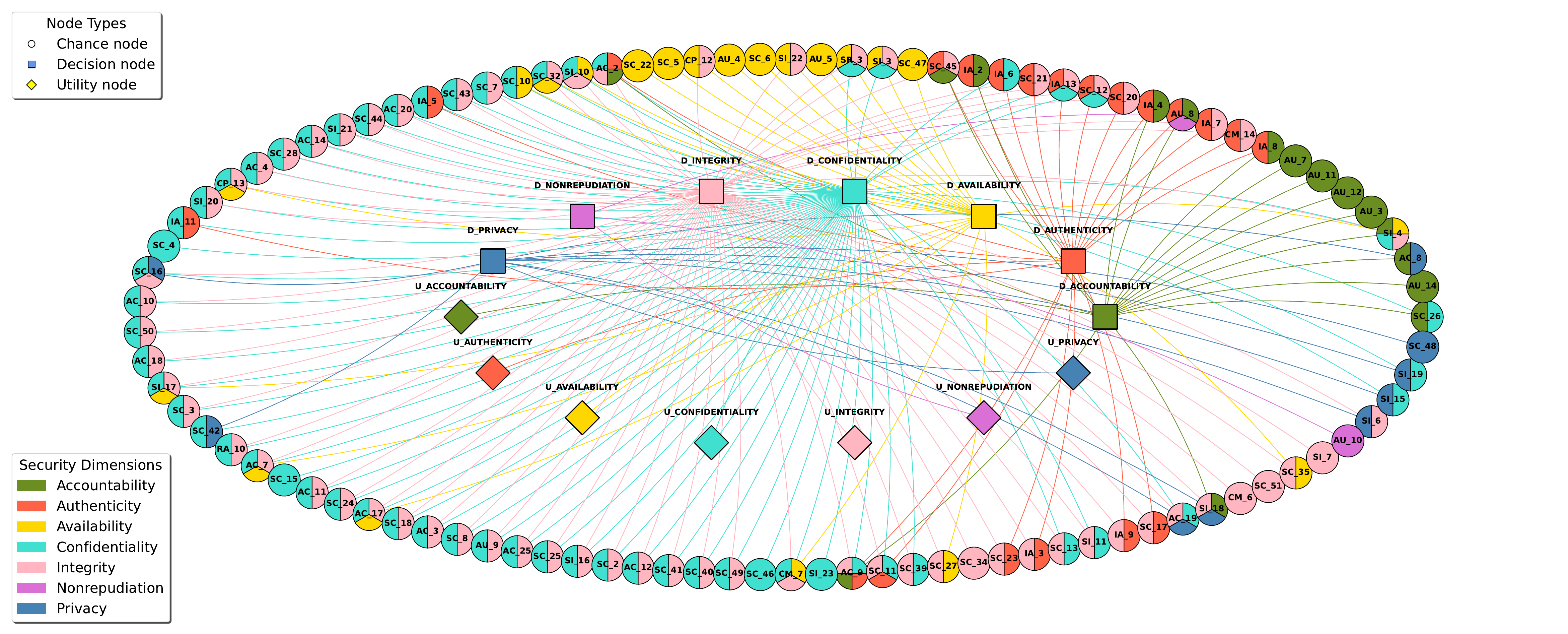}
    \caption{Full MAID showing the contributions of each of the 97 SW-implementable SPCSF to each agent}
    \label{fig:implementation-model:dataset-in-model}
\end{figure*}

\begin{figure}[hbtp]
    \centering
    \includegraphics[width=0.95\columnwidth]{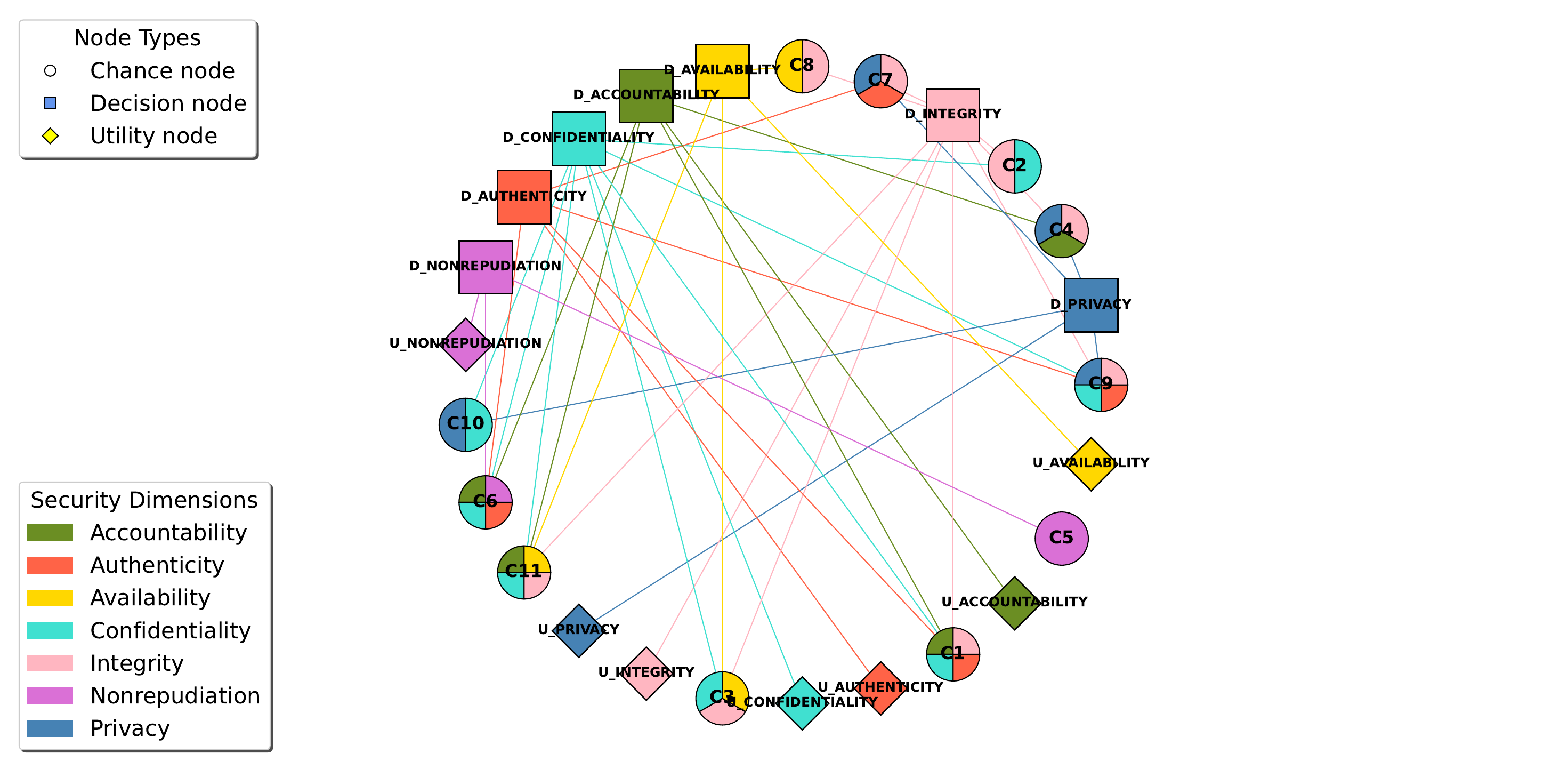}
    \caption{Clustered MAID with contribution per SPCSF to $A$ (k=11 clusters)}
    \label{fig:implementation-model:dataset-in-model-clustered}
\end{figure}

Two modes $M$ of MAID are proposed to model this problem: Fig. \ref{fig:implementation-model:dataset-in-model} shows the full MAID and Fig. \ref{fig:implementation-model:dataset-in-model-clustered} depicts the clustered MAID.
These have $|V| = 111$, $|E| = 203$ and $|V| = 25$, $|E| = 39$ vertices and edges, respectively.
Edges $E$ connect $C$ with $D$ and $D$ with $U$.
Vertices $V$ directly represent (full $M$) or indirectly aggregate (clustered $M$) the 97 SW-implementable SPCSF from the dataset, along with 7 decision and utility nodes to represent the agents.
Each $C$ node is coloured based on the $D$ and $U$ nodes it contributes and relates to.
Each proposed MAID  $\mathcal{M}$ is used for a multi-player, non-zero-sum, simultaneous game with utilities that consider both incentives and penalties.
$\mathcal{M}$ can be used to find suitable joint, optimal cross-agent policy or decision profile $\sigma^{*}$ that cannot be improved by any agent's unilateral deviation; which acts as an approximate equilibrium.

It is important to note that, while obtaining exact optimal equilibria for a full MAID is intractable using exact calculations methods, the clustered MAID heavily brings the computational complexity down; making it still possible to obtain an approximate equilibrium.
Both $M$ are solved in this work with no-regret learning dynamics rather than exact calculations to ensure proper scalability; specifically using the Hedge / Exponential Weights Algorithm (EWA) \cite{Freund_Schapire_1997} and the Exponential-weight algorithm for Exploration and Exploitation (EXP3) \cite{Auer_Cesa-Bianchi_Freund_Schapire_2002} methods. 
Section \ref{sec:implementation_model:complexity-limitations} explains the rationale for this approach, including empirical validation and theoretical complexity analysis.

%
%

\begin{figure}
    \centering
    \includegraphics[width=0.95\columnwidth]{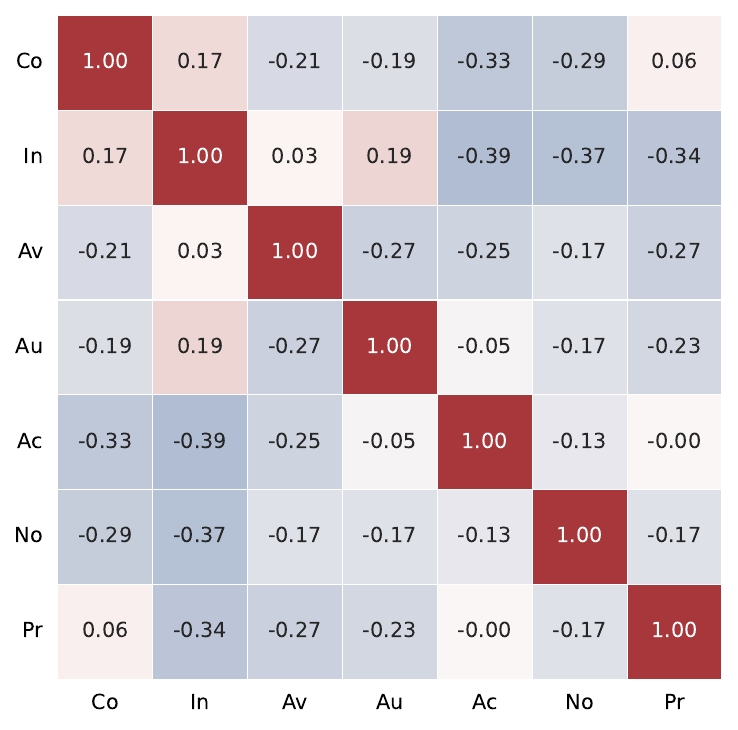}
    \caption{Correlation across agents for clustered MAID}
    \label{fig:implementation-model:dataset-correlations}
\end{figure}

\subsection{Utility definition}
\label{sec:implementation_model:utility-definition}

The utility function is constructed on-the-fly during the MAID generation from the dataset $DS_{8}$.
The utility $U_{a}(\sigma)$ for an agent $a$ evaluated for a joint decision profile $\sigma$ is defined in Eq. \eqref{eq:implementation-model:utility}.

\begin{equation}
\label{eq:implementation-model:utility}
U_{a}(\sigma) = \underbrace{\Big(S_{a}(\sigma) + Q_{a}(\sigma)\Big)}_{\text{Incentives (}\Delta_{a}(\sigma)\text{)}} - \underbrace{\Big(P^{p}_{a}(\sigma) + P^{s}_{a}(\sigma)\Big)}_{\text{Penalties (}\Gamma_{a}(\sigma)\text{)}}
\end{equation}

The utility encodes the motivation (objective) per agent $a$, which is to be maximised as per Eq. \eqref{eq:implementation-model:utility:optimisation-agent}. To do so, it evaluates $U_{a}$ to determine its optimal strategy $\sigma_{a}^*$ against the strategies of other agents $\sigma_{-a}$.
The objective of the system, in Eq. \eqref{eq:implementation-model:utility:optimisation-system}, is to find the optimal joint strategy (discrete) profile $\sigma^*$ or approximate equilibrium; where incentives and penalties are balanced across agents.

\begin{subequations}
\label{eq:implementation-model:utility:optimisation}
\begin{align}
    \sigma_{a}^{*} &= \argmax_{\sigma_{a} \in \mathbb{D}} ~ U_{a}(\sigma_{a}, \sigma_{-a}) \label{eq:implementation-model:utility:optimisation-agent} \\[0.5em]
    \sigma^* &= \{ (a, \sigma_{a}^*) \mid a \in A \} \label{eq:implementation-model:utility:optimisation-system}
\end{align}
\end{subequations}

In practice, $U_{a}$ must be able to guide the agent $a \in A$ to explore the minimal number of SPCSFs that can best serve each security requirement $\tau_{a}$ per agent $a$ and maximise tailored coverage for its associated security dimension.
It considers contributions providing both incentives $\Delta_{a}(\sigma)$ and penalties $\Gamma_{a}(\sigma)$, as per Eq. \eqref{eq:implementation-model:utility:terms}.
The incentives group
(i) the synergy $S_{a}(\sigma)$ between any pair of agents; and
(ii) the overall quality $Q_{a}(\sigma)$ for a chance node in the MAID --- where chance nodes are either a SPCSF cluster or item, depending on the type of MAID $\mathcal{M}$, and prioritising controls with higher coverage towards agents.
The penalties group
(iii) the provisioning penalty $P^{p}_{a}(\sigma)$; and
(iv) the sparsity penalty $P^{s}_{a}(\sigma)$ to maximise accuracy and minimise the number of SPCSFs in use, respectively.

\begin{subequations}
\label{eq:implementation-model:utility:terms}
\begin{align}
    S_{a}(\sigma) &= \sum_{-a \in A \setminus \{a\}} C_{-a}(\sigma) \cdot \rho_{a,-a} \cdot w_s \label{eq:implementation-model:utility:util_synergy} \\[0.5em]
    Q_{a}(\sigma) &= \sum_{\omega \in \Omega} p(\omega, \sigma) \cdot q(\omega, a) \label{eq:implementation-model:utility:util_quality} \\[0.5em]
    P^{p}_{a}(\sigma) &= \lambda(E_{a}, \tau_{a}) \cdot E_{a}^2 \label{eq:implementation-model:utility:util_prov_penalty} \\[0.5em]
    P^{s}_{a}(\sigma) &= \sum_{\omega \in \Omega} p(\omega, \sigma) \cdot \frac{P^{s}_{max}}{|\Omega|} \label{eq:implementation-model:utility:util_sparsity_penalty}
\end{align}
\end{subequations}

As a scaling factor in $S_{a}(\sigma)$ via $C_{a}(\sigma)$ in Eq. \eqref{eq:implementation-model:utility:coverage}, $Q_{a}(\sigma)$, and $P^{s}_{a}(\sigma)$; the weight $p(\omega, \sigma) \in [0, 1]$ indicates how likely each SPCSF $\omega$ is to contribute to cover a given agent under the examined joint policy profile $\sigma$. That is, it represents the need of an SPCSF to satisfy each requirement.
It is worth noting that this weight is obtained deterministically and does not follow a known probability distribution. In particular, Algorithm \ref{alg:implementation-model:maid-generation} maps the agents' possible choices from $\mathbb{D}_{A}$ to numerical requirement weights denoting the required coverage of each agent's associated security dimension. $p(\omega, \sigma)$ takes the maximum weight across all $a \in A$ covered by each SPCSF $\omega$.
Besides this, other weights defined for the incentive ($w_{s}$) and penalty ($W_{u}, \theta_{neg}, P^s_{max}$) utility terms were fine-tuned after manual sensitivity testing. It had been observed that $w_{s} > 0.1$ hoarded SPCSFs to increase synergy at the cost of over-provisioning; that small $W_{u}$ were not penalising enough the under-provisioning and were increased orders of magnitude; that $\theta_{neg}$ should only consider for negative correlation under a minimal threshold (analysed from Fig. \ref{fig:implementation-model:dataset-correlations}) and that $P^s_{max}$ was calculated to be a bit larger than the value delivered by $S_{a}(\sigma)$, yet considerably smaller than that given by $P^{p}_{a}(\sigma)$.
Finally, the $C_{max}$ is used in $S_{a}(\sigma)$ as the upper boundary to limit coverage when aggregating contributions.

\subsubsection{Incentive $S_{a}(\sigma)$: dimension synergy}
\label{sec:implementation_model:dimension-synergy}

The synergy across agents $S_{a}(\sigma)$, in Eq. \eqref{eq:implementation-model:utility:util_synergy}, rewards agent $a$ when the selected SPCSFs provide simultaneous coverage to any other agent $-a \in A \setminus \{a\}$.
This models the nature of collaborative vs competitive coverage; given that when covering an agent $a$, the coverage of another agent $-a$ can be either augmented or diminished.
This is obtained from the correlation matrix $R$ across agents and fosters complementary coverage relationships, measured through the cross-dimensional coverage $C_{-a}(\sigma)$ and the dimensional correlation $\rho_{d,-a}$.
The total synergy is scaled by weight $w_s$ (here, 0.1) 
to avoid dominating the utility function that guides the exploration.

\begin{equation}
\label{eq:implementation-model:utility:coverage}
C_{a}(\sigma) = \min\left(C_{max}, \sum_{\omega \in \Omega} p(\omega, \sigma) \cdot v_{\omega, a}\right)
\end{equation}

Eq. \eqref{eq:implementation-model:utility:coverage} denotes the cross-dimensional coverage $C_{a}(\sigma)$ per dimension $a \in A$.
It is worth noting that $C_{-a}(\sigma)$ is derived by evaluating $C_{a}(\sigma)$ identically, yet for other dimensions $-a$.
$C_{a}(\sigma)$ aggregates the contribution $v_{\omega, a} \in [0,1]$ of SPCSF (chance nodes $\mathcal{N}_\omega$, or simply $\omega \in \Omega$ in $\mathcal{M}$), towards each agent $a$; which is multiplied by the weight $p(\omega, \sigma)$.
In order to account for multiple SPCSFs covering a given $a$ (even exceeding 100\% coverage), the upper bound is set to $C_{max}$ (here, 1).
To compute the final synergy $S_{a}(\sigma)$, the dimensional correlation $\rho_{a,-a} \in [-1, 1]$ expresses the synergy between dimensions $a$ and $-a$, as obtained from $R$.
Fig. \ref{fig:implementation-model:dataset-correlations} provides an example of $R$.
$\rho_{a,-a}$ acts as a weight: (i) if positive, agents are complementary and it yields an incentive for SPCSFs covering also other dimensions; and (ii) otherwise, the agents are in conflict; likely hindering coverage of the current agent, and thus discourages the selection of such SPCSFs.

\subsubsection{Incentive $Q_{a}(\sigma)$: SPCSF quality}
\label{sec:implementation_model:spcsf-quality}

The quality incentive $Q_{a}(\sigma)$, in Eq. \eqref{eq:implementation-model:utility:util_quality}, rewards an agent $a$ that explores SPCSFs with high alignment on both the security baseline and its total score; where the baseline alignment is favoured.
This can act as a tie breaker when multiple SPCSFs offer equivalent coverage.
To obtain this value, it considers the weight $p(\omega, \sigma)$ and multiplies it to the evaluation function $q(\omega, a)$ - which combines a weighted sum of (i) the NIST SP 800-53 baseline, aligning its significance to that defined in the joint profile $\sigma$ for $a$; and (ii) the total score; both available in $DS_{8}$ for each SPCSF.

\subsubsection{Penalty $P^{p}_{a}(\sigma)$: under- and over-provisioning}
\label{sec:implementation_model:provisioning-penalty}

The recommendation of suitable SPCSFs to meet the security requirements $\tau_{a}$ per agent $a$ can be seen as a resource provisioning problem (i.e. provisioning SPCSFs to cover requirements on security dimensions).
This can incur two undesirable situations: under-provisioning (recommending fewer SPCSFs than needed) and over-provisioning (recommending redundant SPCSFs).
From a conservative cybersecurity provisioning approach, under-provisioning increases the vulnerability of the environment; whereas over-provisioning wastes resources (e.g. time or adoption effort) - making the latter preferable to an exposed attack surface.
The provisioning penalty $P^{p}_{a}$, in Eq. \eqref{eq:implementation-model:utility:util_prov_penalty}, is introduced to discourage agents from exploring the decision space in ways that can more easily lead to such conditions.
To identify both conditions, the difference (distance) between the provided coverage $C_{a}(\sigma)$ and requested coverage $\tau_{a}$ of an agent $a$ is calculated.
This is captured by the error $E_{a} = C_{a}(\sigma) - \tau_{a}$, with a negative value indicating under-provisioning; and otherwise, over-provisioning.
The squared error $E_{a}^{2}$ is multiplied by a dynamic weight function $\lambda(E_{a}, \tau_{a})$ that defines a proper coefficient depending on conditions of under-provisioning ($E_{a} < 0$) or over-provisioning ($E_{a} \geq 0$). In the first case, it uses $W_{u}$ (here, 50) to heavily penalise this case; whereas in the second case it uses $W_{o}$, obtained from the correlation matrix $R$ across agents and from a multiplier proportional to the maximum over-provisioning (i.e. exceeding coverage regarding security requirements), which is bounded between $0.1$ and $1.0$.

\begin{equation}
\label{eq:implementation-model:utility:term:penalty:overprovision}
\begin{gathered}
    \mathcal{C}_{neg} = \{ \rho_{i,j} \in R_{tri} \mid \rho_{i,j} < \theta_{neg} \} \\[0.5em]
    V = \begin{cases} 
        \frac{\min(\mathcal{C}_{neg}) + \text{mean}(\mathcal{C}_{neg})}{2} & \text{if } \mathcal{C}_{neg} \neq \emptyset \\
        0 & \text{otherwise}
    \end{cases} \\[0.5em]
    W_{o} = \max\Big(w_{min}, \min\big(w_{max}, w_{base} + \gamma \cdot V\big)\Big)
\end{gathered}
\end{equation}

$W_{o}$, in Eq. \eqref{eq:implementation-model:utility:term:penalty:overprovision}, is dynamically calculated to avoid agents from getting a direct penalty when the available SPCSFs in the dataset lead to over-provisioning.
It is calculated dynamically from the conflicting dimensions $\mathcal{C}_{neg}$, using the upper triangular elements $R_{tri}$ of the security dimensions' correlation matrix $R$ whose negative correlation is below a negative correlation threshold $\theta_{neg}$ (here, -0.05).
It is worth noting that $R$ is also used to calculate $S_{a}$.
From that, $V$ is calculated from the minimal and average values of these conflicts.
Then, $W_{o}$ increases $V$ by $\gamma$ and keeps it bound between $w_{min}$ and $ w_{max}$ around a baseline $w_{base}$.
In this case, the chosen values for these parameters were 2, 0.5, 2 and 2, respectively.
Negative correlations (seen as conflicts between agents) incur less over-provisioning; and thus, $W_{o} \propto 1 / V_{i}$.
In these cases, a higher $V$ denotes that covering multiple dimensions becomes more complicated and likely require more overlapping SPCSFs; thus $W_{o}$ must be more lenient.

\subsubsection{Penalty $P^{s}_{a}(\sigma)$: sparsity}
\label{sec:implementation_model:sparsity-penalty}

The sparsity penalty $P^s_{a}(\sigma)$, in Eq. \eqref{eq:implementation-model:utility:util_sparsity_penalty}, applies L1 regularisation to limit feature selection.
That is, it penalises the selection of unnecessary SPCSFs and instead fosters the selection of a smaller, more meaningful set of SPCSFs that can equally cover each security dimension $a$.
To do so, it uses a global penalty $P^s_{max}$ (here, 2) as the upper boundary evenly across the number of available SPCSFs ($|\Omega|$).
With a smaller number of SPCSFs (e.g. clustered mode), this penalty increases considerably --- since adding a cluster may incur massive over-provisioning depending on the SPCSFs it contains.
Conversely, with a larger number of SPCSFs (e.g. in the full mode), this penalty is reduced to allow selecting more specialised SPCSFs (i.e. those covering less dimensions at the time).
This global penalty is then weighted by $p(\omega, \sigma)$.
Therefore, $P^{s}_{a}(\sigma)$ can adapt to the topology of the MAID $\mathcal{M}$ (and thus, to the selected mode $M$) and the maximum available SPCSFs; guiding towards more accurate decision profiles.

\subsection{Recommendation procedure}
\label{sec:implementation_model:recommendation-procedure}

The MAID model $\mathcal{M}$ is constructed along with the utility function $U_{a}$ per agent $a$ (Algorithm \ref{alg:implementation-model:maid-generation}). This is used for exploring the decision space and recommending SPCSFs (Algorithm \ref{alg:implementation-model:gt-exploration-recommendation}). Notation is available in Table \ref{tab:implementation-model:maid-gt-notation}.

\begin{table}[h!]
  \caption{Notation for Algorithm \ref{alg:implementation-model:maid-generation} and Algorithm \ref{alg:implementation-model:gt-exploration-recommendation}}
  \label{tab:implementation-model:maid-gt-notation}
  \begin{tabular}{cll}
    \cline{1-3}
    \textbf{Term} & \multicolumn{2}{l}{\textbf{Description}} \\
    \cline{1-3}
     \noalign{\vspace{0.2em}}
     \multicolumn{1}{l}{\textbf{Inputs}} & \multicolumn{2}{l}{} \\
     $DS_{8}$ & \multicolumn{2}{l}{Dataset with SPCSFs} \\
     $T$ & \multicolumn{2}{l}{Filtering threshold for $DS_{8}$} \\
     $M$ & \multicolumn{2}{l}{MAID mode (full, clustered)} \\
     $G_{c}$ & \multicolumn{2}{l}{Target no. of clusters for GMM (optional)} \\
     $\Omega$ & \multicolumn{2}{l}{Set of filtered candidate chance nodes} \\
     $A$ & \multicolumn{2}{l}{Agents (security dimensions $\mathcal{D}$)} \\
     $\mathbb{D}_{A}$ & \multicolumn{2}{l}{Decision domain for $A$ ($\sim$ SPCSF baseline)} \\
     $R$ & \multicolumn{2}{l}{security dimension correlation matrix} \\
     $\vec{\tau}$ & \multicolumn{2}{l}{Security requirements per Sec. Dim. $A$} \\
     $\mathcal{M}$ & \multicolumn{2}{l}{MAID model} \\
     $\mathcal{L}$ & \multicolumn{2}{l}{Learning algorithm (Hedge, EXP3)} \\
     $L$ & \multicolumn{2}{l}{Total learning iterations} \\
     $B$ & \multicolumn{2}{l}{Maximum budget limit for SPCSFs} \\
     \noalign{\vspace{0.2em}}
     \multicolumn{1}{l}{\textbf{Variables}} & \multicolumn{2}{l}{} \\
     $\mathcal{N}_d, \mathcal{N}_u, \mathcal{N}_c$ & \multicolumn{2}{l}{Decision, utility and chance nodes} \\
     $E_{a}$ & \multicolumn{2}{l}{Edges for agent $a \in \mathcal{M}$} \\
     $C_{a}$ & \multicolumn{2}{l}{Coverage for agent $a$} \\
     $S_{a}, Q_{a}$ & \multicolumn{2}{l}{Synergy and quality incentives for agent $a$} \\
     $P^{p}_{a}, P^{s}_{a}$ & \multicolumn{2}{l}{Penalties per agent $a$: provisioning, sparsity} \\
     $U_{a}$ & \multicolumn{2}{l}{Utility function for agent $a$} \\
     $W, \vec{p}$ & \multicolumn{2}{l}{Weights and sampling probabilities} \\
     $\mathcal{H}$ & \multicolumn{2}{l}{History space of explored joint profiles} \\
     $\sigma$ & \multicolumn{2}{l}{Joint policy profile} \\
     $\vec{\beta}$ & \multicolumn{2}{l}{Upper bounds extracted from $\sigma^*$} \\
     $F, F_{s}$ & \multicolumn{2}{l}{Raw and ranked list of $\omega$ by adequacy} \\
     $f_\omega$ & \multicolumn{2}{l}{Adequacy (fitness) of $\omega$ to cover $\vec{\tau}$} \\
     $\mathcal{R}$ & \multicolumn{2}{l}{Final recommended set of SPCSFs} \\
     $\vec{c}$ & \multicolumn{2}{l}{Achieved physical coverage vector of $\mathcal{R}$} \\
     \noalign{\vspace{0.2em}}
    \cline{1-3}
  \end{tabular}
\end{table}

\subsubsection{MAID construction from dataset}
\label{sec:implementation_model:recommendation-procedure:maid-constructions}

Algorithm \ref{alg:implementation-model:maid-generation} uses the dataset $DS_{8}$ from Algorithm \ref{alg:implementation-dataset:dataset} along with the dataset filtering threshold $T$, the mode $M$ (i.e. clustered or full) used to represent the data in the MAID, the maximum clusters $G_{c}$ (if needed), the data for SPCSFs $\Omega$, the list of agents $A$ and their domain $\mathbb{D}_{A}$, the correlation matrix $R$ across agents and the requirements per agent $\vec{\tau}$.

\begin{algorithm}[hbtp]
\caption{MAID generation and hybrid utility formulation}
\label{alg:implementation-model:maid-generation}
\begin{algorithmic}[1]
\Require $DS_{8}$, $T$, $M$, $G_{c}$, $\Omega$, $A$, $\mathbb{D}_{A}$, $R$, $\vec{\tau}$
\Ensure $\mathcal{M}$

\LineComment{-3.5}{Filter dataset and generate clusters}
\State $DS_{8f} \gets \mathrm{filter\_implementable}(DS_8, T)$ \label{alg:implementation-model:maid-generation:line:s1}
\If{$M = \mathrm{clustered}$} \label{alg:implementation-model:maid-generation:line:s2}
    \State $\Omega \gets \mathrm{gmm\_g\_bic}(DS_{8f}, G_{c})$ \label{alg:implementation-model:maid-generation:line:s3}
\Else \label{alg:implementation-model:maid-generation:line:s4}
    \State $\Omega \gets DS_{8f}$ \label{alg:implementation-model:maid-generation:line:s5}
\EndIf

\LineComment{-1.5}{Construct MAID with data}
\For{each node $\omega \in \Omega$} \label{alg:implementation-model:maid-generation:line:s6}
    \State $\mathcal{N}_{c, \omega} \gets \mathrm{create\_chance}(\mathrm{normal\_distr}(\omega))$ \label{alg:implementation-model:maid-generation:line:s7}
\EndFor
\For{each agent $a \in A$} \label{alg:implementation-model:maid-generation:line:s8}
    \State $\mathcal{N}_{d, a} \gets \mathrm{create\_decision}(a, \mathbb{D}_{a})$ \label{alg:implementation-model:maid-generation:line:s9}
    \State $E_{a} \gets \big{(}({N}_{c, \omega}, {N}_{d, a}), ({N}_{d, a}, {N}_{u, a})\big{)}$ \label{alg:implementation-model:maid-generation:line:s10}
    \State $C_{a} \gets \mathrm{agent\_coverage}(\Omega, a)$ \label{alg:implementation-model:maid-generation:line:s11}
\EndFor

\LineComment{-1.5}{Hybrid utility function per agent}
\For{each agent $a \in A$} \label{alg:implementation-model:maid-generation:line:s12}
    \State $S_{a} \gets \mathrm{agent\_synergy}(R, C_{-a})$ \label{alg:implementation-model:maid-generation:line:s13}
    \State $Q_{a} \gets \mathrm{spcsf\_quality}(\Omega, \mathcal{N}_{d, a})$ \label{alg:implementation-model:maid-generation:line:s14}
    \State $P^p_{a} \gets \mathrm{provisioning\_penalty}(C_{a}, \tau_{a})$ \label{alg:implementation-model:maid-generation:line:s15}
    \State $P^s_{a} \gets \mathrm{sparsity\_penalty}(\Omega)$ \label{alg:implementation-model:maid-generation:line:s16}
    \State $U_{a} \gets Q_{a} + S_{a} - P^p_{a} - P^s_{a}$ \label{alg:implementation-model:maid-generation:line:s17}
    \State $\mathcal{N}_{u, a} \gets \mathrm{create\_utility}(U_{a})$ \label{alg:implementation-model:maid-generation:line:s18}
\EndFor

\State $\mathcal{M} \gets \mathrm{create\_maid}(\mathcal{N}_{c}, \mathcal{N}_{a}, \mathcal{N}_{u}, U_{A}, E_{A})$ \label{alg:implementation-model:maid-generation:line:s19}

\State \Return $\mathcal{M}$ \label{alg:implementation-model:maid-generation:line:s20}
\end{algorithmic}
\end{algorithm}

First, the dataset $DS_{8}$ is filtered in Algorithm \ref{alg:implementation-model:maid-generation}, line \ref{alg:implementation-model:maid-generation:line:s1}.
This filters SPCSFs according to their total score through a configurable threshold $T$, and ends up with a subset of SPCSFs $DS_{8f}$ that will be used to cover the security requirements $\vec{\tau}$.
Depending on $M$, the set of available, filtered SPCSFs to be used as chance nodes in the MAID $\mathcal{M}$ is further modified, as $\Omega$.
If the mode $M$ is clustered, the GMM clustering process creates clusters containing specific SPCSFs according to their coverage of the agents. This process can either use an optional upper limit for clusters ($G_{c}$) or apply the Bayesian Information Criterion (BIC) to find the optional number of clusters. $G_{c}$ is optional and the default runs GMM with BIC to automatically find this value.
Otherwise, for a full MAID, the whole $DS_{8f}$ is used.
This runs in Algorithm \ref{alg:implementation-model:maid-generation}, lines 
\ref{alg:implementation-model:maid-generation:line:s2}-\ref{alg:implementation-model:maid-generation:line:s5}.

With this data, the vertices $V$ and edges $E$ for the MAID $\mathcal{M}$ can be extracted from the filtered SPCSFs in $\Omega$.
$V$ is defined from the different nodes $\mathcal{N}_c$, $\mathcal{N}_d$ and $\mathcal{N}_u$.
These nodes encode some information as well: chance nodes $\mathcal{N}_c$ comprise its approximate, average total score (Algorithm \ref{alg:implementation-model:maid-generation}, lines \ref{alg:implementation-model:maid-generation:line:s6}-\ref{alg:implementation-model:maid-generation:line:s7}); decision nodes $\mathcal{N}_d$ host the domain $\mathbb{D}$, or values that can be considered (Algorithm \ref{alg:implementation-model:maid-generation}, line \ref{alg:implementation-model:maid-generation:line:s9}); and utility nodes $\mathcal{N}_u$ contain hybrid utility functions $U_{A}$ (Algorithm \ref{alg:implementation-model:maid-generation}, lines \ref{alg:implementation-model:maid-generation:line:s12}-\ref{alg:implementation-model:maid-generation:line:s18}).
It must be noted that chance nodes use here a probabilistic normal distribution around its average total score ($\mu$) and with a deviation ($\sigma$) of 0.1, restricted to the $[0.0, 1.0]$ interval. These represent the uncertainty over exact values to accommodate changes in the origin data, whether from $DS_8$ or its sourced datasets (e.g. changes in 1-2 columns for $DS_{7}.cm$ can get $\sim 10\%$ average variability), and introduce variability to test the framework under varying conditions.
Utility functions base on direct or indirect data from the dataset, e.g. the coverage $C_{a}$ of an agent (Algorithm \ref{alg:implementation-model:maid-generation}, line 
\ref{alg:implementation-model:maid-generation:line:s11}).
The edges $E_{A}$ (constructed in Algorithm \ref{alg:implementation-model:maid-generation}, line 
\ref{alg:implementation-model:maid-generation:line:s10}) express the flow of information by connecting chance nodes to decision nodes, and decision nodes to utility nodes.
Finally, a MAID $\mathcal{M}$ can now be generated for each set of parameters (e.g. mode $M$) in Algorithm \ref{alg:implementation-model:maid-generation}, line 
\ref{alg:implementation-model:maid-generation:line:s19}.

\subsubsection{Game-theoretic exploration and recommendation}
\label{sec:implementation_model:recommendation-procedure:gt-exploration-recomm}

Once MAID $\mathcal{M}$ is constructed, Algorithm \ref{alg:implementation-model:gt-exploration-recommendation} runs two stages: (i) the game-theoretic exploration across the decision space for all agents, invoking each utility action $U_{a}$ to maximise each agent's payoff and obtain the most accurate joint solution or (discrete) profile $\sigma^{*}$ that can meet best the security requirements without exceeding these; and (ii) the recommendation stage itself, which navigates the sorted SPCSFs and select those that contribute to the coverage without exceeding it or exhausting the budget.

\begin{algorithm}[t]
\caption{MAID exploration and SPCSF recommendation}
\label{alg:implementation-model:gt-exploration-recommendation}
\begin{algorithmic}[1]
\Require $\mathcal{M}$, $M$, $\Omega$, $\vec{\tau}$, $\mathcal{L}$, $L$, $B$
\Ensure $\sigma^*$, $\mathcal{R}$

\LineComment{-3.5}{Decision space exploration}
\State $W, L \gets \mathrm{init\_params}(\mathcal{M}, M, \mathcal{L})$ \label{alg:implementation-model:gt-exploration-recommendation:line:s1}
\State $\mathcal{H} \gets \emptyset$ \label{alg:implementation-model:gt-exploration-recommendation:line:s2}
\For{$l = 1 \mathrm{ to } L$} \label{alg:implementation-model:gt-exploration-recommendation:line:s3}
    \State $\sigma^{(l)}, \vec{p} \gets \mathrm{sample\_joint\_profile}(W, \mathcal{L})$ \label{alg:implementation-model:gt-exploration-recommendation:line:s4}
    \State $\mathcal{H} \gets \mathcal{H} \cup \{\sigma^{(l)}\}$ \label{alg:implementation-model:gt-exploration-recommendation:line:s5}
    \For{each agent $a \in A$} \label{alg:implementation-model:gt-exploration-recommendation:line:s6}
        \State $U_{a}^{(l)} \gets U_{a}(\sigma^{(l)})$ \label{alg:implementation-model:gt-exploration-recommendation:line:s7}
        \State $W \gets \mathrm{update\_params}(W, U_{a}^{(l)}, \vec{p}, \mathcal{L})$ \label{alg:implementation-model:gt-exploration-recommendation:line:s8}
    \EndFor
\EndFor

\LineComment{-1.5}{Extract best joint profile from history}
\State $\sigma^* \gets \mathrm{extract\_min\_distance\_profile}(\mathcal{H}, \vec{\tau})$ \label{alg:implementation-model:gt-exploration-recommendation:line:s9}

\LineComment{-1.5}{Recommendation analysis and ranking}
\State $F \gets \emptyset$ \label{alg:implementation-model:gt-exploration-recommendation:line:s10}
\For{each node $\omega \in \Omega$} \label{alg:implementation-model:gt-exploration-recommendation:line:s11}
    \State $f_\omega \gets \mathrm{calculate\_similarity}(\omega, \vec{\tau})$ \label{alg:implementation-model:gt-exploration-recommendation:line:s12}
    \State $F \gets F \cup \{(\omega, f_\omega)\}$ \label{alg:implementation-model:gt-exploration-recommendation:line:s13}
\EndFor
\State $F_{s} \gets \mathrm{sort\_descending}(F)$ \label{alg:implementation-model:gt-exploration-recommendation:line:s14}

\LineComment{-1.5}{Bounded, greedy selection with dynamic pruning}
\State $\mathcal{R} \gets \emptyset$, $\vec{c} \gets \vec{0}$ \label{alg:implementation-model:gt-exploration-recommendation:line:s15}
\State $\vec{\beta} \gets \max(\mathrm{coverage}(\sigma^*), \vec{\tau})$ \label{alg:implementation-model:gt-exploration-recommendation:line:s16}
\While{$\vec{c} < \vec{\tau}$ \textbf{and} $|\mathcal{R}| < B$ \textbf{and} $\exists \omega \in F_{s}$} \label{alg:implementation-model:gt-exploration-recommendation:line:s17}
    \State $\omega^{*} \gets \argmax_{\omega} \big( \Delta\,\mathrm{cov}(\omega, \vec{c}) - \mathrm{pen}_{o}(\omega, \vec{\beta}, W_o) \big)$ \label{alg:implementation-model:gt-exploration-recommendation:line:s18}
    \State $\mathcal{R} \gets \mathcal{R} \cup \{\omega^{*}\}$ \label{alg:implementation-model:gt-exploration-recommendation:line:s19}
    \State $\vec{c} \gets \max(\vec{c}, \mathrm{coverage}(\omega^{*}))$ \label{alg:implementation-model:gt-exploration-recommendation:line:s20}
\EndWhile

\For{each node $r \in \mathcal{R}$} \label{alg:implementation-model:gt-exploration-recommendation:line:s21}
    \If{$\mathrm{coverage}(\mathcal{R} \setminus \{r\}) \ge \vec{\tau}$} \label{alg:implementation-model:gt-exploration-recommendation:line:s22}
        \State $\mathcal{R} \gets \mathcal{R} \setminus \{r\}$ \label{alg:implementation-model:gt-exploration-recommendation:line:s23}
    \EndIf
\EndFor

\State \Return $\sigma^*, \mathcal{R}$ \label{alg:implementation-model:gt-exploration-recommendation:line:s24}
\end{algorithmic}
\end{algorithm}

Based on the MAID $\mathcal{M}$, mode $M$ and learning algorithm $\mathcal{L}$, the initial learning parameters are defined (Algorithm \ref{alg:implementation-model:gt-exploration-recommendation}, lines \ref{alg:implementation-model:gt-exploration-recommendation:line:s1}-\ref{alg:implementation-model:gt-exploration-recommendation:line:s2}). The probability weights $W$ per agent and the number of iterations $L$ per algorithm $\mathcal{L}$ (i.e. Hedge/EWA and EXP3) are calculated, along with the empty history set $\mathcal{H}$ to track explored decisions.

For each iteration $l \in L$, the learning algorithm picks, in a stochastic manner, a joint strategy profile $\sigma^{(l)}$ with values from domain $\mathbb{D}_{A}$ and a probabilistic distribution $\vec{p}$ (Algorithm \ref{alg:implementation-model:gt-exploration-recommendation}, line \ref{alg:implementation-model:gt-exploration-recommendation:line:s4}). This profile represents the collective explored/covered agents at that iteration.
This is recorded as part of the decisions' history (Algorithm \ref{alg:implementation-model:gt-exploration-recommendation}, line \ref{alg:implementation-model:gt-exploration-recommendation:line:s5}).
Then, the utility function $U_{a}$ is invoked per agent considering the current profile $\sigma^{(l)}$; and its result is used to update the probability weights $W$ and sampling distribution $\vec{p}$, in order to steer the stochastic selection of the profile (Algorithm \ref{alg:implementation-model:gt-exploration-recommendation}, lines \ref{alg:implementation-model:gt-exploration-recommendation:line:s6}-\ref{alg:implementation-model:gt-exploration-recommendation:line:s8}).
Finally, and after running all $L$ iterations per algorithm $\mathcal{L}$; the history of explored decisions $\mathcal{H}$ is analysed to extract the most accurate profile $\sigma^*$ from it (Algorithm \ref{alg:implementation-model:gt-exploration-recommendation}, line \ref{alg:implementation-model:gt-exploration-recommendation:line:s9}).
Such profile $\sigma^*$ minimises the distance between the required $\tau_{a}$ and obtained $c_{a}$ degree of coverage to an agent $a$; where this coverage is obtained from the joint profiles previously explored and tracked in $\mathcal{H}$.
With this, the game-theoretic exploration stage produces the coverage profile $\sigma^*$, implemented as a discretised vector that maps each agent $a$ to a degree of coverage $c_{a}$ that can be low, intermediate or high (or none), depending on the amount of coverage. Since this result is met after running the utilities $U_{a}$, as the approximate equilibrium for the agents it indicates which security requirements can be effectively met and to which extent.

From this point onwards, the recommender stage takes place. An initial ranking stage (Algorithm \ref{alg:implementation-model:gt-exploration-recommendation}, lines \ref{alg:implementation-model:gt-exploration-recommendation:line:s10}-\ref{alg:implementation-model:gt-exploration-recommendation:line:s14}) evaluates how well each SPCSF $\omega \in \Omega$ can cover the requested security requirements.
This is calculated by the identified similarity $f_\omega$ between the coverage provided by such SPCSF $\omega$ and the expected total coverage $\vec{\tau}$ (Algorithm \ref{alg:implementation-model:gt-exploration-recommendation}, line \ref{alg:implementation-model:gt-exploration-recommendation:line:s12}) and is used to discourage both under- and over- provisioning.
Then, all items added to $F$ are ranked from most to least similarity, yielding $F_{s}$.

After the ranking, the second part of the recommendation stage ensures that the coverage provided by the approximate equilibrium $\sigma^{*}$ acts as an upper bound; avoiding over-provisioning and allowing to identify the subset of the SPCSFs $\mathcal{R}$ to recommend. The set $\mathcal{R}$ is initialised along with the coverage vector $\vec{c}$.
The best joint profile $\omega^*$ is considered, along with the passed security requirements $\vec{\tau}$, to determine the upper boundary $\vec{\beta}$ for coverage that each agent $a$ must achieve to avoid over-provisioning (Algorithm \ref{alg:implementation-model:gt-exploration-recommendation}, lines \ref{alg:implementation-model:gt-exploration-recommendation:line:s15}-\ref{alg:implementation-model:gt-exploration-recommendation:line:s16}).
With all of that, the recommender runs a loop to accumulate adequate SPCSFs in Algorithm \ref{alg:implementation-model:gt-exploration-recommendation}, lines \ref{alg:implementation-model:gt-exploration-recommendation:line:s17}-\ref{alg:implementation-model:gt-exploration-recommendation:line:s20}.
This process checks whether (i) the requested security requirements $\vec{\tau}$ are still unmet by the current achieved coverage $\vec{c}$, while ensuring that (ii) the current amount of recommendations do not exceed the internally defined budget $B$; it continues taking SPCSFs from those available at the sorted pool $F_{s}$.

\begin{subequations}
\label{eq:implementation-model:gt-exploration-recommendation:rec:inc-pen}
\begin{align}
    \Delta\,\mathrm{cov}(\omega, \vec{c}) &= \sum_{a \in A} \Big( \min(c_{a} + v_{\omega, a}, \tau_{a}) - c_{a} \Big) \label{eq:implementation-model:gt-exploration-recommendation:rec:inc-pen:cov} \\[0.5em]
    \mathrm{pen}_{o}(\omega, \vec{\beta}, W_o) &= W_o \cdot \sum_{a \in A} \max\big(0, (c_{a} + v_{\omega, a}) - \beta_{a}\big) \label{eq:implementation-model:gt-exploration-recommendation:rec:inc-pen:pen}
\end{align}
\end{subequations}

At each iteration, the best candidate SPCSF $\omega^{*}$ is selected, trying to maximise only the useful coverage for requested agents $\Delta \mathrm{ cov}$ and to minimise unnecessary coverage $\mathrm{pen}_{o}$, in the same way as the under- and over- provision constraints in the utility function itself. These conditions are detailed in Eq. \eqref{eq:implementation-model:gt-exploration-recommendation:rec:inc-pen:cov} and Eq. \eqref{eq:implementation-model:gt-exploration-recommendation:rec:inc-pen:pen} respectively; with $c_{a}$ being the accumulated coverage per agent $a$ and $v_{\omega, a}$ being the coverage of SPCSF $\omega$ to agent $a$.
It is worth noting that, if an agent (its represented security dimension) is already fully covered, $\Delta\,\mathrm{cov}$ is zero.
Then, the recommended set $\mathcal{R}$ and the coverage are updated adequately with the best SPCSF $\omega^{*}$.
In a final pass (Algorithm \ref{alg:implementation-model:gt-exploration-recommendation}, lines \ref{alg:implementation-model:gt-exploration-recommendation:line:s21}-\ref{alg:implementation-model:gt-exploration-recommendation:line:s23}), $\mathcal{R}$ is reviewed to ensure the bare minimum SPCSF set is proposed to cover the security requirements $\vec{\tau}$.

\begin{figure*}[htbp]
\begin{minipage}[t]{0.24\textwidth}
    \begin{subfigure}{\textwidth}
        \includegraphics[draft=false, width=\textwidth]{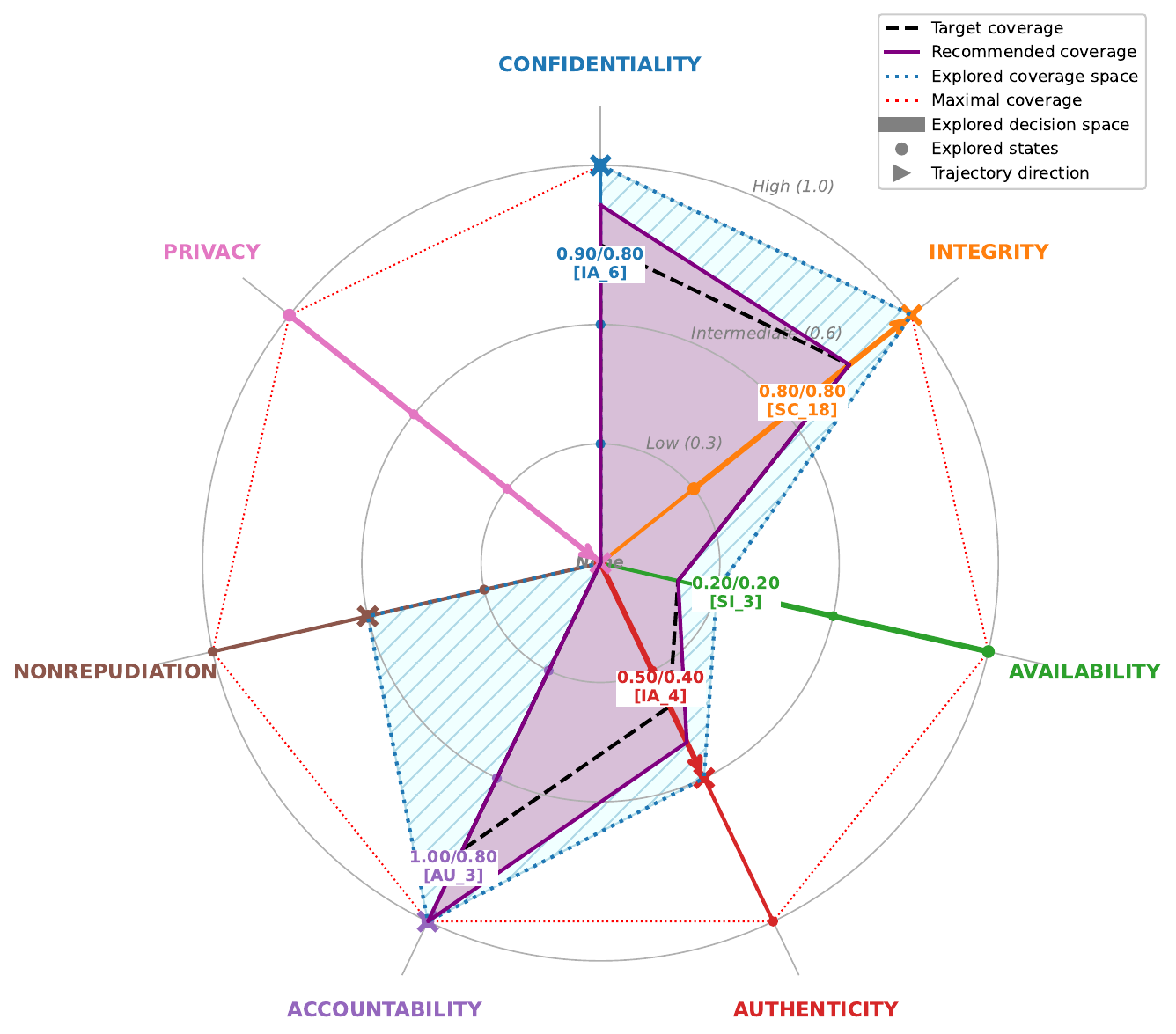}
        \caption{Clustered $\mathcal{M}$ using Hedge}
        \label{fig:implementation-model:gt-recom:clust-hl}
    \end{subfigure}
\end{minipage}
\begin{minipage}[t]{0.24\textwidth}
    \begin{subfigure}{\textwidth}
        \includegraphics[draft=false, width=\textwidth]{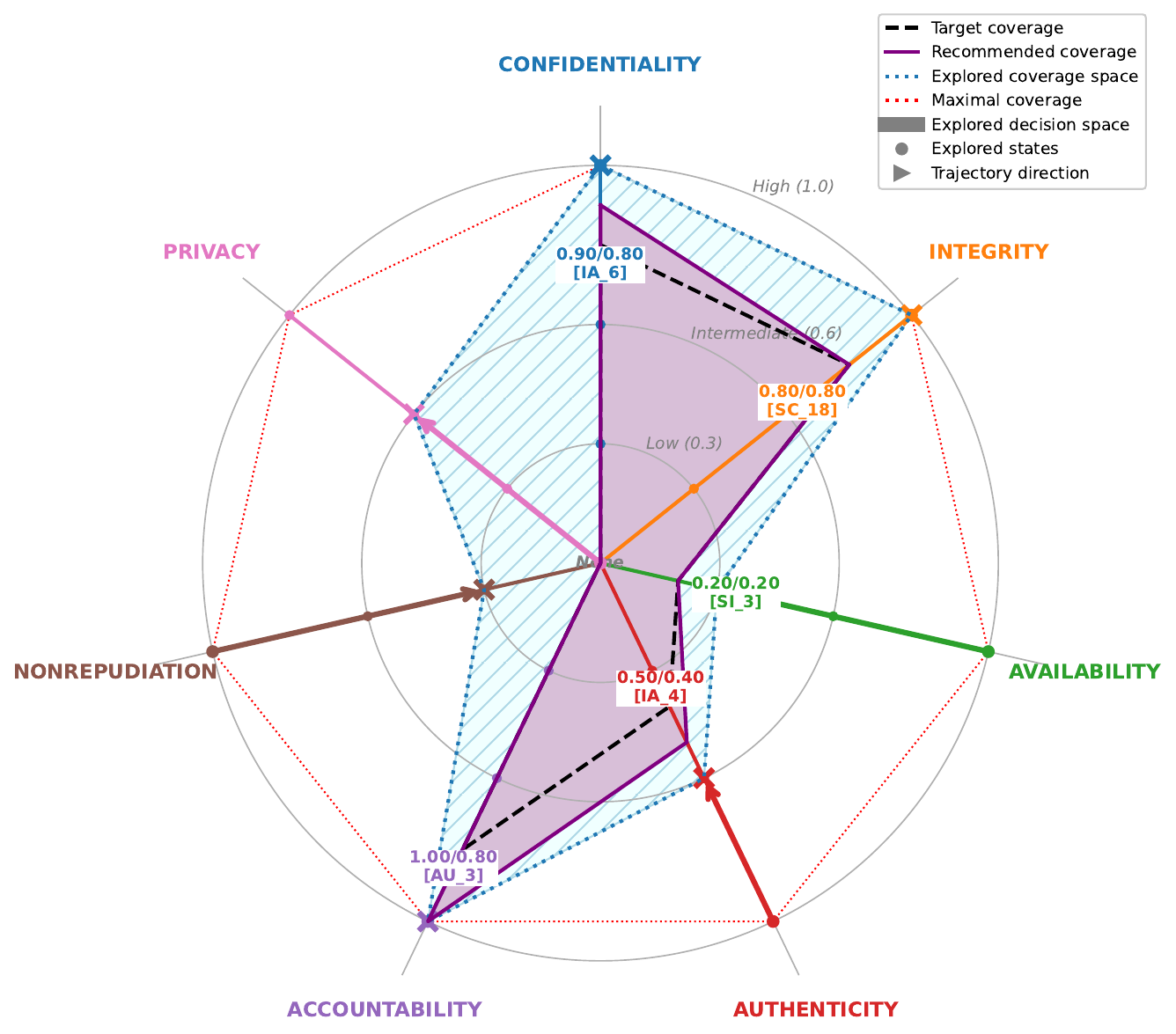}
        \caption{Clustered $\mathcal{M}$ using EXP3}
        \label{fig:implementation-model:gt-recom:clust-e3}
    \end{subfigure}
\end{minipage}
\begin{minipage}[t]{0.24\textwidth}
    \begin{subfigure}{\textwidth}
        \includegraphics[draft=false, width=\textwidth]{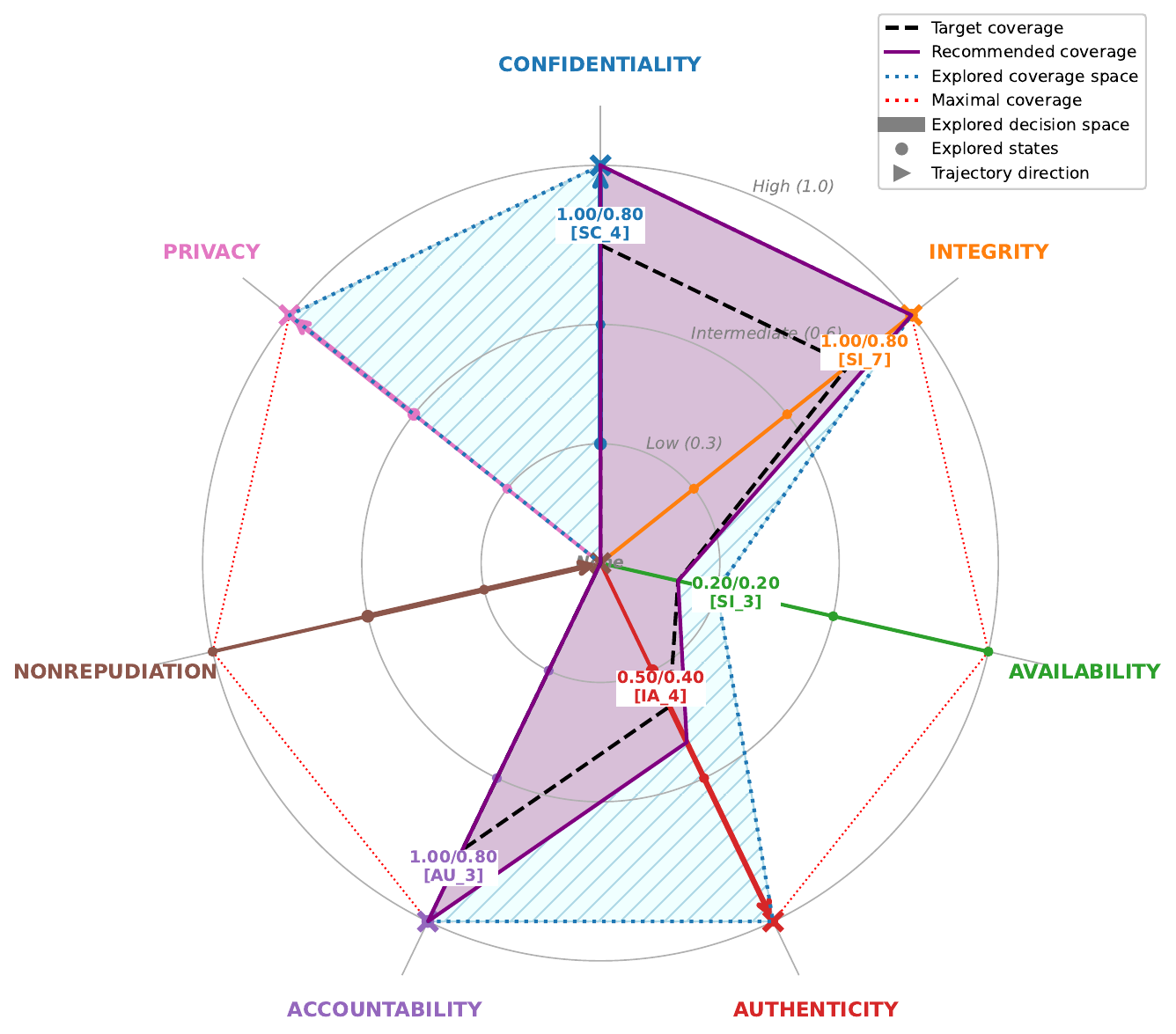}
        \caption{Full $\mathcal{M}$ using Hedge}
        \label{fig:implementation-model:gt-recom:full-hl}
    \end{subfigure}
\end{minipage}
\begin{minipage}[t]{0.24\textwidth}
    \begin{subfigure}{\textwidth}
        \includegraphics[draft=false, width=\textwidth]{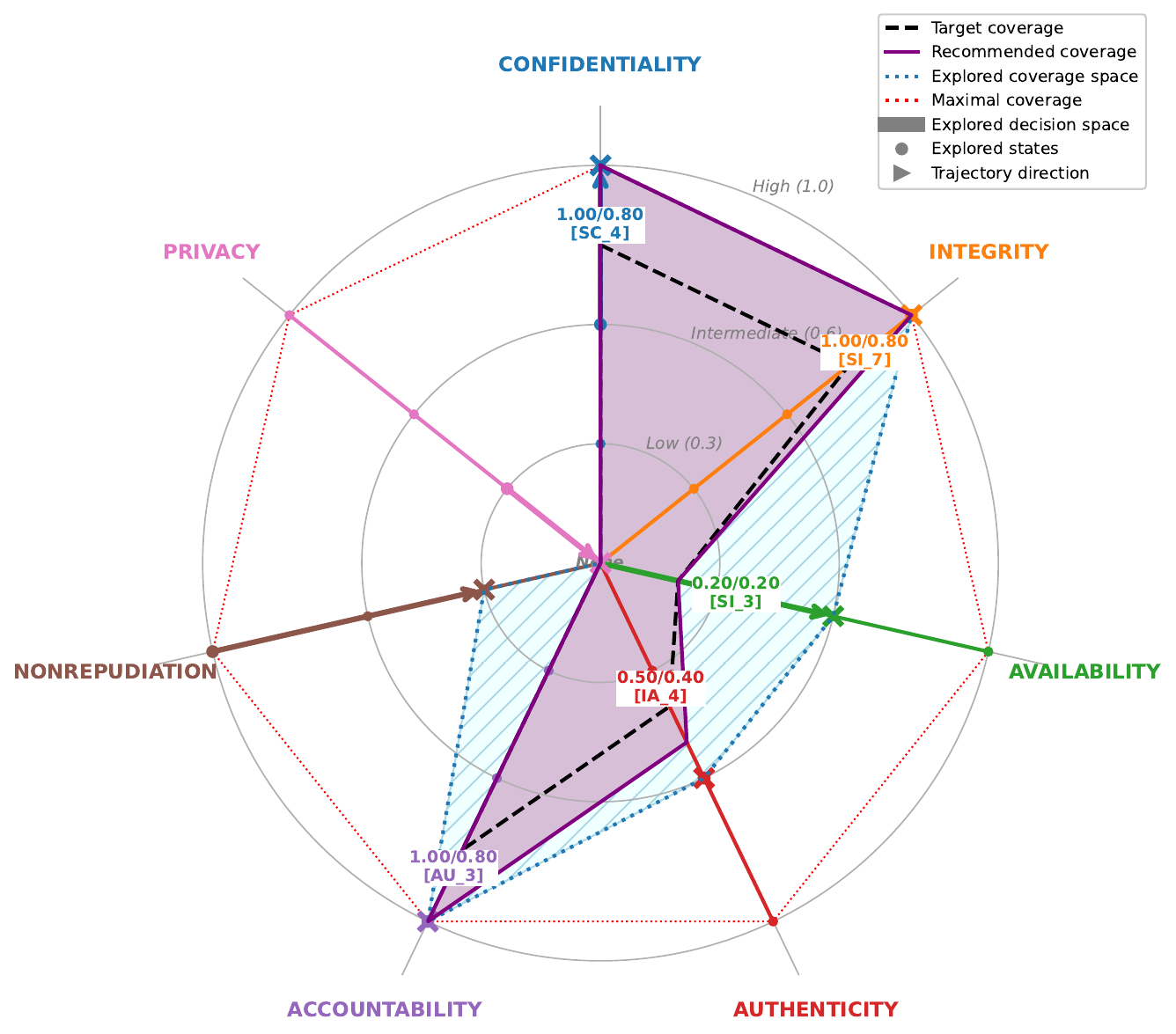}
        \caption{Full $\mathcal{M}$ using EXP3}
        \label{fig:implementation-model:gt-recom:full-e3}
    \end{subfigure}
\end{minipage}
\caption{MAID $\mathcal{M}$ exploration and recommendation}
\label{fig:implementation-model:gt-recom}
\end{figure*}

Results for an exploration and recommendation stage (Algorithm \ref{alg:implementation-model:gt-exploration-recommendation}) using a selected requirements profile chosen at random are shown in Fig. \ref{fig:implementation-model:gt-recom}, as evaluated for both full (Fig. \ref{fig:implementation-model:dataset-in-model}) and clustered (Fig. \ref{fig:implementation-model:dataset-in-model-clustered}) modes and applying two different learning algorithms $\mathcal{L}$ to minimise the regret over the selected decisions: Hedge Learning and EXP3.
In the radar plot, each agent is located in a vertex and coloured differently.
The explored space is identified by dots, with arrows indicating the direction of the exploration considering the best joint profile (approximate equilibrium).
The security requirements $\vec{\tau}$ are shown in a bold dashed polygon, whose shape is the approximation to the ideal, exact fit that indicates no under- and over-provisioning.
The best joint discrete profile (approximate equilibrium) obtained from the exploration stage is shaded in light blue and its contour is dotted. The crosses at each agent denote its perimeter or boundaries.
Finally, the polygon in purple shade and solid contour reflects the degree of coverage for the agents suggested in the final, recommending stage.

The best joint profile is selected out of the history of explored profiles $\mathcal{H}$, where each learning algorithm $\mathcal{L}$ runs different iterations $L$, suited to their behaviour. As in the utility function, the criteria to select it heavily penalises under-provisioning, and in case of similar coverage between two decision profiles, it favours those that minimise over-provisioning. This can be observed in the blue shades, which are always larger than the purple shades. The purple shade will more often than not have minor over-provisioning to ensure covering at least the desired degree of security. Conversely, it also can incur under-provisioning if the SPCSF dataset ($DS_8$) is too small (i.e. a high threshold $T$); which is considerably more problematic.
In the execution shown in Fig. \ref{fig:implementation-model:gt-recom}, the recommendation between the clustered and full MAID returned slightly different SPCSFs.
It is worth noting that no under-provisioning is found in that execution until filtering the dataset with $T = 0.4$ (i.e. total score of at least 0.4).
Here, the accountability agent is penalised due to related SPCSFs being filtered out due to its lower total score; considerably reducing the original 16 identified SW-implementable SPCSFs from Table \ref{tab:implementation-dataset:mapping:ciatunp}. This is comparable to executions with other requirements (Section \ref{sec:evaluation:use-case}).

\subsection{Complexity limitations}
\label{sec:implementation_model:complexity-limitations}

Finding exact equilibria faces two primary computational limitations: (i) generating and storing CPDs and (ii) evaluating the MAID subgames across a large decision space.
In fact, these make exact calculation intractable for models requiring extensive data combinations, such as the full MAID (Fig. \ref{fig:implementation-model:dataset-in-model}). More compact models, such as the clustered MAID (Fig. \ref{fig:implementation-model:dataset-in-model-clustered}) alleviate these limitations and allow computing some exact equilibria, yet very slowly.
This section explains these findings and the motivation for no-regret online learning dynamics, which allow to bypass the intractability and obtain an approximate equilibrium considerably faster.

\subsubsection{Space: storing probabilistic data}

In BN and derived approaches using exact inference, the conditioned Bayesian probabilities per decision node are computed based on its parent chance nodes, as per $\mathbb{P}(D \mid Pa(D))$, and used to populate its CPDs.
In the full MAID $\mathcal{M}$, in Fig. \ref{fig:implementation-model:dataset-in-model}, the in-degree (number of parents) of decision nodes ranges from 2 to 67.
Eq. \eqref{eq:implementation-model:cpd:calculation:upper-bound} and \eqref{eq:implementation-model:cpd:calculation:disk-gb} determine the upper bound for the number of CPD entries $\mathcal{C}_{e}$ and the associated memory size $\mathcal{C}_{s}$, respectively.
These scale exponentially with $m_p = |max(Pa(D))|$, but also grow with $\mathbb{D}_{V}$ --- namely, the maximum number of parents and decision domain per decision node.

\begin{subequations}
\label{eq:implementation-model:cpd:calculation}
\begin{align}
    \mathcal{C}_{e} &\le |V| \cdot |\mathbb{D}_{V}|^{m_{p} } \mid m_{p} = \text{max(}|Pa(V)|\text{)} \label{eq:implementation-model:cpd:calculation:upper-bound} \\[0.5em]
    \mathcal{C}_{s} &= \frac{4}{{10^{9}}} \cdot \sum_{v_=1}^{|V|} (|k|+|\mathbb{D}_{v}|-1) \cdot (|\mathbb{D}_{k}|^{|k|}) \mid k = Pa(v) \label{eq:implementation-model:cpd:calculation:disk-gb}
\end{align}
\end{subequations}

\begin{table}[hbtp]
    \caption{Dimension and upper space bound required per MAID}
    \setlength{\tabcolsep}{4pt}
    \centering
    \scalebox{0.85}{
    \begin{tabular}{cccccc}
    \toprule
    \multicolumn{4}{c}{\textbf{$\mathcal{M}$ type, size ($N$, $|V|, |E|$), $Pa(V)$}} & \multicolumn{2}{c}{\textbf{Space used by CPD}} \\
    \cmidrule(l){1-4} \cmidrule(l){5-6}
    \small{\textbf{DAG content}} & \small{\textbf{$N$}} & \small{\textbf{$|V|, |E|$}} & \small{\textbf{$Pa(V)$}} & \small{\textbf{$\mathcal{C}_{e}$}} & \small{\textbf{$\mathcal{C}_{s}$ (GB)}} \\
    \cmidrule(r){1-1} \cmidrule(l){2-3} \cmidrule(l){4-4} \cmidrule(l){5-6}
    Full ($DS_{8}$) & 300 & 314, 644 & 2-181 & $3.14 \cdot 10^{183}$ & $7.6 \cdot 10^{174}$ \\
    score\_total > 0.0 & 285 & 299, 612 & 2-174 & $2.99 \cdot 10^{176}$ & $7.32 \cdot 10^{167}$ \\
    SW-implem. & 97 & 111, 203 & 2-67 & $1.11 \cdot 10^{69}$ & $3.04 \cdot 10^{60}$ \\
    score\_total > 0.4 & 37 & 51, 87 & 0-28 & $5.1 \cdot 10^{29}$ & $1.48 \cdot 10^{21}$ \\
    \bottomrule
    \end{tabular}
	}
    \label{tab:implementation-model:definition:reduction}
\end{table}

Table \ref{tab:implementation-model:definition:reduction} summarises the application of Eq. \eqref{eq:implementation-model:cpd:calculation:upper-bound} and Eq. \eqref{eq:implementation-model:cpd:calculation:disk-gb} to $\mathcal{M}$ based on the iteratively filtered dataset $DS_{8}$ based on score constraints and the SW-implementation criteria, which reduces $m_{p}$.
It shows the vertices $|V|$ and edges $|E|$ per $\mathcal{M}$ generated from each filtered dataset --- where $|V| = N + 14$ to accommodate the decision and utility nodes per security dimension. It also finds the $\mathcal{C}_{e}$, equivalent to the minimum independent parameters required to represent all joint CPDs \cite{Koller_Friedman_2010}; and the size in GBs in $\mathcal{C}_{s}$.
$\mathcal{C}_{e}$ assumes $|\mathbb{D}_{V}| = 10$, since agents can decide on values from the score deciles (Fig. \ref{fig:implementation-dataset:dataset-score-decils}).  
In the best case (i.e. $DS_{8}$ reduced to half after filtering it with threshold $T=0.4$), $\mathcal{C}_{e}$ is still around the order of $10^{33}$ entries. Even if also reducing decisions to $|\mathbb{D}_{V}| = 4$, in the best case $\mathcal{C}_{e}$ would yield around the order of $10^{20}$ entries. Thus, finding the exact equilibrium is intractable for the full MAID.
Since exact inference on BN models is NP-hard \cite{Cooper_1990, Systems_2013}, processing all entries for such a dense decision or policy space incurs a combinatorial explosion that requires prohibitive amounts of memory.
One way of bringing the complexity of the MAID considerably down is by decreasing the in-degree of decision nodes $D$; e.g. applying (child-friendly) parent divorcing \cite{Rohrbein_Eggert_Korner_2009} or grouping chance nodes (SPCSFs) together, e.g. identifying latent variables \cite{Spirtes_2000} or clustering based on similar contribution from SPCSFs towards agents.
The clustering approach is examined here applying Gaussian Mixture Model (GMM) along with the Bayesian Information Criterion (BIC), to perform soft clustering and group SPCSFs with similar properties into latent variables (clusters).
The resulting clusters group chance nodes into $k=11$ groups, as shown in Fig. \ref{fig:implementation-model:dataset-in-model-clustered}.
This drastically reduces $D$'s in-degree and the number of edges, limiting CPDs to a computationally tractable size. $\mathcal{C}_{e}$ would reduce in the order of $10^9$ and $10^5$ for $|\mathbb{D}_{V}|$ equal to 10 and 4, respectively.

\subsubsection{Computation: exploring the decision space}

Besides memory limitations, a game-theoretic system must evaluate expected utilities $EU_d(\sigma)$ across the agents' joint strategy profiles $\sigma$ to find optimal equilibria $\sigma^{*}$.
Table \ref{tab:implementation_model:gt_complexities} summarises the upper complexity bounds to converge on such optimal outcome in terms of (i) search space, understood as the (joint or individual) profiles $\sigma$; and (ii) of time complexity across these methods (iterations).
Notation refers to the number of agents $|A|$, decision domain per agent $\mathbb{D}_{A}$, number of iterations $L$, maximum number of parents (in-degree) in the decision $D$ nodes from the MAID $\mathcal{M}$ as $m_{p} = \text{max(}|Pa(V)|\text{)}$ and $s$ being the worst-case exponent used by the implemented LP solver.

\begin{table}[htpb]
  \caption{Space and time upper complexity bounds for game theory methods}
  \centering
  \resizebox{0.95\columnwidth}{!}{
    \renewcommand{\arraystretch}{1.3}
    \begin{tabular}{lll}
      \toprule
      \textbf{Method} & \textbf{Space ($|\sigma|$)} & \textbf{Time (iter.)} \\
      \midrule
      \textbf{Nash Eq.} \cite{Daskalakis:EECS-2008-107, Daskalakis_2009} & \multirow{4}{*}{$O(|\mathbb{D}_A|^{|A|})$} & $O(|\mathbb{D}_A|^{|A|})$ \\
      \textbf{CE} \cite{Papadimitriou_2005} \& \textbf{CCE} \cite{Blum_Mansour_2007} & & $O((|A| \cdot |\mathbb{D}_A|^{m_p})^s)$ \\
      \textbf{BRD} \cite{Fabrikant_2004} & & $O(|\mathbb{D}_A|^{|A|})$ \\
      \textbf{MMR} \cite{Savage_1951} & & $O((|A| \cdot |\mathbb{D}_{A}|^{m_p})^s)$ \\
      \textbf{Fictitious Play} \cite{Brandt_2010} & $O(|A| \cdot |\mathbb{D}_{A}|^{m_p})$ & $O(L \cdot |A| \cdot |\mathbb{D}_A|^{m_p})$ \\
      \textbf{EWA} \cite{Freund_Schapire_1997} \& \textbf{EXP3} \cite{Auer_et_al_2002} & $O(|A| \cdot |\mathbb{D}_{A}|)$ & $O(L \cdot |A| \cdot |\mathbb{D}_{A}|)$ \\
      \bottomrule
    \end{tabular}
  }
  \label{tab:implementation_model:gt_complexities}
\end{table}

Finding an exact Nash Equilibrium \cite{Nash_1951} is PPAD-complete, even for 2-player games \cite{Daskalakis_2009}, and its complexity scales exponentially with $|A|$.
More efficient approaches exist for graphical game models like MAIDs \cite{Kearns_Littman_Singh_2001} that exploit conditional independence \cite{Koller_Milch_2003} and can bring complexity down to polynomial time when bound to $|\mathbb{D}_{A}| = 2$ and $|A| = 3$ \cite{Daskalakis:EECS-2008-107} and exploiting the relevance graph \cite {Hammond_Fox_Everitt_Carey_Abate_Wooldridge_2023}; yet increasing agents or domain is also intractable on commodity hardware.
Correlated Equilibrium (CE) \cite{Aumann_1974} and Coarse Correlated Equilibrium (CCE) \cite{Blum_Mansour_2007} offer more computationally tractable alternatives for multi-agent games.
CE can be solved via Linear Programming (LP) \cite{H_RoughgardenTim_2008,Papadimitriou_2005}, but that requires generating utility matrices that scale exponentially with $|A|$ and can still be PPAD-complete under certain constraints \cite{Bernasconi_Castiglioni_Celli_Farina_2025}.
Best-Response Dynamics (BRD) is Polynomial Local Search (PLS)-complete \cite{Fabrikant_2004} with exponential worst-case time, like Nash.
Min-Max Regret (MMR) \cite{Savage_1951} and Fictitious Play (FP) \cite{Brandt_2010} require significant memory to persist large utility matrices as in LP implementations (MMR) or a potentially large number of historical probabilistic actions (FP).

To overcome time and memory complexities, no-regret learning dynamics work in a decentralised approach that allows each agent to independently focus on minimising its cumulative regret (i.e. a poor action taken from $\mathbb{D}_{A}$ that diverges from the best possible one) through $L$ iterations.
These can ensure, over time, that agents obtain an average reward that is asymptotically as good as that of any fixed strategy from the explored decision space.
This mechanism does not require loading the full decision strategy space into memory, which can converge rapidly to CCE \cite{Cesa-Bianchi_Lugosi_2006} and present upper boundaries on memory and time complexities that are considerably more bounded than other methods.
The convergence of these methods towards optimal strategies is determined by the upper bounds for regrets $O(\sqrt{T \cdot \ln N})$ and $O(\sqrt{T \cdot N \cdot \ln N})$, respectively; with $T = L$ as the number of iterations and $N = |\mathbb{D}_{A}|$ as the available actions/strategies per agent.
The selection of learning algorithms depends on factors like the comprehensiveness of the MAID $\mathcal{M}$ or the assured convergence of each algorithm. No-regret learning algorithms are very well-suited for this model, given its reduced complexity and fast convergence to CCE.


\section{Evaluation}
\label{sec:evaluation}

The Security DSS was evaluated in two ways: according to its internal behaviour and against a baseline.
On the one hand, its internal behaviour used different game-theoretic and recommender system's metrics. The metrics used for the full evaluation are described and interpreted first in Section \ref{sec:evaluation:internal}. A use case is proposed in Section \ref{sec:evaluation:use-case} to illustrate specific requirements passed to the DSS and analyse its outcomes (Table \ref{tab:evaluation:use-case:req-vs-coverage}) and aggregated statistical values for a subset of dataset sizes (Table \ref{tab:evaluation:use-case}).
On the other hand, the two baseline implementations were compared against the proposed work in Section \ref{sec:evaluation:comparison-baseline}. It shows first a summarised evaluation, aggregating data across a number of dataset sizes, modes and learning algorithms in Table \ref{tab:evaluation:baseline-vs-proposed:summary-t00-t03}.
Full details for the internal evaluation of the proposed recommender and the comparison against the greedy, filtered baseline are available in Table \ref{tab:evaluation:full-results} and Table \ref{tab:evaluation:comparison:results-0_0_0_8}, respectively.
Both tables summarise statistical values across different filtering iterations of the dataset (using thresholds $T$) for each metric, where each cell shows their mean ± SD.

It is important to note that the threshold $T$ (between 0.0 and 0.8) directly influences the size of dataset $DS_{8}$, since it filters those $N$ SPCSFs whose score\_total is equal or higher than $T$. A bigger threshold yields fewer SPCSFs ($T \propto 1 / N$).
Consequently, a good recommending accuracy expects a sufficiently large $N$, as better SPCSFs can be recommended from a larger pool.

\subsection{Methodology}
\label{sec:evaluation:methodology}

The following metrics are considered: (i) performance (time taken to produce the recommendation); (ii) efficiency (minimisation of resources used to cover the requirements); (iii) accuracy (recommendation fitting with respect to requirements); and (iv) equity (degree of satisfaction per requirement without actively penalising others).
The recommender was run $m \cdot n \cdot o$ times in nested iterations for both validations; with (i) $m$ being the range of thresholds ($T$); (ii) $n$ being the number of iterations (100); and (iii) $o$ each of the modes ($M$), i.e. "clustered" or "full".
The security requirements are randomly generated within the $n$ loop to ensure different constraints considered in the recommendation, which also introduces considerable variability in multiple metrics.

The proposed Security DSS (comprised by Algorithms \ref{alg:implementation-model:maid-generation} and \ref{alg:implementation-model:gt-exploration-recommendation}) explores the decision space within each mode $M$ in the $o$ loop by using two learning algorithms $\mathcal{L}$ (Hedge / EWA and EXP3) to minimise the regret on the selected choice. 
These repeat explorations for 200 and 400 iterations, respectively.
This work is also compared in Section \ref{sec:evaluation:comparison-baseline} against a greedy filtered recommender that implements the last sorting and filtering stage from Algorithm \ref{alg:implementation-model:gt-exploration-recommendation} (lines \ref{alg:implementation-model:gt-exploration-recommendation:line:s15}-\ref{alg:implementation-model:gt-exploration-recommendation:line:s23}).

\subsection{Internal evaluation}
\label{sec:evaluation:internal}

The box plots in the following subsections render average values and represent the IQR.

\subsubsection{Performance}
\label{sec:evaluation:internal:performance}

The performance of the system evaluates the time taken by the different stages from the Security DSS. That is, to (i) generate the MAID first; (ii) explore the decision space to look for a joint decision profile as approximate equilibrium; and (iii) perform the recommendation filtering and ranking.

Times are shown in Table \ref{tab:evaluation:full-results}.
Both the time to generate the MAID $\mathcal{M}$ and the recommendations $\mathcal{R}$ are negligible in comparison with the learning time $\mathcal{L}$ for the game-theoretic exploration of the space; which is mainly caused by the hundreds of iterations required to reach enough decision profiles.
The MAID generation time is considerably slower for the clustered mode with respect to the full mode; as expected given that the clustered mode always run the GMM clustering process. This difference in time reduces with a higher $T$ and less available SPCSFs, as there are less operations to perform.
The recommendation time for the full mode, on the other hand, approximately doubles that of the clustered mode; given the larger number of chance nodes $\omega$ it handles directly. This also decreases along with fewer SPCSFs.

\subsubsection{Efficiency}
\label{sec:evaluation:internal:efficiency}

The efficiency of the system is measured by (i) the Price of Anarchy (PoA) from Eq. \eqref{eq:evaluation:internal:poa}, which measures the efficiency of the game-theoretic equilibrium; as well as the (ii) minimum number of SPCSFs, which heavily depends on the dataset size and is also limited by an internal budget control to avoid recommending unnecessary controls when security requirements are covered -- and thus limit over-provisioning.

\begin{equation}
\label{eq:evaluation:internal:poa}
    \text{PoA} = \frac{\max_{\sigma \in \mathcal{H}} \text{Welfare}(\sigma)}{\text{Welfare}(\sigma^*)}
\end{equation}

\begin{figure}[h]
    \centering
    \includegraphics[width=0.95\columnwidth]{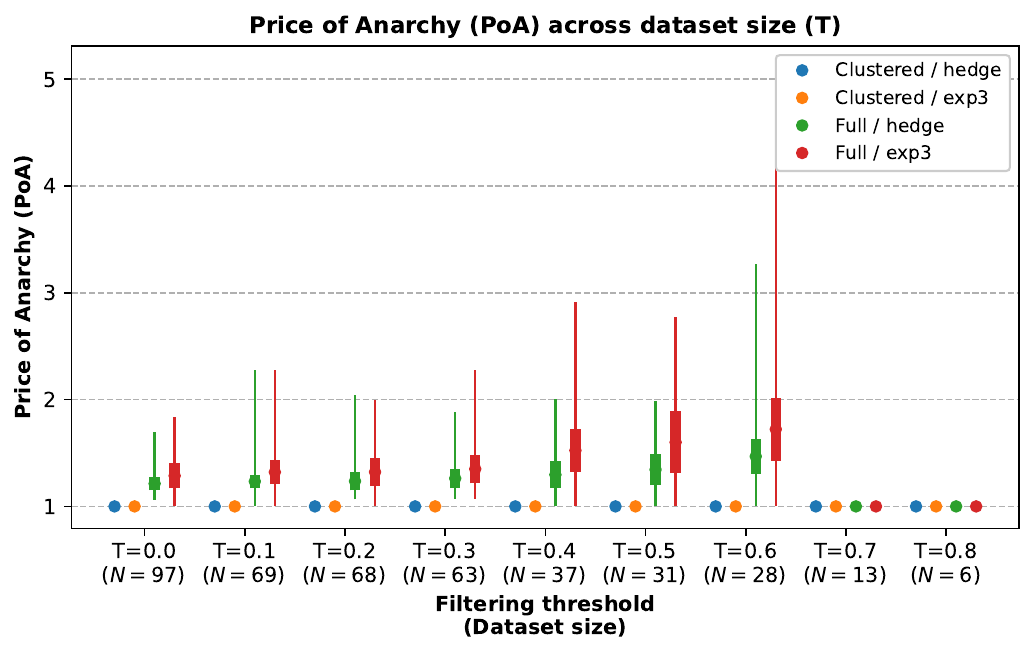}
    \caption{PoA from T=0.0 to T=0.8, 100 iterations}
    \label{fig:evaluation:internal:poa}
\end{figure}

The PoA checks how the system efficiency degrades with selfish behaviour from the agents through the ratio between the optimal solution (maximum welfare of all explored strategies $\sigma \in \mathcal{H}$) and the achieved equilibrium (welfare of the optimal joint profile $\sigma^{*}$).
The perfect ratio, and minimum boundary, is 1.
Fig. \ref{fig:evaluation:internal:poa} showcases the distribution of the PoA values across different dataset sizes; and averaged values are provided in Table \ref{tab:evaluation:full-results}.
There, PoA stays equal or very close to 1 for both learning algorithms in the clustered mode; and grows with the full (large) MAID for both learning algorithms until $T = 0.5$, especially increased for EXP3 and suffering from spikes in some outliers.

\begin{figure}[h]
    \centering
    \includegraphics[width=0.95\columnwidth]{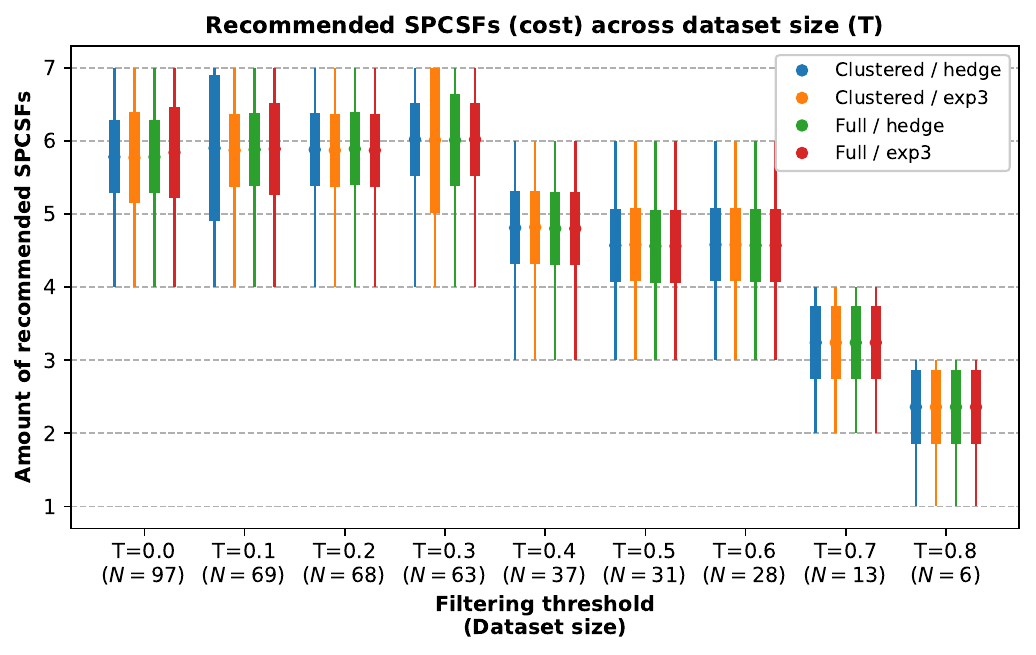}
    \caption{No. of recommended SPCSFs from T=0.0 to T=0.8, 100 iterations}
    \label{fig:evaluation:internal:controls-recommended}
\end{figure}

On the other hand, the minimum number of SPCSFs is computed to stop recommending SPCSFs the moment it satisfies the requirements, as well as re-iterating to minimise unwanted over-provisioning; while ensuring there is no avoidable under-provisioning.
Fig. \ref{fig:evaluation:internal:controls-recommended} plots the number of recommended SPCSFs with different T (i.e. dataset sizes).
A higher number of SPCSFs allows the Security DSS to recommend more of them to better fit the security requirements; yet that amount must be limited to controls that strictly cover these requirements, without adding more (i.e. over-provisioning).
This number ranges between 3 and 6 SPCSFs, even with the smallest dataset ($T = 0.8$, with 6 SPCSFs).
When there are few SPCSFs available, the IoU and satisfaction coverage tend to be smaller.
This figure shows that between $T = 0.0$ and $T = 0.2$, where there is higher satisfaction and accuracy (Section \ref{sec:evaluation:accuracy}), at least 6\%-9\% of the original or mildly filtered dataset $DS_{8}$ is enough to deliver accurate results.

\subsubsection{Accuracy}
\label{sec:evaluation:accuracy}

The accuracy is measured by (i) the satisfaction ratios, which measure how well each requirement to cover a security dimension is achieved, and is obtained through the ratio $s_{r} = $ recommended / requested $\mid s_{r} \in [0, 1]$; and (ii) the Normalised Discounted Cumulative Gain (NDCG) and (iii) the coverage ratios.

NDCG, in Eq. \eqref{eq:evaluation:internal:ndcg:base}, is a well-known ranking quality metric that measures how well an algorithm sorts items based on relevance.
To do so, it compares the top $K$ items ($K$ being the cut-off point) of the ranking to the ideal order (based on relevance) and provides a ratio, where 1 means full alignment (higher quality or accuracy).
It is important to point out that the relevance criteria ($rel_{i}$) per item $i$ can differ depending on the purpose of the system. For instance, it can assume the ground truth as the set of items with highest total score, in descending order; or it can assume as ground truth those items that are the most precise in term of a particular use case.

\begin{equation}
\begin{gathered}
\label{eq:evaluation:internal:ndcg:base}
\text{DCG}@K = \sum_{i=1}^{K} \frac{rel_{i}}{\log_{2}(i+1)} \\[0.2em]
\text{NDCG}@K = \frac{\text{DCG}@K ~ \text{(obtained)}}{\text{DCG}@K ~ \text{(ideal)}}
\end{gathered}
\end{equation}

In this case, both approaches were considered to generate the $DCG@K$ ideal value (ground truth), using a subset of SPCSFs ordered by relevance: (i) of those with highest total score ($rel_i^1 = f(\text{score\_total})$); and (ii) of those that avoid under-provisioning and minimise over-provisioning, accommodating the most to the exact requirements ($rel_i^2 = f(\text{coverage})$).
Both values are provided in Table \ref{tab:evaluation:full-results} as "NDCG (Score)" and "NDCG (Min)", respectively.
The first one, purely based on the "score\_total" column from $DS_{8}$ (where the top 20 high-scoring SPCSFs are shown in Table \ref{tab:implementation-dataset:composition:top-20-contribution-impl}), yields 82\%-99\% on the clustered MAID and 52\%-99\% for the full MAID; both increasing along with $T$ (i.e. with fewer SPCSFs to select from for the ranking).
The second one, mostly focused on the "coverage\_*" columns from $DS_{8}$ (denoting contribution per SPCSF to an agent $a$), the values range between 99\%-100\% in the clustered MAID and 96\%-100\% for the full MAID.
Given that both the utility function from the exploration stage, and the criteria of the recommender stage both point at ensuring there is no under-provisioning and avoiding over-provisioning, suggesting a minimal set of SPCSFs, it is natural that the second NDCG is much more aligned and stable.
Another measurement comes from the coverage ratios. These are evaluated both using linear and geometric metrics and consider the security requirement array ($\vec{\tau}$), the security coverage array ($a_{i}$) and the number of agents ($|A|$). The best value these can achieve is 1, as it indicates full satisfaction/coverage.

On the one hand, the considered linear metrics are as follows:
\begin{itemize}
    \item The satisfaction coverage, in Eq. \eqref{eq:evaluation:linear_sat}, measures the provided coverage per agent, ensuring its satisfaction to avoid under- or over provisioning.
    \item The resource efficiency, in Eq. \eqref{eq:evaluation:linear_eff}, measures over-\\provisioning, as an unnecessary bloat that adds nothing to cover the security requirements.
    \item The weighted Jaccard similarity index, or Intersection over Union (IoU), calculates how well the generated array (or shape) for the recommendation aligns with respect to the array (or shape) of the provided security requirements. This is computed in Eq. \eqref{eq:evaluation:jaccard-continuous-index} and where $x_{i} = a_{i}, y_{i} = {\tau_{i}}$ as a 1D array; yet the formula can also be used for a 2D plane -- which is, after all, the graphical representation of the covered requirements.
\end{itemize}

\begin{subequations}
\label{eq:linear_metrics}
\begin{align}
    \text{Satisfaction coverage} &= \frac{\sum_{a=1}^{|A|} \min(c_{a}, \tau_{a})}{\sum_{a=1}^{|A|} \tau_{a}} \label{eq:evaluation:linear_sat} \\[1ex]
    \text{Resource efficiency} &= \frac{\sum_{a=1}^{|A|} \min(c_{a}, \tau_{a})}{\sum_{a=1}^{|A|} c_{a}} \label{eq:evaluation:linear_eff} \\[1ex]
    \text{Jaccard (IoU)} &= \frac{\sum_{a=1}^{|A|} \min(c_{a}, \tau_{a})}{\sum_{a=1}^{|A|} \max(c_{a}, \tau_{a})} \label{eq:evaluation:jaccard-continuous-index}
\end{align}
\end{subequations}

\begin{figure}[h]
    \centering
    \includegraphics[width=0.95\columnwidth]{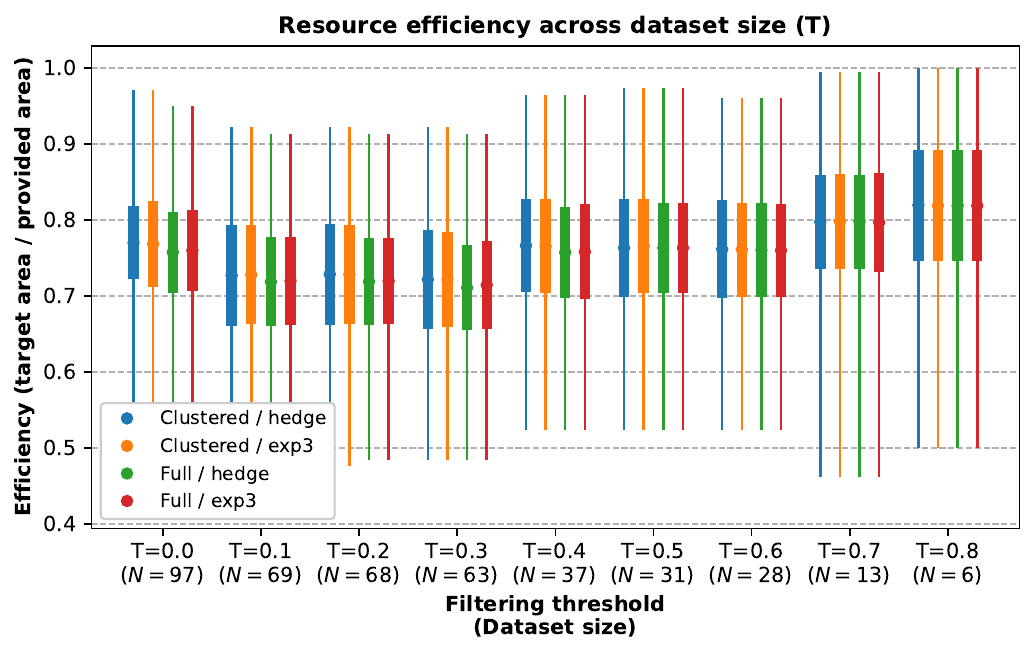}
    \caption{Resource efficiency from T=0.0 to T=0.8, 100 iterations}
    \label{fig:evaluation:accuracy:res-eff}
\end{figure}

The average resource efficiency (Fig. \ref{fig:evaluation:accuracy:res-eff}) stays relatively constant, between 72\% and 82\%. This hints that the over-provisioning varies between 18\% and 28\%.

When the recommended SPCSFs are more specialised and better fit few expected security requirements, less over-provisioning is expected. The high variability observed as $T$ increases could be derived from less specialised controls that may be providing extra, unrequested coverage.
An example of this is visible in the single runs shown in Fig. \ref{fig:implementation-model:gt-recom} and Fig. \ref{fig:evaluation:use-case}; where some SPCSFs over-contribute and negatively impact this metric.

\begin{figure}[h]
    \centering
    \includegraphics[width=0.95\columnwidth]{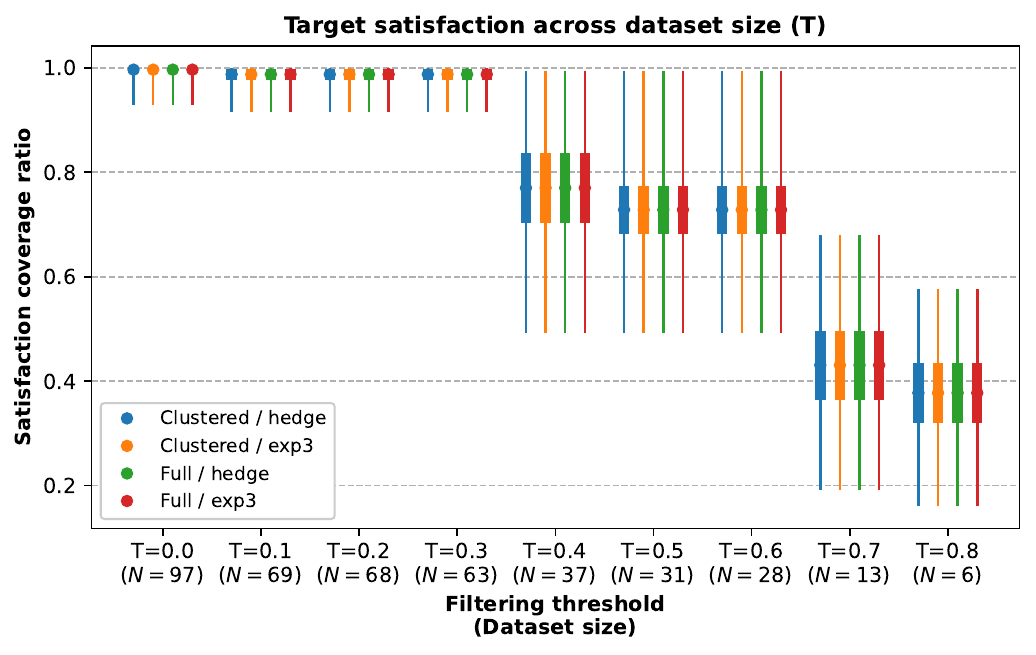}
    \caption{Satisfaction coverage from T=0.0 to T=0.8, 100 iterations}
    \label{fig:evaluation:accuracy:sat-cov}
\end{figure}

The satisfaction coverage, shown in Fig. \ref{fig:evaluation:accuracy:sat-cov}, proves that until $T = 0.3$, all security requirements are almost fully covered at 99\%-100\%. Beyond it, and due to the smaller datasets and lesser availability of SPCSFs, this coverage drops to 73\% on $T = 0.5$ (31 SPCSFs) and to $\sim$40\% from $T = 0.7$ (6-13 SPCSFs).

\begin{figure}[h]
    \centering
    \includegraphics[width=0.95\columnwidth]{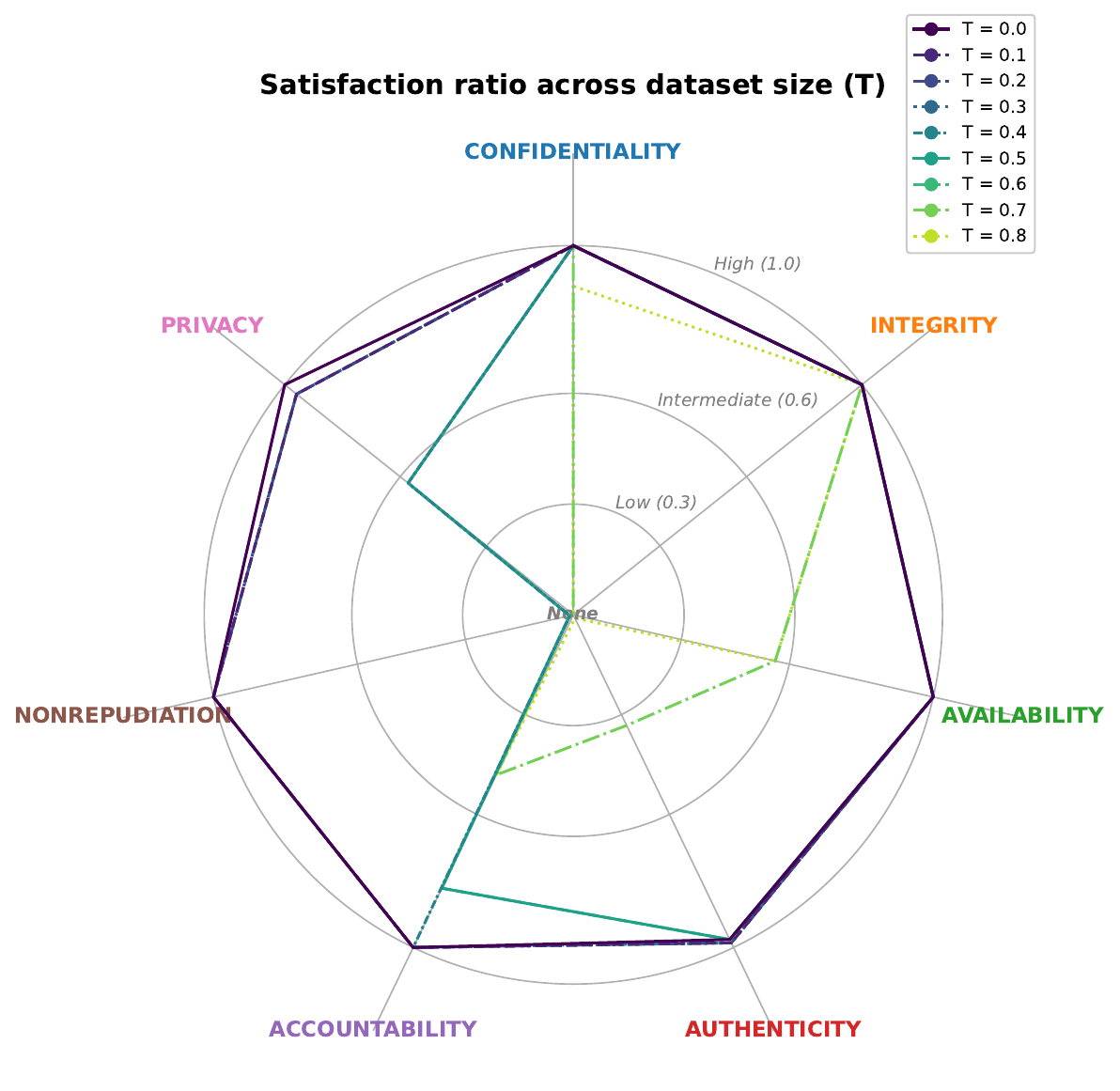}
    \caption{Satisfaction ratio from T=0.0 to T=0.8, 100 iterations}
    \label{fig:evaluation:accuracy:sat-ratio}
\end{figure}

Another way to look at how well the coverage is satisfied is the satisfaction ratio, which simply divides the recommended coverage between the requested one, per agent $a$. See Fig. \ref{fig:evaluation:accuracy:sat-ratio}, showing the progression of the average satisfaction per $a$ across each threshold $T$.
This plot provides detail on which $a$ are most affected by the reduction of the dataset size as $T$ increases: except for confidentiality, integrity and accountability, other SPCSFs are gradually affected or even drastically - the more with under-covered security dimensions for SPCSFs, as for non-repudiation and privacy (Table \ref{tab:implementation-dataset:mapping:ciatunp}).

\begin{figure}[h]
    \centering
    \includegraphics[width=0.95\columnwidth]{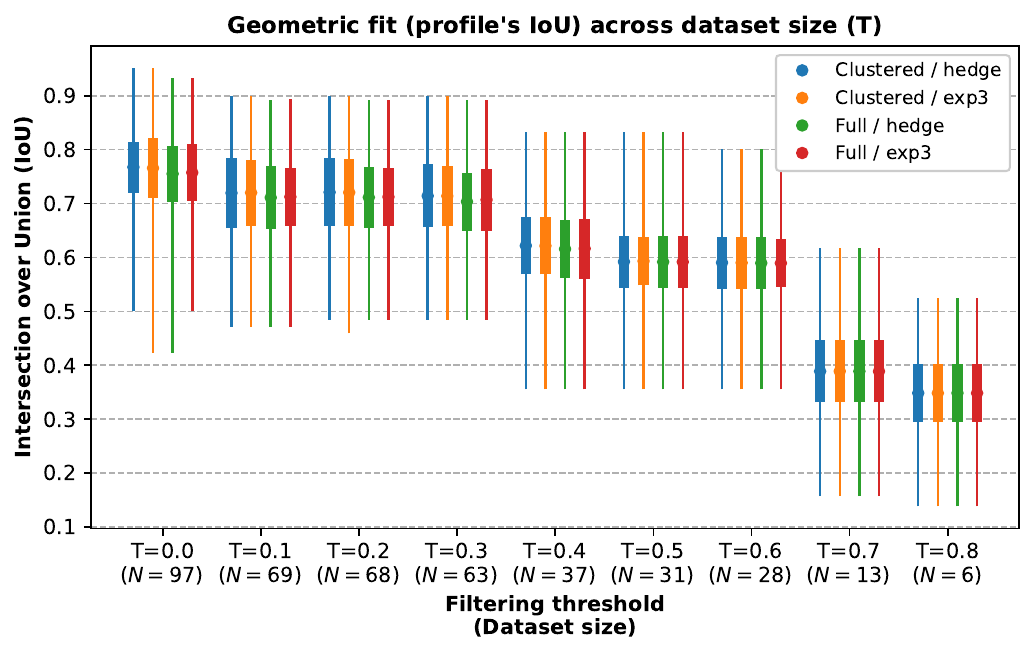}
    \caption{Jaccard / IoU metric from T=0.0 to T=0.8, 100 iterations}
    \label{fig:evaluation:accuracy:iou}
\end{figure}

Fig. \ref{fig:evaluation:accuracy:iou} depicts the IoU, diminishing steadily from the full dataset with 77\% ($T = 0.0$, with 97 SPCSFs) towards the minimal dataset ($T = 0.8$, with 6 SPCSFs), with 35\%. At $T = 0.5$, the intersection of the recommended VS requested security requirements drops below 60\%, impacted already by the lack of SPCSFs in the dataset with enough total score and dimension coverage.

On the other hand, the considered geometric metrics are:
\begin{itemize}
    \item The area coverage, in Eq. \eqref{eq:evaluation:area_cov}, assesses the percentage of the area of the requested security requirements that is also covered by the area of the recommended security controls. This indirectly measures under-provisioning; which is one of the key (utility) terms to minimise.
    \item The area efficiency, in Eq. \eqref{eq:evaluation:area_eff}, indirectly measures over-provisioning; as it identifies which area from the recommended security controls contributes to fill/intersect the area from the requested security requirements.
\end{itemize}

\begin{subequations}
\label{eq:evaluation:geometric_metrics}
\begin{align}
    \text{Area coverage} &= \frac{\text{Area}(\min(\vec{c}, \vec{\tau}))}{\text{Area}(\vec{\tau})} \label{eq:evaluation:area_cov} \\[1ex]
    \text{Area efficiency} &= \frac{\text{Area}(\min(\vec{c}, \vec{\tau}))}{\text{Area}(\vec{c})} \label{eq:evaluation:area_eff}
\end{align}
\end{subequations}

Since this is a 2D plane, the area of each polygon with $|A|$ and formed by a vector $\vec{x}$ is computed for both metrics with Eq. \eqref{eq:evaluation:radar_area}.
Other common operations are the intersection vector between each element of the security coverage array ($a_{i}$) and the target array $\vec{\tau}$, with $\min(\vec{a}, \vec{\tau})$.

\begin{equation}
    \text{Area}(\vec{x}) = \frac{1}{2} \sin\left(\frac{2\pi}{|A|}\right) \sum_{i=1}^{|A|} x_i \cdot x_{i+1} \mid x_{|A|+1} = x_1 \label{eq:evaluation:radar_area}
\end{equation}

\begin{figure}[h]
    \centering
    \includegraphics[width=0.95\columnwidth]{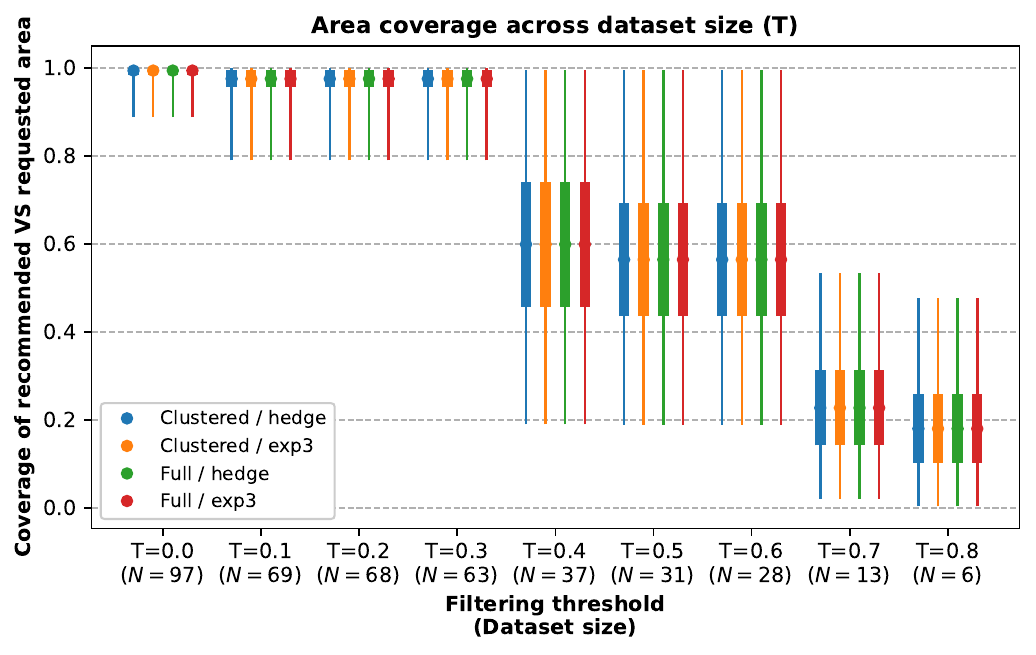}
    \caption{Area coverage from T=0.0 to T=0.8, 100 iterations}
    \label{fig:evaluation:accuracy:area-coverage}
\end{figure}

The area coverage is shown in Fig. \ref{fig:evaluation:accuracy:area-coverage}. It follows a trend that is relatively similar to that in the satisfaction coverage (Fig. \ref{fig:evaluation:accuracy:sat-cov}), demonstrating how between $T = 0.0$ and $T = 0.3$ the security recommendations are almost perfectly aligned with the security requirements and how it drops around $\sim$50\% accuracy starting in $T = 0.5$, where only 32\% of the initial implementable SPCSFs are then available.

\begin{figure}[h]
    \centering
    \includegraphics[width=0.95\columnwidth]{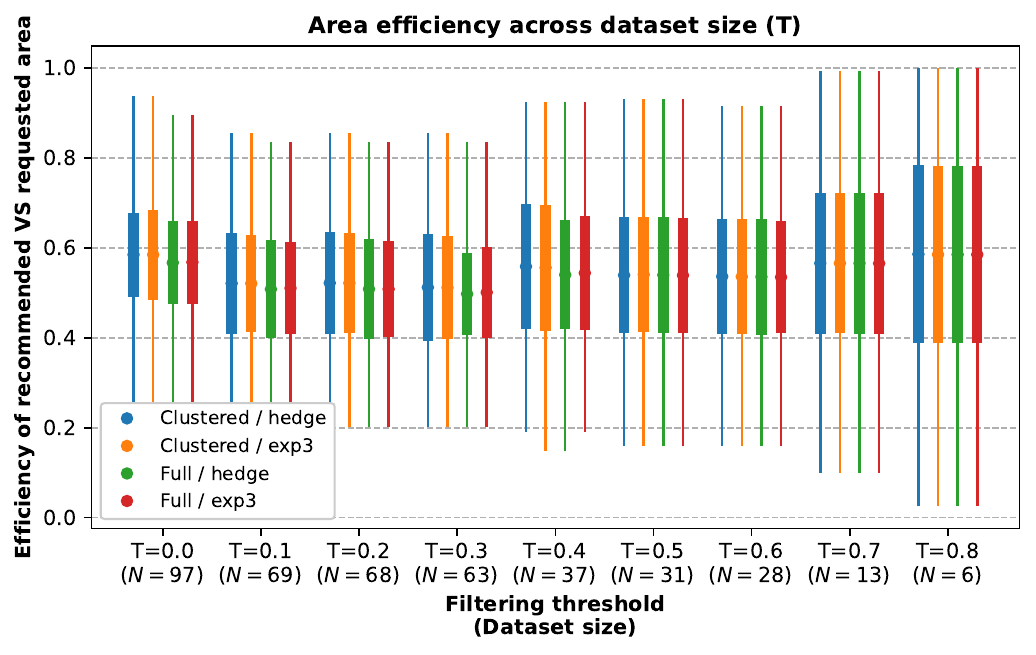}
    \caption{Area efficiency from T=0.0 to T=0.8, 100 iterations}
    \label{fig:evaluation:accuracy:area-efficiency}
\end{figure}

Fig. \ref{fig:evaluation:accuracy:area-efficiency} plots the area efficiency as highly variable (highest SD in Table \ref{tab:evaluation:full-results}).
This lesser availability of SPCSFs for suggestion is likely introducing extra, not needed coverage on each of the requested agents.

\subsubsection{Equity}
\label{sec:evaluation:equity}

The fairness or equity of the system is assessed by the Jain's index in Eq. \eqref{eq:evaluation:jain-index}. This index is commonly used to evaluate resource consumption in networks (e.g. assigned network flows), and provides a ratio between the sum of the resources to share ($x_{i}$), powered to two, and the number of users ($n$) and the quadratic sum of the same resources.
In this context, $n = |A|$ and resources are satisfaction ratios.

\begin{equation}
\label{eq:evaluation:jain-index}
    \mathcal{J}(x_{1}, ..., x_{n}) = \frac{\left(\sum_{i=1}^{n} x_i\right)^2}{n \cdot \sum_{i=1}^{n} x_i^2}
\end{equation}

The outcome denotes the equity of the system in that the provided recommendations do not sacrifice too much to cover a given agent at the expense of others. This situation is more evident upon negative synergies across clusters or SPCSFs. The perfect ratio, and maximum boundary, is 1; with the minimum being $1 / n$.

\begin{figure}[h]
    \centering
    \includegraphics[width=0.95\columnwidth]{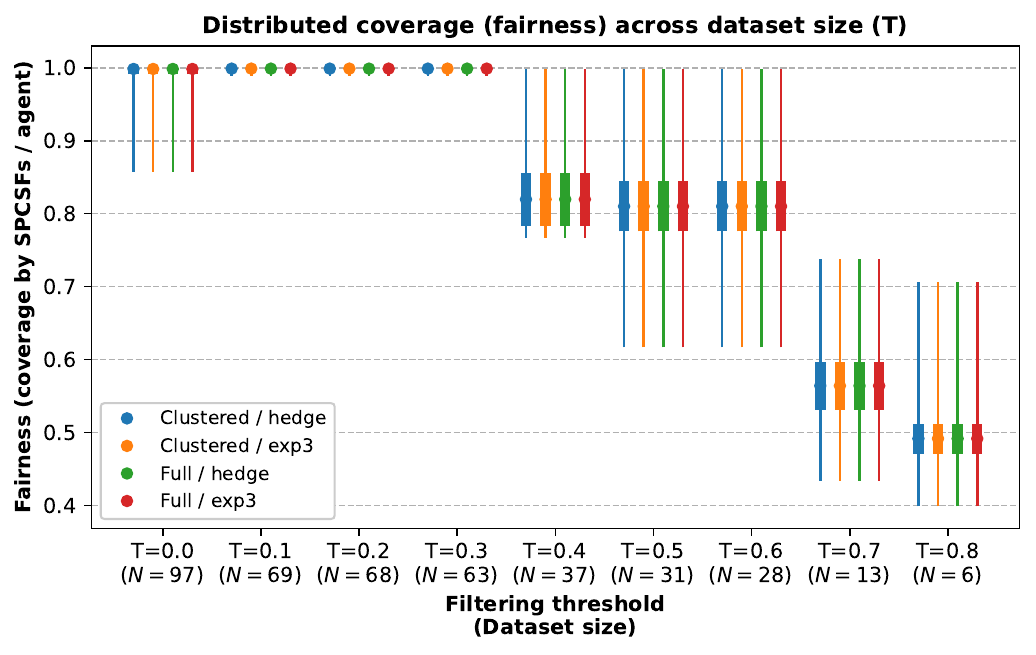}
    \caption{Jain's fairness from T=0.0 to T=0.8, 100 iterations}
    \label{fig:evaluation:accuracy:jains-fairness}
\end{figure}

In this case, the Jain's fairness (Fig. \ref{fig:evaluation:accuracy:jains-fairness}) is stable at 1 until $T = 0.3$, after which it stabilises at 0.81-0.82 until $T = 0.6$ and degrades until reaching 0.49.

\subsection{Use case evaluation}
\label{sec:evaluation:use-case}

The 7 considered security dimensions are already considered as part of the 8 security goals, extended from the CIA triad in the Reference Model of Information Assurance \& Security (RMIAS) or IAS-Octave \cite{Cherdantseva_Hilton_2013} framework. The authors provide a qualitative analysis of an organisation's information security policies, qualitatively mapping a subset of these to the CIA triad and the accountability security dimensions.
Similarly, authors from \cite{Ayedh_M_Wahab_Idris_2023} analyse a Bring Your Own Device (BYOD) environment and suggest that both the CIA, accountability and auditability (potentially considered as a group under authenticity), privacy and non-repudiation security dimensions are required to ensure traceability of users' devices for post-incident analyses.

Based on the above, this section first performs a high-level analysis of an IT enterprise environment to extract its requirements in terms of security dimensions, roughly quantifying these. These are used as a specific use case to validate against the proposed recommender.
A common, non-critical IT enterprise delivers solutions that are often protected by Intellectual Property Rights (IPR), leverages private/public cloud resources to carry out the work and may allow BYOD to some extent. The security of the data generated is subject to IPR and guided through institutional security policies. Likewise, the proper functioning and security of the hardware and tools in use is dictated by security policies that monitor and alert about malfunctioning, filter specific applications or actions within a system and continuously monitor the security posture of both internal and BYOD devices to grant (Zero Trust) access to network segments and the resources accessible by these.
In this sense, such an environment emphasises data confidentiality to conform to the IPR internal regulation. It also requires maintaining good availability and integrity for both internal and potentially customer-exposed data and services, as well as a degree of privacy when processing human data to cover regulatory requirements. Authenticity, accountability and non-repudiation are needed to minimise unauthorised access to such resources and to identify who did what when analysing incidents.
Considering this, an operator setting up the supporting environment for such organisation may quantify security requirements $\vec{\tau}$ as those from Table \ref{tab:evaluation:use-case:req-vs-coverage}.

\begin{figure*}[htbp]
\begin{minipage}[t]{0.24\textwidth}
    \begin{subfigure}{\textwidth}
        \includegraphics[draft=false, width=\textwidth]{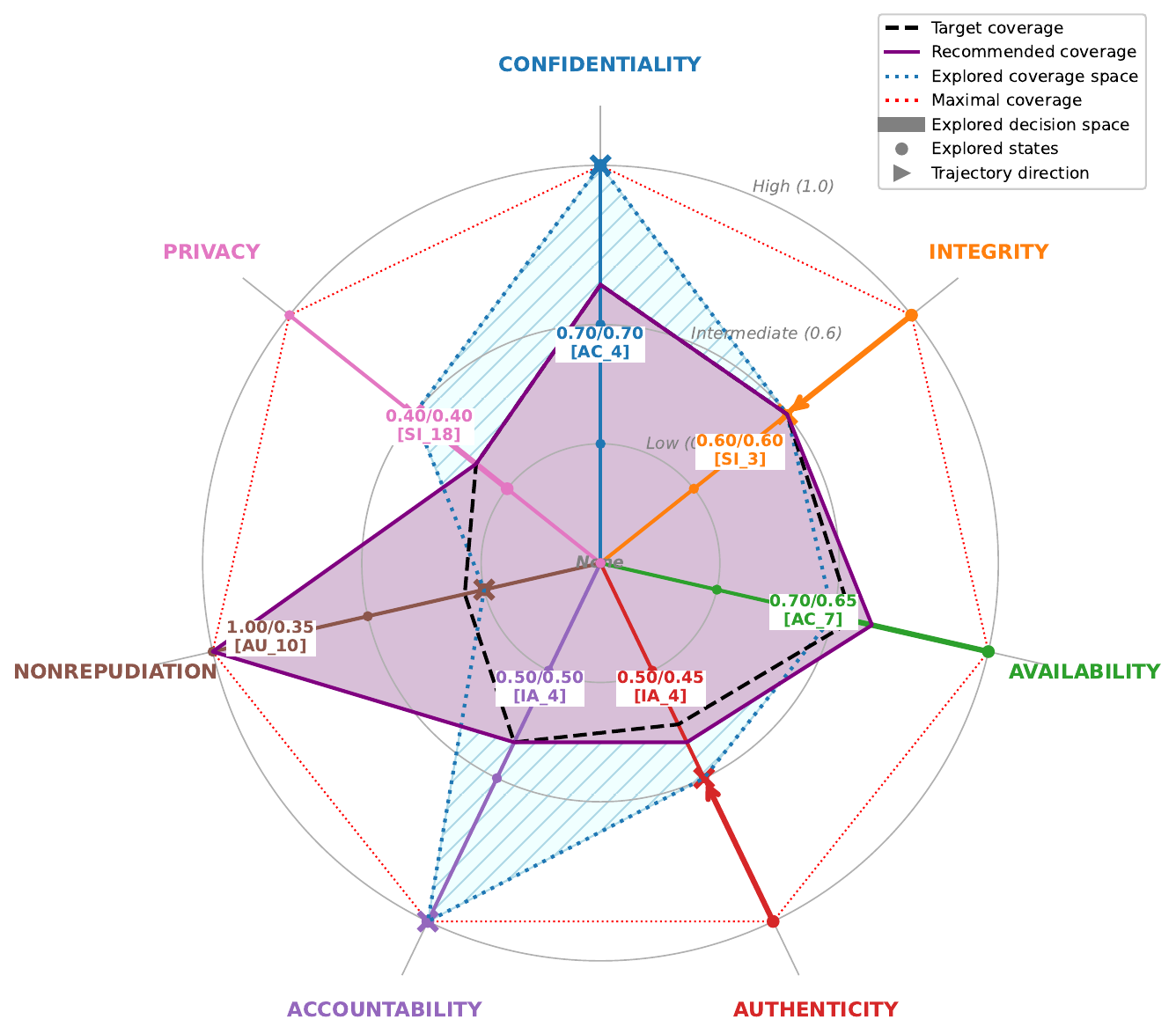}
        \caption{Clustered $\mathcal{M}$  ($T=0.0$)}
        \label{fig:evaluation:use-case:clust-exp3-t00}
    \end{subfigure}
\end{minipage}
\begin{minipage}[t]{0.24\textwidth}
    \begin{subfigure}{\textwidth}
        \includegraphics[draft=false, width=\textwidth]{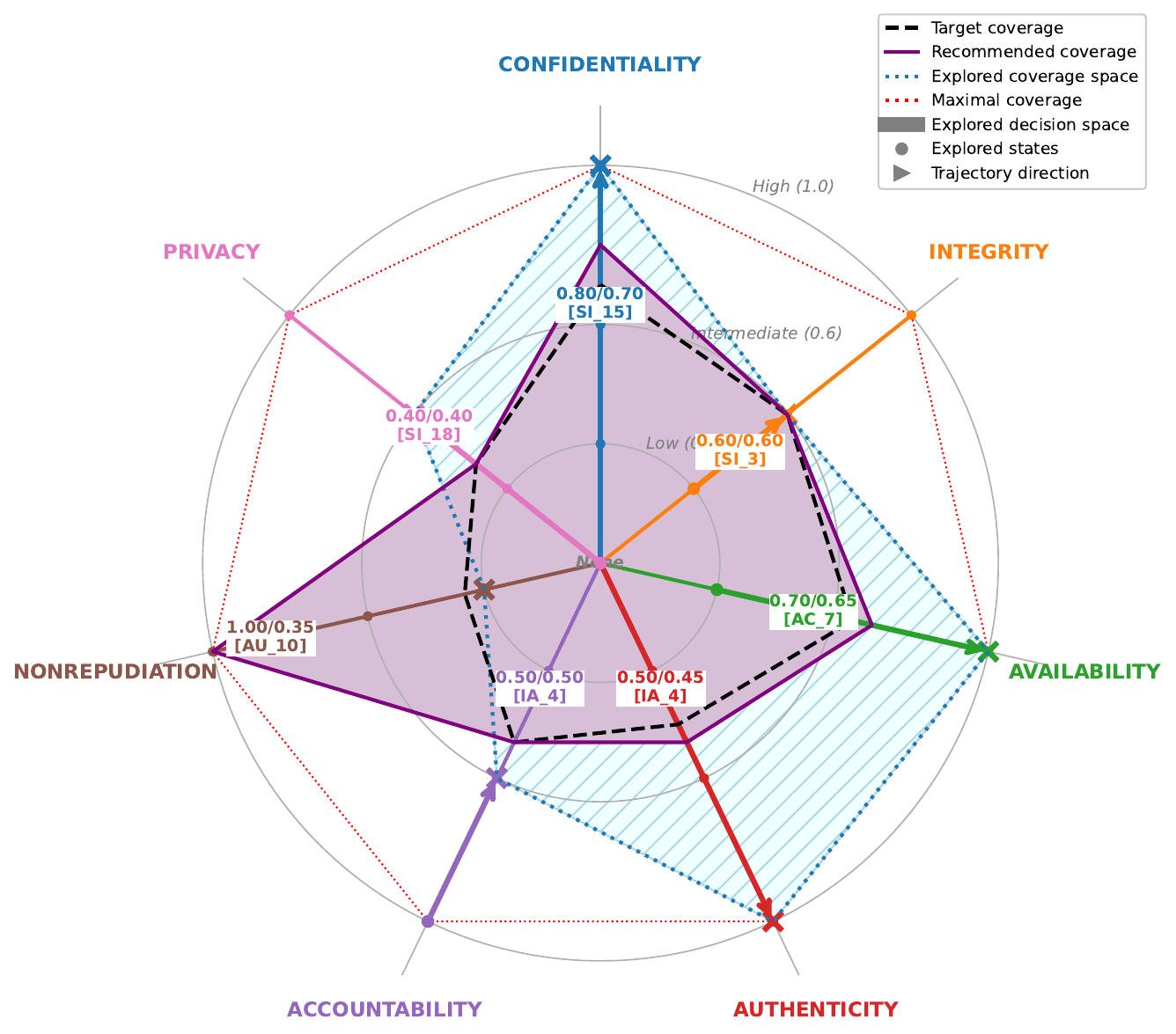}
        \caption{Full $\mathcal{M}$ ($T=0.0$)}
        \label{fig:evaluation:use-case:full-exp3-t00}
    \end{subfigure}
\end{minipage}
\begin{minipage}[t]{0.24\textwidth}
    \begin{subfigure}{\textwidth}
        \includegraphics[draft=false, width=\textwidth]{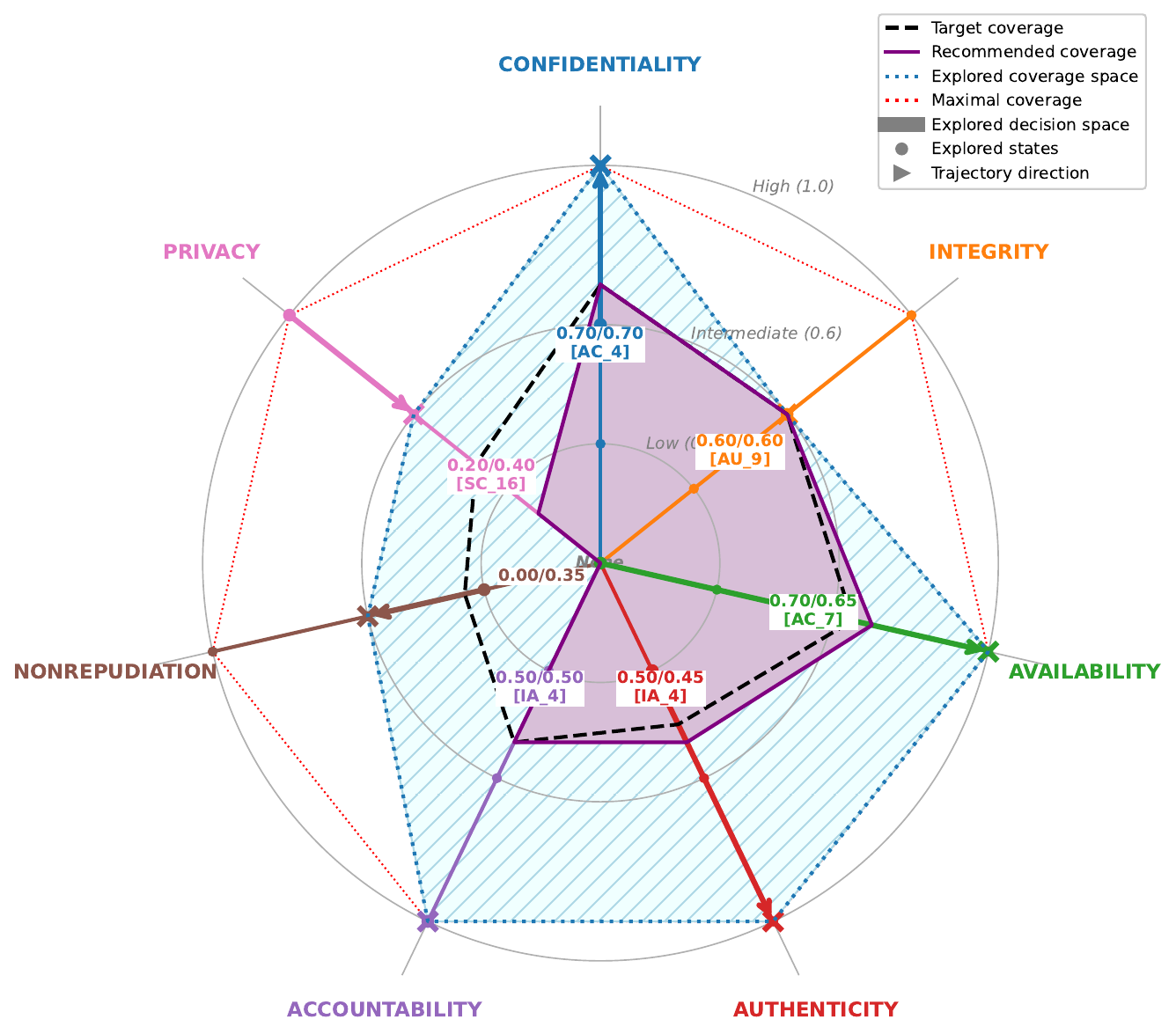}
        \caption{Clustered $\mathcal{M}$ ($T=0.4$)}
        \label{fig:evaluation:use-case:clust-exp3-t04}
    \end{subfigure}
\end{minipage}
\begin{minipage}[t]{0.24\textwidth}
    \begin{subfigure}{\textwidth}
        \includegraphics[draft=false, width=\textwidth]{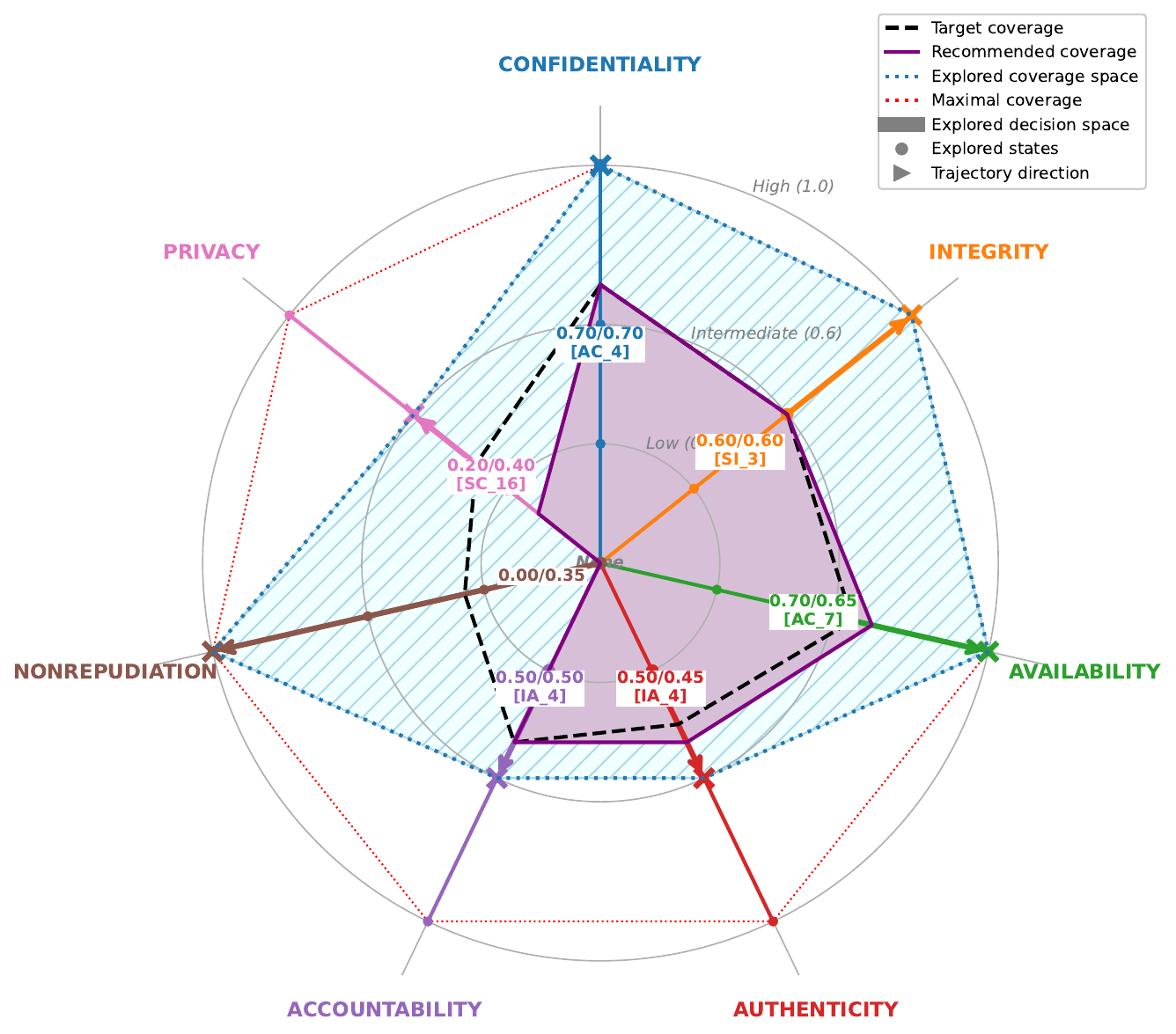}
        \caption{Full $\mathcal{M}$  ($T=0.4$)}
        \label{fig:evaluation:use-case:full-exp3-t04}
    \end{subfigure}
\end{minipage}
\caption{MAID $\mathcal{M}$ exploration and recommendation for the use case with EXP3 as $\mathcal{L}$}
\label{fig:evaluation:use-case}
\end{figure*}

\begin{table}[htbp]
  \centering
  \caption{Required ($\vec{\tau}$) and covered ($\vec{c}$) security dimensions per $M$ and $T$.}
  \label{tab:evaluation:use-case:req-vs-coverage}
  \resizebox{\columnwidth}{!}{%
  \begin{tabular}{l c c c c c}
      \toprule
      \multirow{2}{*}{\textbf{Dim.}} & \multirow{2}{*}{\textbf{Req. ($\tau$)}} & \multicolumn{2}{c}{\textbf{$\vec{c}\quad (T=0.0)$}} & \multicolumn{2}{c}{\textbf{$\vec{c} \quad (T=0.4)$}} \\
      \cmidrule(lr){3-4} \cmidrule(l){5-6}
      & & \textbf{$M$ = Clustered} & \textbf{$M$ = Full} & \textbf{$M$ = Clustered} & \textbf{$M$ = Full} \\
      \midrule
      C  & 0.70 & 0.70 (AC-4) & 0.80 (SI-15)& 0.70 (AC-4) & 0.70 (AC-4) \\
      I & 0.60 & 0.60 (SI-3) & 0.60 (SI-3) & 0.60 (AU-9) & 0.60 (SI-3) \\
      A & 0.65 & 0.70 (AC-7) & 0.70 (AC-7) & 0.70 (AC-7) & 0.70 (AC-7) \\
      Au & 0.45 & 0.50 (IA-4) & 0.50 (IA-4) & 0.50 (IA-4) & 0.50 (IA-4) \\
      Ac & 0.50 & 0.50 (IA-4) & 0.50 (IA-4) & 0.50 (IA-4) & 0.50 (IA-4) \\
      Nr & 0.35 & 1.00 (AU-10) & 1.00 (AU-10) & - & - \\
      Pr & 0.40 & 0.40 (SI-18) & 0.40 (SI-18) & 0.20 (SC-16) & 0.20 (SC-16)\\
      \bottomrule
  \end{tabular}%
  } 
\end{table}

The recommender is run once with the indicated requirements, testing both modes $M$ and methods $\mathcal{L}$ and using both the full pool of SPCSFs ($T=0.0$) and a constrained one ($T=0.4$) to assess the changing behaviour.
Fig. \ref{fig:evaluation:use-case} and Table \ref{tab:evaluation:use-case:req-vs-coverage} represent the DSS outcomes in graphical and tabular form: the suggested SPCSFs and the coverage $\vec{c}$ per security dimension. This is shown here across modes $M$ and $T$ values whilst pinning $\mathcal{L}$ to EXP3 for brevity.
Among the proposed SPCSFs are those preserving integrity (\textit{SI-3: Malicious Code Protection}, \textit{AU-9: Protection of Audit Information}), availability (\textit{AC-7: Unsuccessful Logon Attempts}) or authenticity (\textit{IA-4: Identifier Management}). 
Table \ref{tab:evaluation:use-case} shows averaged values for relevant metrics across both learning methods $\mathcal{L}$.

\begin{table}[htbp]
    \centering
    \caption{Statistics for single use case executions (aggregated $\mathcal{L}$ per $M$).}
    \label{tab:evaluation:use-case:results-summary}
    \resizebox{\columnwidth}{!}{%
    \begin{tabular}{lcccc}
        \toprule
        & \multicolumn{2}{c}{\textbf{$T=0.0$}} & \multicolumn{2}{c}{\textbf{$T=0.4$}} \\
        \cmidrule(lr){2-3} \cmidrule(l){4-5}
        \textbf{Metric} & \textbf{$M$ = Clustered} & \textbf{$M$ = Full} & \textbf{$M$ = Clustered} & \textbf{$M$ = Full} \\
        \midrule
        Time ($\mathcal{L}$) & 3.4853 & 27.2730 & 1.1592 & 10.7252 \\
        Time ($\mathcal{R}$) & 0.0026 & 0.0031 & 0.0013 & 0.0019 \\
        Fairness & 1 & 1 & 0.8231 & 0.8231 \\
        Cost ($|\mathcal{R}|$) & 6 & 6 & 5 & 5 \\
        Sat. Cov. & 1 & 1 & 0.8493 & 0.8493 \\
        Res. Eff. & 0.8295 & 0.8111 & 0.9688 & 0.9688 \\
        Prof. IoU & 0.82955 & 0.8111 & 0.8267 & 0.8267 \\
        NDCG (Min) & 0.9945 & 0.9618 & 0.9963 & 0.9817 \\
        \bottomrule
    \end{tabular}
    }
  \label{tab:evaluation:use-case}
\end{table}

This use case is specially illustrative as it exhibits both over- and under- provisioning --- where the latter is even more critical to avoid, as indicated in Section \ref{sec:implementation_model:provisioning-penalty}.
Over-provisioning occurs at $T=0.0$, with all available SW-implementable SPCSFs ($N=97$): Fig. \ref{fig:evaluation:use-case:clust-exp3-t00} and Fig. \ref{fig:evaluation:use-case:full-exp3-t00} show a spike at Nr and offering maximum coverage, well beyond the explored decision space and the targeted security requirements that directly or indirectly act as bounds. Only 2 SPCSFs cover Nr (Table \ref{tab:implementation-dataset:mapping:ciatunp}), namely AU-8 and AU-10; which offer coverage of 0.2 and 1 and have a total score of 0.071 and 0.392, respectively. AU-10 can thus avoid under-provisioning.
Under-provisioning happens at $T=0.4$, with $N=37$ SPCSFs: there, Fig. \ref{fig:evaluation:use-case:clust-exp3-t04} and Fig. \ref{fig:evaluation:use-case:full-exp3-t04} show a lack of coverage on Nr because AU-10's total score is just below the filtered $T$. Pr also suffers from under-provision at this point, since the only 2 SPCSFs after the score filtering are AC-19 and SC-16 with coverage of 0.1 and 0.2, respectively. With such a limited pool of SPCSFs, these cannot be fully met.

\subsection{Evaluation against baseline}
\label{sec:evaluation:comparison-baseline}

A greedy recommender that is bounded and filtered was implemented as baseline for the evaluation. Similar to the dynamic pruning stage at Algorithm \ref{alg:implementation-model:gt-exploration-recommendation} (lines \ref{alg:implementation-model:gt-exploration-recommendation:line:s15}-\ref{alg:implementation-model:gt-exploration-recommendation:line:s23}), it also applies a budget $B$ and upper bounds $\vec{\beta}$ (the direct security requirements $\vec{\tau}$ ) to limit over-provisioning.
This baseline first collects the coverage and score per SPCSF from the dataset, then sorts these using that score and iterates through these while the requirements are still not met and the budget is not exhausted. Only SPCSFs that increase the existing coverage while not exceeding the requirement bounds are added. Ultimately, it performs another pass to filter out redundant controls and thus suggest a minimal subset.

\begin{table}[htbp]
    \centering
    \caption{Summarised evaluation results (mode = all, $T$ = 0.0 to $T$ = 0.3).}
    \label{tab:evaluation:baseline-vs-proposed:summary-t00-t03}
    \resizebox{\columnwidth}{!}{%
    \begin{tabular}{l c c c}
        \toprule
        \textbf{{Metric}} & \textbf{{Greedy (naive)}} & \textbf{{Greedy (filtered)}} & \textbf{{Proposed}} \\
        \midrule
        Time ($\mathcal{R}$) & 0.0017 $\pm$ 0.00 & 0.0019 $\pm$ 0.00 & 0.0020 $\pm$ 0.00 \\
        Fairness & 0.8632 $\pm$ 0.11 & 0.9952 $\pm$ 0.02 & 0.9987 $\pm$ 0.01 \\
        Cost ($|\mathcal{R}|$) & 45.1137 $\pm$ 16.75 & 6.0175 $\pm$ 0.81 & 5.8925 $\pm$ 0.84 \\
        Sat. Cov. & 0.7766 $\pm$ 0.17 & 0.9857 $\pm$ 0.03 & 0.9896 $\pm$ 0.02 \\
        Res. Eff. & 0.5456 $\pm$ 0.11 & 0.6953 $\pm$ 0.10 & 0.7319 $\pm$ 0.09 \\
        Prof. IoU & 0.4658 $\pm$ 0.12 & 0.6878 $\pm$ 0.10 & 0.7258 $\pm$ 0.09 \\
        NDCG (Min) & 0.8544 $\pm$ 0.11 & 0.8488 $\pm$ 0.11 & 0.9860 $\pm$ 0.01 \\
        \bottomrule
    \end{tabular}
    } 
\end{table}

Table \ref{tab:evaluation:baseline-vs-proposed:summary-t00-t03} roughly compares two greedy recommenders with (i) a naive one and (ii) a bounded, filtered one used as baseline; and (iii) the recommender proposed in this work. It shows the aggregated average and SD values resulting from 100 iterations and across all full and clustered modes, displaying a subset of relevant metrics and thresholds, where $T=0.3$ is the turning point for all recommending approaches (Section \ref{sec:evaluation:analysis}).
The proposed recommender shows improved values for all metrics, except for the recommendation time $\mathcal{R}$, since it has to run extra steps on e.g. the received profile $\sigma^{*}$ and the mode $M$ processing (clustered vs full).
It is also worth noting that, whilst directly using $\vec{\tau}$ as $\vec{\beta}$ is enough to maximise coverage, it does actively ignore the identified synergies across security dimensions and the overall agreement, calculated in the utility functions (Section \ref{sec:implementation_model:utility-definition}) to also minimise over-provisioning --- which is not sufficiently addressed by the filtering logic during the recommendation stage.

Table \ref{tab:evaluation:comparison:results-0_0_0_8} compiles the detailed baseline values and deviation for each metric and threshold, and it is compared against the corresponding values for Hedge and EXP3 in the rows named $\Delta_{1}$ and $\Delta_{2}$, respectively. Improving $\uparrow$ and worsening $\downarrow$ indicators are placed next to each value.
The time (performance, $\mathcal{R}$) taken by both approaches is in the order of milliseconds, where the baseline halves that of the proposed recommender. This is so because the latter needs multiple passes to fine-tune SPCSF selection and limit SPCSF over-provisioning when issuing recommendations, whilst maintaining a high accuracy.
Such accuracy is consistently higher for all dataset sizes ($T$) of the proposed recommender compared to that of the baseline and is as a trade-off for the slight increase in time.

\subsection{Analysis}
\label{sec:evaluation:analysis}

Regarding the internal evaluation (Section \ref{sec:evaluation:internal}), the satisfaction ratio, area coverage and fairness are of 97\%-100\% between $T = 0.0$ and $T = 0.2$ (i.e. with 68-97 SPCSFs). There, 6 SPCSFs are needed to achieve a high accuracy.
Between $T = 0.3$ and $T = 0.4$ (37-63 SPCSFs), there is already a significant drop in the satisfaction and area coverage along with the IoU. However, the satisfaction coverage of the security requirements is still good for $T = 0.3$ (99\%).
$T = 0.3$ acts as the turning point where many metrics degrade for both the proposed recommender and the baseline.
Starting from $T = 0.5$, the dataset only provides 31 SPCSFs; with the satisfaction coverage dropping below 80\% and the IoU dropping below 60\%, denoting severe under-provisioning; which must be avoided.

Whilst there are some side benefits to a small number of SPCSFs (higher resource efficiency with less over-pro\-vi\-sio\-ning, a PoA that gradually reduces to its theoretical target and a steadily increasing NDCG), such a small dataset is clearly not enough to fulfil the requirements.
Summing up, the dataset $DS_{8}$ should not be heavily filtered and rather rely on adequate (likely customised) budget controls that can limit the recommendation of SPCSFs to ensure the operator does not incur too much effort to achieve the desired degree of security.
With at least 63 SPCSFs, the satisfaction coverage, satisfaction ratio and NDCG stay close or equal to 100\%.

When compared against the baseline (Section \ref{sec:evaluation:comparison-baseline}), this work shows a consistent improvement in area efficiency, resource efficiency and profile IoU (up to $21.43\%$, $10.59\%$ and $10.70\%$ for $T=0.0$), with improvements up to $T \le 0.3$.
The provisioning-conscious ranking NDCG score, which recommends minimal SPCSF sets, is naturally higher for all $T$ (up to $27.09\%$ in $T = 0.6$).
Also, the cost (numbers of controls used) is slightly smaller until $T \le 0.3$, reducing up to $6.03\%$ for $T = 0.0$.
This suggests some better adjustment and minimal over-provisioning than that from the baseline.


\section{Discussion and conclusion}
\label{sec:discussion-conclusion}

This section discusses the overall idea and behaviour of the proposed dataset and DSS, considering their assumptions and limitations; and concludes by summarising the contributions and future lines of work to consider.

\subsection{Discussion}
\label{sec:discussion-conclusion:discussion}

The proposed DSS is targeted at system operators with limited or no cybersecurity background, aiming to guide them on the initial security setup.
It thus assumes that cybersecurity experts are not in the loop to further curate or validate the dataset -- although it relies on well-known frameworks and adaptable data processing to incorporate new partial scores and recalculate the total one ultimately used by the recommender.

The dataset is constructed by correlating, transforming (i.e. often normalising) and aggregating multiple sources, keeping its average as partial and total scores. The latter is used as a normal distribution in the MAID chance nodes to account for potential minor variability in the scores from the list of SPCSFs (Section \ref{sec:implementation_model:recommendation-procedure:maid-constructions}).

From the evaluations in Section \ref{sec:evaluation} it can be extracted that  DSS performs best when it can select at least from $N = 63$ SPCSFs ($T = 0.0$ to $T = 0.3$), achieving an accuracy (measured as area coverage and satisfaction ratio) of 97\%-99\% and 99\%-100\%, respectively.
$T = 0.3$ is seen as the turning point, where these drop to 60\% and 80\% but the system works well.
After this, a massive drop in area coverage suggests under-provisioning problems, which should be avoided as it noticeably degrades the recommendation due to the narrower coverage of the remaining SPCSFs towards the security dimensions.
For best results, an organisation should cover at least 63 SPCSFs; although with at least 28 SPCSFs it could still deliver reasonably well for well-covered security dimensions (e.g. others than accountability and privacy, as per Section \ref{sec:evaluation:use-case}).

\subsection{Limitations}
\label{sec:discussion-conclusion:limitations}

This section documents how some assumptions from the dataset generation and the recommendation algorithm introduce limitations that must be considered.

\subsubsection{Data-related limitations}
\label{sec:discussion-conclusion:limitations:dataset}

The dataset is obtained by correlating existing peer-reviewed sources of different types and sources (Table \ref{tab:implementation-dataset:structure}).
When correlating data of different granularity and size, such aggregation can incur loss of detail. This happens for $DS_{2*}$ \cite{Liu_Shore_Yeoh_Jiang_Zeadally_2025}, which maps expert-provided scores to 36 security processes in their taxonomy and to other frameworks (e.g. NIST CSF). The translation between NIST CSF to NIST SP 800-53 rev5 relates 20 categories from the former to 300 SPCSFs from the latter; which introduces repeated scores for different types of SPCSFs and reduces its variability and discriminative potential.
Another aspect to consider is that the dataset is explicitly biased towards SW-implementable SPCFs, which inherit higher scores from the source datasets and result in a higher total score which keeps these available across higher $T$ during the dataset filtering.

Interpreting large amounts of descriptions for the SPCSFs from the OSCAL catalogue \cite{NIST_OSCAL_800_53_r5_repo} and online resources \cite{CSF_Tools_NIST_Special_Publication_800_53_Revision_5_2022} was delegated to a GenAI model (Gemini 2.5 Pro) to create $DS_{4}$ and identify how much a SPCSF contributes to preserve a security dimension.
The reader may assess the validity of this mapping by reviewing $DS_{8}$ the "coverage\_justification" column, explaining the rationale behind the assignment. An alternative is to perform a deterministic, replicable text parsing and analysis, with coverage being inferred through the frequency of keywords and their weight towards security dimensions based on authoritative sources like OSCAL \cite{NIST_OSCAL_800_53_r5_repo}, CNSSI 1253 \cite{CNSSI_1253_Security_Categorization_and_Control_Selection_for_National_Security_Systems_2022} or community-backed resources \cite{SCF_Free_Metaframework_2026}.
However, it is worth noting that human subjectivity can be brought into the process directly (e.g. expert survey, as in $DS_{2*}$) or indirectly (through GenAI or assumed keywords and weights for text analysis).
A highly robust approach would possibly involve formal modelling of security dimensions and protected resources, running sandboxed emulated or real attacks and mitigating mechanisms and observing the impact of both.

Besides, the dataset provides only up to the granularity of SPCSFs. It would be, however, interesting to also relate SPCSFs to categories of mechanisms, and mechanisms to specific frameworks or tools. \cite{FernandezMartinez_Siddiqui_Daza_2025} shows an example with custom security controls, categories and mechanisms.

\subsubsection{Recommender-related limitations}
\label{sec:discussion-conclusion:limitations:recommender}

The recommender process also runs on some assumptions that impact its outcome.
The DSS algorithm discretises the decision space into four categories (Section \ref{sec:implementation_model:recommendation-procedure:gt-exploration-recomm}) that align with traditional security baselines used in InfoSec and regulatory sources. This decreases the exploration space for a game-theoretic approach at the cost of less granularity in this stage, thus requiring a fine-tuning (recommending) stage to provide fine-grained coverage and identify attribution of SPCSFs to security dimensions. Combining (i) fine-grained profiles (e.g. based in coverage deciles) and (ii) a suitable application of the Shapley value (assuming a coalition-based game) in the utility functions to extract attribution could possibly remove the second recommending and pruning stage.
Regarding its output, it does not return a pure deterministic outcome across varying $M$ values (although it seems to do so for $\mathcal{L}$), as sometimes different SPCSFs can be suggested depending on the number of clusters and their arrangements. See Table \ref{tab:evaluation:use-case:req-vs-coverage} and Fig \ref{fig:evaluation:use-case}. On the other hand, coverage and scoring are here simplified to few attributes; yet a more advanced model should consider the nuances of the resources to protect and adequately expand the representing features or dimensions.

\subsection{Conclusion and future work}
\label{sec:discussion-conclusion:conclusion-future}

This work presented a scalable, accurate and fair security DSS to recommend NIST SP 800-53 rev5 security and privacy controls (SPCSFs), leveraging for decision a dataset extended from academic and InfoSec sources; which can be further adapted by users with cybersecurity exposure and extended by researchers.
It is implemented as a game-theoretic MAID, where multiple agents -representing the security dimensions to cover- apply no-regret learning to each utility function, maximising its payoff.
The outcome is the recommended subset of SPCSFs that best meet the provided security requirements from the available SPCSF pool.

Future steps to evaluate in this research currently consider three fronts: (i) easing the extraction of the user-defined coverage for the security requirements from other kinds of datasets, such as those containing a subset of relevant threats for an organisation; (ii) evaluating games modelling the adversarial behaviour based on its profile (i.e. applying specific tactics and techniques), potentially integrating the results of real or simulated campaigns as feedback to the modelling of the behaviour for the defenders; as well as (iii) improving the granularity and completeness of recommendations through an open source-derived taxonomy relating SW-implementable mechanisms for defenders related to both NIST SP 800-53 rev5 and MITRE ATT\&CK.


\section*{Generative AI tools usage declaration}

During the preparation of this work, the authors used Gemini 2.5/3 Pro, GPT 5.2 and Google NotebookLM for academic literature exploration, InfoSec interpretation, code prototyping, LaTeX editing, figure rendering and language revision.
GenAI tools were used in a limited and supervised manner.
All results were reviewed, corrected, and validated by the authors.
The authors take full responsibility for the content of this article.


\section*{CRediT authorship contribution statement}

\textbf{Carolina Fernández-Martínez}: writing -- review \& editing, writing – original draft, visualisation, validation, software, resources, project administration, methodology, investigation, formal analysis, data curation, conceptualisation.
\textbf{Shuaib Siddiqui}: writing -- review \& editing, project administration, funding acquisition, supervision, conceptualisation.
\textbf{Vanesa Daza}: writing -- review \& editing, supervision, conceptualisation.


\section*{Code and data availability}

The dataset generated in Section \ref{sec:implementation_dataset} and the code used to generate it are available at \cite{Paper_Dataset_2026}.
Both the dataset and the code used in Sections \ref{sec:implementation_model} and \ref{sec:evaluation} are available at \cite{SecDSS_MAID_GT_Fernandez_Martinez}.


\section*{Declaration of competing interest}

The authors declare that they have no known competing financial interests or personal relationships that could have appeared to influence the work reported in this paper.


\section*{Funding}

This work was supported by the grant COALESCE-6G PID2024-163028OB-I00, funded by MICIU/AEI/10.13039/501100011033/FEDER, EU.


\section*{Acknowledgements}

The authors would like to thank Estela Carmona-Cejudo for her advice on improving the mathematical formulation and the consistency and correctness of the utility function; as well as Amir Ansari for his insights on the MAID modelling and working sessions on BN and probabilistic-related validation.
The authors also acknowledge the Spanish Recovery, Transformation and Resilience Plan through the European Union (Next Generation) under grant AEI-PID2021-128521OB-I00.


\appendix
\gdef\thesection{Appendices}
\setcounter{figure}{0}
\gdef\thefigure{\Alph{section}.\arabic{figure}}
\setcounter{table}{0}
\gdef\thetable{\Alph{section}.\arabic{table}}
\section{}

Table \ref{tab:evaluation:full-results} details the full evaluation of the recommender, whereas Table \ref{tab:evaluation:comparison:results-0_0_0_8} shows the comparison against the greedy, filtered baseline. These can be found as appendices.

\begin{landscape}
\thispagestyle{empty}
\begin{table}[p]
    \centering
        \centering
        \caption{Full evaluation results per threshold, mode and method ($T = 0.0$ to $T = 0.8$).}
        \setlength{\tabcolsep}{3pt}
        \small
        \begin{tabular}{cccccccccccccccccc}
        \toprule
        \multicolumn{4}{c}{\textbf{Configuration}} & \multicolumn{3}{c}{\textbf{Time}} & \multicolumn{3}{c}{\textbf{Cost \& Fairness}} & \multicolumn{8}{c}{\textbf{Accuracy Metrics}} \\
        \cmidrule(r){1-4} \cmidrule(lr){5-7} \cmidrule(lr){8-10} \cmidrule(l){11-18}
        \textbf{$T$} & \textbf{N} & \textbf{Mode} & \textbf{Method} & \textbf{$\mathcal{M}$} & \textbf{$\mathcal{L}$} & \textbf{$\mathcal{R}$} & \textbf{PoA} & \textbf{Fairness} & \textbf{Cost ($|\mathcal{R}|$)} & \textbf{Sat. Ratio} & \textbf{Sat. Cov.} & \textbf{Res. Eff.} & \textbf{Prof. IoU} & \textbf{Area Cov.} & \textbf{Area Eff.} & \textbf{NDCG (Score)} & \textbf{NDCG (Min)} \\
        \cmidrule(r){1-1} \cmidrule(lr){2-2} \cmidrule(lr){3-3} \cmidrule(lr){4-4} \cmidrule(lr){5-5} \cmidrule(lr){6-6} \cmidrule(lr){7-7} \cmidrule(lr){8-8} \cmidrule(lr){9-9} \cmidrule(lr){10-10} \cmidrule(lr){11-11} \cmidrule(lr){12-12} \cmidrule(lr){13-13} \cmidrule(lr){14-14} \cmidrule(lr){15-15} \cmidrule(lr){16-16} \cmidrule(lr){17-17} \cmidrule(l){18-18}
            \multirow{4}{*}{0.0} & \multirow{4}{*}{97} & \multirow{2}{*}{Clustered} & Hedge & 0.1085 $\pm$ 0.02 & 4.4649 $\pm$ 0.09 & 0.0017 $\pm$ 0.00 & 1.00 $\pm$ 0.00 & 1.00 $\pm$ 0.01 & 5.78 $\pm$ 0.91 & \multirow{4}{*}{1.00 $\pm$ 0.02} & 1.00 $\pm$ 0.01 & 0.77 $\pm$ 0.08 & 0.77 $\pm$ 0.08 & 0.99 $\pm$ 0.02 & 0.58 $\pm$ 0.14 & 0.82 $\pm$ 0.08 & 0.99 $\pm$ 0.01 \\
         &  &  & EXP3 & 0.1085 $\pm$ 0.01 & 2.3449 $\pm$ 0.05 & 0.0018 $\pm$ 0.00 & 1.00 $\pm$ 0.00 & 1.00 $\pm$ 0.01 & 5.77 $\pm$ 0.92 &  & 1.00 $\pm$ 0.01 & 0.77 $\pm$ 0.09 & 0.77 $\pm$ 0.09 & 0.99 $\pm$ 0.02 & 0.58 $\pm$ 0.14 & 0.82 $\pm$ 0.08 & 0.99 $\pm$ 0.01 \\
         &  & \multirow{2}{*}{Full} & Hedge & 0.0079 $\pm$ 0.00 & 35.6978 $\pm$ 0.49 & 0.0036 $\pm$ 0.00 & 1.21 $\pm$ 0.13 & 1.00 $\pm$ 0.01 & 5.78 $\pm$ 0.88 &  & 1.00 $\pm$ 0.01 & 0.76 $\pm$ 0.09 & 0.76 $\pm$ 0.08 & 0.99 $\pm$ 0.02 & 0.57 $\pm$ 0.13 & 0.52 $\pm$ 0.17 & 0.96 $\pm$ 0.01 \\
         &  &  & EXP3 & 0.0079 $\pm$ 0.00 & 18.0905 $\pm$ 0.27 & 0.0035 $\pm$ 0.00 & 1.29 $\pm$ 0.16 & 1.00 $\pm$ 0.01 & 5.84 $\pm$ 0.90 &  & 1.00 $\pm$ 0.01 & 0.76 $\pm$ 0.08 & 0.76 $\pm$ 0.08 & 0.99 $\pm$ 0.02 & 0.57 $\pm$ 0.13 & 0.52 $\pm$ 0.17 & 0.96 $\pm$ 0.01 \\
        \midrule
        \multirow{4}{*}{0.1} & \multirow{4}{*}{69} & \multirow{2}{*}{Clustered} & Hedge & 0.0933 $\pm$ 0.01 & 2.8956 $\pm$ 0.02 & 0.0014 $\pm$ 0.00 & 1.00 $\pm$ 0.00 & 1.00 $\pm$ 0.00 & 5.90 $\pm$ 0.83 & \multirow{4}{*}{0.99 $\pm$ 0.02} & 0.99 $\pm$ 0.02 & 0.73 $\pm$ 0.10 & 0.72 $\pm$ 0.09 & 0.97 $\pm$ 0.04 & 0.52 $\pm$ 0.15 & 0.95 $\pm$ 0.04 & 1.00 $\pm$ 0.00 \\
         &  &  & EXP3 & 0.0937 $\pm$ 0.01 & 1.5554 $\pm$ 0.02 & 0.0014 $\pm$ 0.00 & 1.00 $\pm$ 0.00 & 1.00 $\pm$ 0.00 & 5.87 $\pm$ 0.84 &  & 0.99 $\pm$ 0.02 & 0.73 $\pm$ 0.09 & 0.72 $\pm$ 0.09 & 0.97 $\pm$ 0.04 & 0.52 $\pm$ 0.15 & 0.95 $\pm$ 0.04 & 1.00 $\pm$ 0.00 \\
         &  & \multirow{2}{*}{Full} & Hedge & 0.0069 $\pm$ 0.00 & 25.2836 $\pm$ 0.20 & 0.0024 $\pm$ 0.00 & 1.23 $\pm$ 0.17 & 1.00 $\pm$ 0.00 & 5.88 $\pm$ 0.82 &  & 0.99 $\pm$ 0.02 & 0.72 $\pm$ 0.09 & 0.71 $\pm$ 0.09 & 0.97 $\pm$ 0.04 & 0.51 $\pm$ 0.14 & 0.67 $\pm$ 0.09 & 0.98 $\pm$ 0.01 \\
         &  &  & EXP3 & 0.0068 $\pm$ 0.00 & 12.8225 $\pm$ 0.09 & 0.0023 $\pm$ 0.00 & 1.32 $\pm$ 0.21 & 1.00 $\pm$ 0.00 & 5.89 $\pm$ 0.84 &  & 0.99 $\pm$ 0.02 & 0.72 $\pm$ 0.09 & 0.71 $\pm$ 0.09 & 0.97 $\pm$ 0.04 & 0.51 $\pm$ 0.14 & 0.67 $\pm$ 0.09 & 0.98 $\pm$ 0.01 \\
        \midrule
        \multirow{4}{*}{0.2} & \multirow{4}{*}{68} & \multirow{2}{*}{Clustered} & Hedge & 0.0963 $\pm$ 0.01 & 2.1793 $\pm$ 0.02 & 0.0013 $\pm$ 0.00 & 1.00 $\pm$ 0.00 & 1.00 $\pm$ 0.00 & 5.88 $\pm$ 0.79 & \multirow{4}{*}{0.99 $\pm$ 0.02} & 0.99 $\pm$ 0.02 & 0.73 $\pm$ 0.09 & 0.72 $\pm$ 0.09 & 0.97 $\pm$ 0.04 & 0.52 $\pm$ 0.15 & 0.95 $\pm$ 0.04 & 1.00 $\pm$ 0.00 \\
         &  &  & EXP3 & 0.0963 $\pm$ 0.01 & 1.1889 $\pm$ 0.01 & 0.0014 $\pm$ 0.00 & 1.00 $\pm$ 0.00 & 1.00 $\pm$ 0.00 & 5.87 $\pm$ 0.81 &  & 0.99 $\pm$ 0.02 & 0.73 $\pm$ 0.10 & 0.72 $\pm$ 0.09 & 0.97 $\pm$ 0.04 & 0.52 $\pm$ 0.15 & 0.95 $\pm$ 0.04 & 1.00 $\pm$ 0.00 \\
         &  & \multirow{2}{*}{Full} & Hedge & 0.0067 $\pm$ 0.00 & 24.9448 $\pm$ 0.22 & 0.0023 $\pm$ 0.00 & 1.24 $\pm$ 0.14 & 1.00 $\pm$ 0.00 & 5.89 $\pm$ 0.80 &  & 0.99 $\pm$ 0.02 & 0.72 $\pm$ 0.09 & 0.71 $\pm$ 0.09 & 0.97 $\pm$ 0.04 & 0.51 $\pm$ 0.14 & 0.68 $\pm$ 0.08 & 0.98 $\pm$ 0.01 \\
         &  &  & EXP3 & 0.0068 $\pm$ 0.00 & 12.6591 $\pm$ 0.12 & 0.0022 $\pm$ 0.00 & 1.32 $\pm$ 0.19 & 1.00 $\pm$ 0.00 & 5.87 $\pm$ 0.80 &  & 0.99 $\pm$ 0.02 & 0.72 $\pm$ 0.09 & 0.71 $\pm$ 0.09 & 0.97 $\pm$ 0.04 & 0.51 $\pm$ 0.14 & 0.68 $\pm$ 0.08 & 0.98 $\pm$ 0.01 \\
        \midrule
        \multirow{4}{*}{0.3} & \multirow{4}{*}{63} & \multirow{2}{*}{Clustered} & Hedge & 0.0807 $\pm$ 0.01 & 3.2711 $\pm$ 0.03 & 0.0013 $\pm$ 0.00 & 1.00 $\pm$ 0.00 & 1.00 $\pm$ 0.00 & 6.02 $\pm$ 0.80 & \multirow{4}{*}{0.99 $\pm$ 0.02} & 0.99 $\pm$ 0.02 & 0.72 $\pm$ 0.10 & 0.71 $\pm$ 0.09 & 0.97 $\pm$ 0.04 & 0.51 $\pm$ 0.15 & 0.94 $\pm$ 0.04 & 1.00 $\pm$ 0.00 \\
         &  &  & EXP3 & 0.0825 $\pm$ 0.01 & 1.7434 $\pm$ 0.01 & 0.0014 $\pm$ 0.00 & 1.00 $\pm$ 0.00 & 1.00 $\pm$ 0.00 & 6.01 $\pm$ 0.81 &  & 0.99 $\pm$ 0.02 & 0.72 $\pm$ 0.10 & 0.71 $\pm$ 0.09 & 0.97 $\pm$ 0.04 & 0.51 $\pm$ 0.15 & 0.94 $\pm$ 0.04 & 1.00 $\pm$ 0.00 \\
         &  & \multirow{2}{*}{Full} & Hedge & 0.0067 $\pm$ 0.00 & 23.1164 $\pm$ 0.15 & 0.0022 $\pm$ 0.00 & 1.26 $\pm$ 0.15 & 1.00 $\pm$ 0.00 & 6.01 $\pm$ 0.82 &  & 0.99 $\pm$ 0.02 & 0.71 $\pm$ 0.09 & 0.70 $\pm$ 0.09 & 0.97 $\pm$ 0.04 & 0.50 $\pm$ 0.14 & 0.69 $\pm$ 0.08 & 0.98 $\pm$ 0.01 \\
         &  &  & EXP3 & 0.0067 $\pm$ 0.00 & 11.7480 $\pm$ 0.08 & 0.0021 $\pm$ 0.00 & 1.35 $\pm$ 0.23 & 1.00 $\pm$ 0.00 & 6.02 $\pm$ 0.79 &  & 0.99 $\pm$ 0.02 & 0.71 $\pm$ 0.09 & 0.71 $\pm$ 0.09 & 0.97 $\pm$ 0.04 & 0.50 $\pm$ 0.15 & 0.69 $\pm$ 0.08 & 0.98 $\pm$ 0.01 \\
        \midrule
        \multirow{4}{*}{0.4} & \multirow{4}{*}{37} & \multirow{2}{*}{Clustered} & Hedge & 0.0733 $\pm$ 0.01 & 1.4277 $\pm$ 0.02 & 0.0010 $\pm$ 0.00 & 1.00 $\pm$ 0.00 & 0.82 $\pm$ 0.04 & 4.81 $\pm$ 0.80 & \multirow{4}{*}{0.80 $\pm$ 0.06} & 0.77 $\pm$ 0.09 & 0.77 $\pm$ 0.09 & 0.62 $\pm$ 0.09 & 0.60 $\pm$ 0.18 & 0.56 $\pm$ 0.17 & 0.98 $\pm$ 0.01 & 1.00 $\pm$ 0.00 \\
         &  &  & EXP3 & 0.0724 $\pm$ 0.01 & 0.8154 $\pm$ 0.02 & 0.0010 $\pm$ 0.00 & 1.00 $\pm$ 0.00 & 0.82 $\pm$ 0.04 & 4.82 $\pm$ 0.80 &  & 0.77 $\pm$ 0.09 & 0.77 $\pm$ 0.09 & 0.62 $\pm$ 0.09 & 0.60 $\pm$ 0.18 & 0.56 $\pm$ 0.17 & 0.98 $\pm$ 0.01 & 1.00 $\pm$ 0.00 \\
         &  & \multirow{2}{*}{Full} & Hedge & 0.0060 $\pm$ 0.00 & 13.7731 $\pm$ 0.20 & 0.0018 $\pm$ 0.00 & 1.30 $\pm$ 0.21 & 0.82 $\pm$ 0.04 & 4.80 $\pm$ 0.82 &  & 0.77 $\pm$ 0.09 & 0.76 $\pm$ 0.09 & 0.62 $\pm$ 0.09 & 0.60 $\pm$ 0.18 & 0.54 $\pm$ 0.16 & 0.83 $\pm$ 0.05 & 0.99 $\pm$ 0.01 \\
         &  &  & EXP3 & 0.0062 $\pm$ 0.00 & 7.0427 $\pm$ 0.11 & 0.0017 $\pm$ 0.00 & 1.52 $\pm$ 0.34 & 0.82 $\pm$ 0.04 & 4.80 $\pm$ 0.82 &  & 0.77 $\pm$ 0.09 & 0.76 $\pm$ 0.09 & 0.62 $\pm$ 0.09 & 0.60 $\pm$ 0.18 & 0.54 $\pm$ 0.16 & 0.83 $\pm$ 0.05 & 0.99 $\pm$ 0.01 \\
        \midrule
        \multirow{4}{*}{0.5} & \multirow{4}{*}{31} & \multirow{2}{*}{Clustered} & Hedge & 0.0764 $\pm$ 0.01 & 2.6171 $\pm$ 0.07 & 0.0010 $\pm$ 0.00 & 1.00 $\pm$ 0.00 & 0.81 $\pm$ 0.04 & 4.57 $\pm$ 0.90 & \multirow{4}{*}{0.77 $\pm$ 0.10} & 0.73 $\pm$ 0.10 & 0.76 $\pm$ 0.09 & 0.59 $\pm$ 0.09 & 0.56 $\pm$ 0.19 & 0.54 $\pm$ 0.17 & 0.98 $\pm$ 0.01 & 1.00 $\pm$ 0.00 \\
         &  &  & EXP3 & 0.0799 $\pm$ 0.01 & 1.4106 $\pm$ 0.04 & 0.0011 $\pm$ 0.00 & 1.00 $\pm$ 0.00 & 0.81 $\pm$ 0.04 & 4.58 $\pm$ 0.90 &  & 0.73 $\pm$ 0.10 & 0.77 $\pm$ 0.09 & 0.59 $\pm$ 0.09 & 0.56 $\pm$ 0.19 & 0.54 $\pm$ 0.17 & 0.98 $\pm$ 0.01 & 1.00 $\pm$ 0.00 \\
         &  & \multirow{2}{*}{Full} & Hedge & 0.0067 $\pm$ 0.00 & 11.8394 $\pm$ 0.19 & 0.0017 $\pm$ 0.00 & 1.34 $\pm$ 0.23 & 0.81 $\pm$ 0.04 & 4.56 $\pm$ 0.90 &  & 0.73 $\pm$ 0.10 & 0.76 $\pm$ 0.09 & 0.59 $\pm$ 0.09 & 0.56 $\pm$ 0.19 & 0.54 $\pm$ 0.17 & 0.87 $\pm$ 0.03 & 0.99 $\pm$ 0.01 \\
         &  &  & EXP3 & 0.0070 $\pm$ 0.00 & 6.0813 $\pm$ 0.11 & 0.0016 $\pm$ 0.00 & 1.60 $\pm$ 0.39 & 0.81 $\pm$ 0.04 & 4.56 $\pm$ 0.90 &  & 0.73 $\pm$ 0.10 & 0.76 $\pm$ 0.09 & 0.59 $\pm$ 0.09 & 0.56 $\pm$ 0.19 & 0.54 $\pm$ 0.17 & 0.87 $\pm$ 0.03 & 0.99 $\pm$ 0.01 \\
        \midrule
        \multirow{4}{*}{0.6} & \multirow{4}{*}{28} & \multirow{2}{*}{Clustered} & Hedge & 0.0688 $\pm$ 0.01 & 2.5249 $\pm$ 0.03 & 0.0009 $\pm$ 0.00 & 1.00 $\pm$ 0.00 & 0.81 $\pm$ 0.04 & 4.58 $\pm$ 0.89 & \multirow{4}{*}{0.77 $\pm$ 0.10} & 0.73 $\pm$ 0.10 & 0.76 $\pm$ 0.09 & 0.59 $\pm$ 0.09 & 0.56 $\pm$ 0.19 & 0.54 $\pm$ 0.17 & 0.98 $\pm$ 0.01 & 1.00 $\pm$ 0.00 \\
         &  &  & EXP3 & 0.0689 $\pm$ 0.01 & 1.3660 $\pm$ 0.02 & 0.0010 $\pm$ 0.00 & 1.00 $\pm$ 0.00 & 0.81 $\pm$ 0.04 & 4.58 $\pm$ 0.89 &  & 0.73 $\pm$ 0.10 & 0.76 $\pm$ 0.09 & 0.59 $\pm$ 0.09 & 0.56 $\pm$ 0.19 & 0.54 $\pm$ 0.17 & 0.98 $\pm$ 0.01 & 1.00 $\pm$ 0.00 \\
         &  & \multirow{2}{*}{Full} & Hedge & 0.0059 $\pm$ 0.00 & 10.4795 $\pm$ 0.09 & 0.0015 $\pm$ 0.00 & 1.47 $\pm$ 0.36 & 0.81 $\pm$ 0.04 & 4.57 $\pm$ 0.89 &  & 0.73 $\pm$ 0.10 & 0.76 $\pm$ 0.09 & 0.59 $\pm$ 0.09 & 0.56 $\pm$ 0.19 & 0.54 $\pm$ 0.17 & 0.89 $\pm$ 0.03 & 1.00 $\pm$ 0.00 \\
         &  &  & EXP3 & 0.0060 $\pm$ 0.00 & 5.3762 $\pm$ 0.05 & 0.0015 $\pm$ 0.00 & 1.72 $\pm$ 0.74 & 0.81 $\pm$ 0.04 & 4.57 $\pm$ 0.89 &  & 0.73 $\pm$ 0.10 & 0.76 $\pm$ 0.09 & 0.59 $\pm$ 0.08 & 0.56 $\pm$ 0.19 & 0.53 $\pm$ 0.17 & 0.89 $\pm$ 0.03 & 1.00 $\pm$ 0.00 \\
        \midrule
        \multirow{4}{*}{0.7} & \multirow{4}{*}{13} & \multirow{2}{*}{Clustered} & Hedge & 0.0388 $\pm$ 0.00 & 1.4616 $\pm$ 0.03 & 0.0008 $\pm$ 0.00 & 1.00 $\pm$ 0.00 & 0.56 $\pm$ 0.05 & 3.24 $\pm$ 0.70 & \multirow{4}{*}{0.48 $\pm$ 0.14} & 0.43 $\pm$ 0.10 & 0.80 $\pm$ 0.10 & 0.39 $\pm$ 0.09 & 0.23 $\pm$ 0.12 & 0.57 $\pm$ 0.21 & 0.99 $\pm$ 0.01 & 1.00 $\pm$ 0.00 \\
         &  &  & EXP3 & 0.0406 $\pm$ 0.01 & 0.8309 $\pm$ 0.02 & 0.0008 $\pm$ 0.00 & 1.00 $\pm$ 0.00 & 0.56 $\pm$ 0.05 & 3.24 $\pm$ 0.70 &  & 0.43 $\pm$ 0.10 & 0.80 $\pm$ 0.09 & 0.39 $\pm$ 0.09 & 0.23 $\pm$ 0.12 & 0.57 $\pm$ 0.21 & 0.99 $\pm$ 0.01 & 1.00 $\pm$ 0.00 \\
         &  & \multirow{2}{*}{Full} & Hedge & 0.0057 $\pm$ 0.00 & 5.2253 $\pm$ 0.08 & 0.0013 $\pm$ 0.00 & 1.00 $\pm$ 0.00 & 0.56 $\pm$ 0.05 & 3.24 $\pm$ 0.70 &  & 0.43 $\pm$ 0.10 & 0.80 $\pm$ 0.10 & 0.39 $\pm$ 0.09 & 0.23 $\pm$ 0.12 & 0.57 $\pm$ 0.21 & 0.96 $\pm$ 0.02 & 1.00 $\pm$ 0.00 \\
         &  &  & EXP3 & 0.0056 $\pm$ 0.00 & 2.7286 $\pm$ 0.05 & 0.0012 $\pm$ 0.00 & 1.00 $\pm$ 0.00 & 0.56 $\pm$ 0.05 & 3.24 $\pm$ 0.70 &  & 0.43 $\pm$ 0.10 & 0.80 $\pm$ 0.10 & 0.39 $\pm$ 0.09 & 0.23 $\pm$ 0.12 & 0.57 $\pm$ 0.21 & 0.96 $\pm$ 0.02 & 1.00 $\pm$ 0.00 \\
        \midrule
        \multirow{4}{*}{0.8} & \multirow{4}{*}{6} & \multirow{2}{*}{Clustered} & Hedge & 0.0255 $\pm$ 0.00 & 1.4788 $\pm$ 0.05 & 0.0007 $\pm$ 0.00 & 1.00 $\pm$ 0.00 & 0.49 $\pm$ 0.05 & 2.36 $\pm$ 0.64 & \multirow{4}{*}{0.42 $\pm$ 0.13} & 0.38 $\pm$ 0.09 & 0.82 $\pm$ 0.10 & 0.35 $\pm$ 0.08 & 0.18 $\pm$ 0.10 & 0.59 $\pm$ 0.25 & 0.99 $\pm$ 0.00 & 1.00 $\pm$ 0.00 \\
         &  &  & EXP3 & 0.0261 $\pm$ 0.01 & 0.8406 $\pm$ 0.02 & 0.0007 $\pm$ 0.00 & 1.00 $\pm$ 0.00 & 0.49 $\pm$ 0.05 & 2.36 $\pm$ 0.64 &  & 0.38 $\pm$ 0.09 & 0.82 $\pm$ 0.10 & 0.35 $\pm$ 0.08 & 0.18 $\pm$ 0.10 & 0.58 $\pm$ 0.25 & 0.99 $\pm$ 0.00 & 1.00 $\pm$ 0.00 \\
         &  & \multirow{2}{*}{Full} & Hedge & 0.0053 $\pm$ 0.00 & 2.6235 $\pm$ 0.06 & 0.0012 $\pm$ 0.00 & 1.00 $\pm$ 0.00 & 0.49 $\pm$ 0.05 & 2.36 $\pm$ 0.64 &  & 0.38 $\pm$ 0.09 & 0.82 $\pm$ 0.10 & 0.35 $\pm$ 0.08 & 0.18 $\pm$ 0.10 & 0.58 $\pm$ 0.25 & 0.99 $\pm$ 0.00 & 1.00 $\pm$ 0.00 \\
         &  &  & EXP3 & 0.0054 $\pm$ 0.00 & 1.4255 $\pm$ 0.04 & 0.0011 $\pm$ 0.00 & 1.00 $\pm$ 0.00 & 0.49 $\pm$ 0.05 & 2.36 $\pm$ 0.64 &  & 0.38 $\pm$ 0.09 & 0.82 $\pm$ 0.10 & 0.35 $\pm$ 0.08 & 0.18 $\pm$ 0.10 & 0.58 $\pm$ 0.25 & 0.99 $\pm$ 0.00 & 1.00 $\pm$ 0.00 \\
        \bottomrule
    \end{tabular}
        \label{tab:evaluation:full-results}
\end{table}
\end{landscape}

\begin{table*}[htbp]
    \centering
    \caption{Comparative evaluation between SPCSF recommendation methods ($T=0.0$ to $0.8$).}
    \label{tab:evaluation:comparison:results-0_0_0_8}
    \setlength{\tabcolsep}{3pt}
    \small
    \scalebox{0.85}{%
    \begin{tabular}{c c l  c c c c c c c c c c}
        \toprule
                \multicolumn{3}{c}{\textbf{Configuration}} & \textbf{Time} & \multicolumn{2}{c}{\textbf{Cost \& fairness}} & \multicolumn{7}{c}{\textbf{Accuracy metrics}} \\
        \cmidrule(r){1-3} \cmidrule(r){4-4} \cmidrule(lr){5-6} \cmidrule(l){7-13}
        \textbf{$T$} & \textbf{Mode} & \textbf{Method} & \textbf{$\mathcal{R}$} & \textbf{Fairness} & \textbf{Cost ($|\mathcal{R}|$)} & \textbf{Sat. Cov.} & \textbf{Res. Eff.} & \textbf{Prof. IoU} & \textbf{Area Cov.} & \textbf{Area Eff.} & \textbf{NDCG (Score)} & \textbf{NDCG (Min)} \\
        \midrule
        \multirow{6}{*}{0.0} & \multirow{3}{*}{Clustered} & Baseline & 0.0012 $\pm$ 0.00 & 1.00 $\pm$ 0.02 & 6.14 $\pm$ 0.77 & 0.99 $\pm$ 0.01 & 0.70 $\pm$ 0.10 & 0.69 $\pm$ 0.10 & 0.99 $\pm$ 0.02 & 0.48 $\pm$ 0.14 & 1.00 $\pm$ 0.00 & 0.84 $\pm$ 0.13 \\
         &  & $\Delta_1$ (vs Hedge) & +43.13\% ($\downarrow$) & +0.16\% ($\uparrow$) & -5.86\% ($\uparrow$) & +0.14\% ($\uparrow$) & +10.59\% ($\uparrow$) & +10.70\% ($\uparrow$) & +0.29\% ($\uparrow$) & +21.43\% ($\uparrow$) & -17.72\% ($\downarrow$) & +17.74\% ($\uparrow$) \\
         &  & $\Delta_2$ (vs EXP3) & +48.76\% ($\downarrow$) & +0.16\% ($\uparrow$) & -6.03\% ($\uparrow$) & +0.14\% ($\uparrow$) & +10.33\% ($\uparrow$) & +10.44\% ($\uparrow$) & +0.29\% ($\uparrow$) & +21.40\% ($\uparrow$) & -17.72\% ($\downarrow$) & +17.74\% ($\uparrow$) \\
        \cmidrule(l){2-13}
         & \multirow{3}{*}{Full} & Baseline & 0.0036 $\pm$ 0.00 & 1.00 $\pm$ 0.01 & 6.14 $\pm$ 0.79 & 0.99 $\pm$ 0.01 & 0.70 $\pm$ 0.10 & 0.69 $\pm$ 0.10 & 0.99 $\pm$ 0.02 & 0.48 $\pm$ 0.15 & 1.00 $\pm$ 0.00 & 0.85 $\pm$ 0.13 \\
         &  & $\Delta_1$ (vs Hedge) & -1.39\% ($\uparrow$) & +0.01\% ($\uparrow$) & -5.86\% ($\uparrow$) & +0.14\% ($\uparrow$) & +8.58\% ($\uparrow$) & +8.68\% ($\uparrow$) & +0.29\% ($\uparrow$) & +17.16\% ($\uparrow$) & -47.53\% ($\downarrow$) & +14.06\% ($\uparrow$) \\
         &  & $\Delta_2$ (vs EXP3) & -3.23\% ($\uparrow$) & +0.01\% ($\uparrow$) & -4.89\% ($\uparrow$) & +0.14\% ($\uparrow$) & +8.91\% ($\uparrow$) & +9.02\% ($\uparrow$) & +0.29\% ($\uparrow$) & +17.41\% ($\uparrow$) & -47.53\% ($\downarrow$) & +14.06\% ($\uparrow$) \\
        \midrule
        \multirow{6}{*}{0.1} & \multirow{3}{*}{Clustered} & Baseline & 0.0010 $\pm$ 0.00 & 0.99 $\pm$ 0.03 & 5.79 $\pm$ 0.81 & 0.99 $\pm$ 0.02 & 0.70 $\pm$ 0.10 & 0.69 $\pm$ 0.09 & 0.98 $\pm$ 0.04 & 0.48 $\pm$ 0.14 & 1.00 $\pm$ 0.00 & 0.85 $\pm$ 0.10 \\
         &  & $\Delta_1$ (vs Hedge) & +43.17\% ($\downarrow$) & +0.56\% ($\uparrow$) & +1.90\% ($\downarrow$) & -0.10\% ($\downarrow$) & +4.27\% ($\uparrow$) & +4.21\% ($\uparrow$) & -0.34\% ($\downarrow$) & +8.20\% ($\uparrow$) & -5.42\% ($\downarrow$) & +17.97\% ($\uparrow$) \\
         &  & $\Delta_2$ (vs EXP3) & +48.32\% ($\downarrow$) & +0.56\% ($\uparrow$) & +1.38\% ($\downarrow$) & -0.10\% ($\downarrow$) & +4.36\% ($\uparrow$) & +4.30\% ($\uparrow$) & -0.34\% ($\downarrow$) & +8.17\% ($\uparrow$) & -5.42\% ($\downarrow$) & +17.97\% ($\uparrow$) \\
        \cmidrule(l){2-13}
         & \multirow{3}{*}{Full} & Baseline & 0.0027 $\pm$ 0.00 & 0.99 $\pm$ 0.03 & 5.92 $\pm$ 0.84 & 0.99 $\pm$ 0.02 & 0.69 $\pm$ 0.10 & 0.68 $\pm$ 0.10 & 0.98 $\pm$ 0.04 & 0.47 $\pm$ 0.14 & 1.00 $\pm$ 0.00 & 0.84 $\pm$ 0.13 \\
         &  & $\Delta_1$ (vs Hedge) & -11.74\% ($\uparrow$) & +0.56\% ($\uparrow$) & -0.68\% ($\uparrow$) & -0.10\% ($\downarrow$) & +4.38\% ($\uparrow$) & +4.34\% ($\uparrow$) & -0.34\% ($\downarrow$) & +8.15\% ($\uparrow$) & -32.88\% ($\downarrow$) & +16.18\% ($\uparrow$) \\
         &  & $\Delta_2$ (vs EXP3) & -15.64\% ($\uparrow$) & +0.56\% ($\uparrow$) & -0.51\% ($\uparrow$) & -0.10\% ($\downarrow$) & +4.58\% ($\uparrow$) & +4.53\% ($\uparrow$) & -0.34\% ($\downarrow$) & +8.61\% ($\uparrow$) & -32.88\% ($\downarrow$) & +16.18\% ($\uparrow$) \\
        \midrule
        \multirow{6}{*}{0.2} & \multirow{3}{*}{Clustered} & Baseline & 0.0009 $\pm$ 0.00 & 0.99 $\pm$ 0.02 & 5.93 $\pm$ 0.82 & 0.98 $\pm$ 0.06 & 0.71 $\pm$ 0.10 & 0.70 $\pm$ 0.10 & 0.97 $\pm$ 0.09 & 0.50 $\pm$ 0.16 & 1.00 $\pm$ 0.00 & 0.85 $\pm$ 0.11 \\
         &  & $\Delta_1$ (vs Hedge) & +49.62\% ($\downarrow$) & +0.39\% ($\uparrow$) & -0.84\% ($\uparrow$) & +0.74\% ($\uparrow$) & +3.03\% ($\uparrow$) & +3.53\% ($\uparrow$) & +0.81\% ($\uparrow$) & +4.72\% ($\uparrow$) & -5.05\% ($\downarrow$) & +17.63\% ($\uparrow$) \\
         &  & $\Delta_2$ (vs EXP3) & +50.60\% ($\downarrow$) & +0.39\% ($\uparrow$) & -1.01\% ($\uparrow$) & +0.74\% ($\uparrow$) & +2.97\% ($\uparrow$) & +3.48\% ($\uparrow$) & +0.81\% ($\uparrow$) & +4.68\% ($\uparrow$) & -5.05\% ($\downarrow$) & +17.63\% ($\uparrow$) \\
        \cmidrule(l){2-13}
         & \multirow{3}{*}{Full} & Baseline & 0.0026 $\pm$ 0.00 & 1.00 $\pm$ 0.02 & 6.08 $\pm$ 0.77 & 0.98 $\pm$ 0.03 & 0.70 $\pm$ 0.10 & 0.69 $\pm$ 0.10 & 0.97 $\pm$ 0.05 & 0.49 $\pm$ 0.15 & 1.00 $\pm$ 0.00 & 0.85 $\pm$ 0.11 \\
         &  & $\Delta_1$ (vs Hedge) & -9.09\% ($\uparrow$) & +0.25\% ($\uparrow$) & -3.13\% ($\uparrow$) & +0.34\% ($\uparrow$) & +3.16\% ($\uparrow$) & +3.41\% ($\uparrow$) & +0.40\% ($\uparrow$) & +4.43\% ($\uparrow$) & -32.33\% ($\downarrow$) & +14.82\% ($\uparrow$) \\
         &  & $\Delta_2$ (vs EXP3) & -12.74\% ($\uparrow$) & +0.25\% ($\uparrow$) & -3.45\% ($\uparrow$) & +0.34\% ($\uparrow$) & +3.29\% ($\uparrow$) & +3.55\% ($\uparrow$) & +0.40\% ($\uparrow$) & +4.42\% ($\uparrow$) & -32.33\% ($\downarrow$) & +14.82\% ($\uparrow$) \\
        \midrule
        \multirow{6}{*}{0.3} & \multirow{3}{*}{Clustered} & Baseline & 0.0010 $\pm$ 0.00 & 0.99 $\pm$ 0.02 & 6.07 $\pm$ 0.87 & 0.98 $\pm$ 0.05 & 0.69 $\pm$ 0.10 & 0.68 $\pm$ 0.10 & 0.96 $\pm$ 0.08 & 0.47 $\pm$ 0.14 & 1.00 $\pm$ 0.00 & 0.85 $\pm$ 0.11 \\
         &  & $\Delta_1$ (vs Hedge) & +34.59\% ($\downarrow$) & +0.43\% ($\uparrow$) & -0.82\% ($\uparrow$) & +1.24\% ($\uparrow$) & +4.43\% ($\uparrow$) & +5.20\% ($\uparrow$) & +1.99\% ($\uparrow$) & +8.89\% ($\uparrow$) & -6.45\% ($\downarrow$) & +16.99\% ($\uparrow$) \\
         &  & $\Delta_2$ (vs EXP3) & +40.28\% ($\downarrow$) & +0.43\% ($\uparrow$) & -0.99\% ($\uparrow$) & +1.24\% ($\uparrow$) & +4.41\% ($\uparrow$) & +5.17\% ($\uparrow$) & +1.99\% ($\uparrow$) & +8.82\% ($\uparrow$) & -6.45\% ($\downarrow$) & +16.99\% ($\uparrow$) \\
        \cmidrule(l){2-13}
         & \multirow{3}{*}{Full} & Baseline & 0.0025 $\pm$ 0.00 & 0.99 $\pm$ 0.02 & 6.07 $\pm$ 0.82 & 0.98 $\pm$ 0.03 & 0.69 $\pm$ 0.10 & 0.68 $\pm$ 0.09 & 0.96 $\pm$ 0.05 & 0.47 $\pm$ 0.14 & 1.00 $\pm$ 0.00 & 0.86 $\pm$ 0.11 \\
         &  & $\Delta_1$ (vs Hedge) & -11.19\% ($\uparrow$) & +0.39\% ($\uparrow$) & -0.99\% ($\uparrow$) & +0.76\% ($\uparrow$) & +3.14\% ($\uparrow$) & +3.66\% ($\uparrow$) & +1.26\% ($\uparrow$) & +6.36\% ($\uparrow$) & -30.62\% ($\downarrow$) & +13.90\% ($\uparrow$) \\
         &  & $\Delta_2$ (vs EXP3) & -15.26\% ($\uparrow$) & +0.39\% ($\uparrow$) & -0.82\% ($\uparrow$) & +0.76\% ($\uparrow$) & +3.65\% ($\uparrow$) & +4.18\% ($\uparrow$) & +1.26\% ($\uparrow$) & +7.10\% ($\uparrow$) & -30.62\% ($\downarrow$) & +13.90\% ($\uparrow$) \\
        \midrule
        \multirow{6}{*}{0.4} & \multirow{3}{*}{Clustered} & Baseline & 0.0006 $\pm$ 0.00 & 0.81 $\pm$ 0.05 & 5.06 $\pm$ 0.81 & 0.78 $\pm$ 0.11 & 0.71 $\pm$ 0.10 & 0.59 $\pm$ 0.11 & 0.66 $\pm$ 0.19 & 0.50 $\pm$ 0.16 & 1.00 $\pm$ 0.00 & 0.81 $\pm$ 0.15 \\
         &  & $\Delta_1$ (vs Hedge) & +73.65\% ($\downarrow$) & +0.58\% ($\uparrow$) & -4.94\% ($\uparrow$) & -1.15\% ($\downarrow$) & +8.29\% ($\uparrow$) & +5.27\% ($\uparrow$) & -9.45\% ($\downarrow$) & +11.01\% ($\uparrow$) & -1.92\% ($\downarrow$) & +23.59\% ($\uparrow$) \\
         &  & $\Delta_2$ (vs EXP3) & +74.91\% ($\downarrow$) & +0.58\% ($\uparrow$) & -4.74\% ($\uparrow$) & -1.15\% ($\downarrow$) & +8.17\% ($\uparrow$) & +5.22\% ($\uparrow$) & -9.45\% ($\downarrow$) & +10.53\% ($\uparrow$) & -1.92\% ($\downarrow$) & +23.59\% ($\uparrow$) \\
        \cmidrule(l){2-13}
         & \multirow{3}{*}{Full} & Baseline & 0.0015 $\pm$ 0.00 & 0.81 $\pm$ 0.05 & 4.89 $\pm$ 0.82 & 0.78 $\pm$ 0.11 & 0.75 $\pm$ 0.09 & 0.62 $\pm$ 0.11 & 0.66 $\pm$ 0.19 & 0.54 $\pm$ 0.14 & 1.00 $\pm$ 0.00 & 0.88 $\pm$ 0.11 \\
         &  & $\Delta_1$ (vs Hedge) & +22.12\% ($\downarrow$) & +0.58\% ($\uparrow$) & -1.84\% ($\uparrow$) & -1.15\% ($\downarrow$) & +1.60\% ($\uparrow$) & -0.20\% ($\downarrow$) & -9.45\% ($\downarrow$) & +0.02\% ($\uparrow$) & -16.68\% ($\downarrow$) & +12.79\% ($\uparrow$) \\
         &  & $\Delta_2$ (vs EXP3) & +15.05\% ($\downarrow$) & +0.58\% ($\uparrow$) & -1.84\% ($\uparrow$) & -1.15\% ($\downarrow$) & +1.73\% ($\uparrow$) & -0.08\% ($\downarrow$) & -9.45\% ($\downarrow$) & +0.76\% ($\uparrow$) & -16.68\% ($\downarrow$) & +12.79\% ($\uparrow$) \\
        \midrule
        \multirow{6}{*}{0.5} & \multirow{3}{*}{Clustered} & Baseline & 0.0006 $\pm$ 0.00 & 0.81 $\pm$ 0.04 & 4.66 $\pm$ 0.77 & 0.74 $\pm$ 0.09 & 0.73 $\pm$ 0.10 & 0.58 $\pm$ 0.09 & 0.61 $\pm$ 0.17 & 0.52 $\pm$ 0.15 & 1.00 $\pm$ 0.00 & 0.85 $\pm$ 0.10 \\
         &  & $\Delta_1$ (vs Hedge) & +82.16\% ($\downarrow$) & +0.14\% ($\uparrow$) & -1.93\% ($\uparrow$) & -1.52\% ($\downarrow$) & +4.12\% ($\uparrow$) & +1.96\% ($\uparrow$) & -6.93\% ($\downarrow$) & +4.04\% ($\uparrow$) & -1.84\% ($\downarrow$) & +18.12\% ($\uparrow$) \\
         &  & $\Delta_2$ (vs EXP3) & +95.80\% ($\downarrow$) & +0.14\% ($\uparrow$) & -1.72\% ($\uparrow$) & -1.52\% ($\downarrow$) & +4.43\% ($\uparrow$) & +2.15\% ($\uparrow$) & -6.93\% ($\downarrow$) & +4.32\% ($\uparrow$) & -1.84\% ($\downarrow$) & +18.12\% ($\uparrow$) \\
        \cmidrule(l){2-13}
         & \multirow{3}{*}{Full} & Baseline & 0.0012 $\pm$ 0.00 & 0.81 $\pm$ 0.04 & 4.51 $\pm$ 0.75 & 0.74 $\pm$ 0.09 & 0.74 $\pm$ 0.10 & 0.59 $\pm$ 0.09 & 0.61 $\pm$ 0.17 & 0.54 $\pm$ 0.15 & 1.00 $\pm$ 0.00 & 0.88 $\pm$ 0.10 \\
         &  & $\Delta_1$ (vs Hedge) & +38.98\% ($\downarrow$) & +0.32\% ($\uparrow$) & +1.11\% ($\downarrow$) & -1.52\% ($\downarrow$) & +2.77\% ($\uparrow$) & +0.71\% ($\uparrow$) & -6.93\% ($\downarrow$) & +0.73\% ($\uparrow$) & -12.64\% ($\downarrow$) & +12.54\% ($\uparrow$) \\
         &  & $\Delta_2$ (vs EXP3) & +32.94\% ($\downarrow$) & +0.32\% ($\uparrow$) & +1.11\% ($\downarrow$) & -1.52\% ($\downarrow$) & +2.77\% ($\uparrow$) & +0.71\% ($\uparrow$) & -6.93\% ($\downarrow$) & +0.62\% ($\uparrow$) & -12.64\% ($\downarrow$) & +12.54\% ($\uparrow$) \\
        \midrule
        \multirow{6}{*}{0.6} & \multirow{3}{*}{Clustered} & Baseline & 0.0005 $\pm$ 0.00 & 0.81 $\pm$ 0.03 & 4.74 $\pm$ 0.84 & 0.73 $\pm$ 0.10 & 0.74 $\pm$ 0.10 & 0.58 $\pm$ 0.09 & 0.59 $\pm$ 0.19 & 0.51 $\pm$ 0.16 & 1.00 $\pm$ 0.00 & 0.79 $\pm$ 0.11 \\
         &  & $\Delta_1$ (vs Hedge) & +80.89\% ($\downarrow$) & +0.24\% ($\uparrow$) & -3.38\% ($\uparrow$) & -0.18\% ($\downarrow$) & +2.80\% ($\uparrow$) & +1.60\% ($\uparrow$) & -3.98\% ($\downarrow$) & +4.50\% ($\uparrow$) & -2.07\% ($\downarrow$) & +27.09\% ($\uparrow$) \\
         &  & $\Delta_2$ (vs EXP3) & +92.75\% ($\downarrow$) & +0.24\% ($\uparrow$) & -3.38\% ($\uparrow$) & -0.18\% ($\downarrow$) & +2.73\% ($\uparrow$) & +1.56\% ($\uparrow$) & -3.98\% ($\downarrow$) & +4.44\% ($\uparrow$) & -2.07\% ($\downarrow$) & +27.09\% ($\uparrow$) \\
        \cmidrule(l){2-13}
         & \multirow{3}{*}{Full} & Baseline & 0.0011 $\pm$ 0.00 & 0.81 $\pm$ 0.04 & 4.57 $\pm$ 0.82 & 0.73 $\pm$ 0.10 & 0.76 $\pm$ 0.10 & 0.59 $\pm$ 0.09 & 0.59 $\pm$ 0.19 & 0.54 $\pm$ 0.16 & 1.00 $\pm$ 0.00 & 0.88 $\pm$ 0.10 \\
         &  & $\Delta_1$ (vs Hedge) & +40.67\% ($\downarrow$) & +0.24\% ($\uparrow$) & 0.00\% & -0.18\% ($\downarrow$) & +0.53\% ($\uparrow$) & -0.21\% ($\downarrow$) & -3.98\% ($\downarrow$) & -0.90\% ($\downarrow$) & -11.11\% ($\downarrow$) & +13.54\% ($\uparrow$) \\
         &  & $\Delta_2$ (vs EXP3) & +32.56\% ($\downarrow$) & +0.24\% ($\uparrow$) & 0.00\% & -0.18\% ($\downarrow$) & +0.49\% ($\uparrow$) & -0.23\% ($\downarrow$) & -3.98\% ($\downarrow$) & -1.05\% ($\downarrow$) & -11.11\% ($\downarrow$) & +13.54\% ($\uparrow$) \\
        \midrule
        \multirow{6}{*}{0.7} & \multirow{3}{*}{Clustered} & Baseline & 0.0002 $\pm$ 0.00 & 0.55 $\pm$ 0.06 & 3.19 $\pm$ 0.77 & 0.41 $\pm$ 0.12 & 0.78 $\pm$ 0.12 & 0.37 $\pm$ 0.11 & 0.22 $\pm$ 0.15 & 0.57 $\pm$ 0.21 & 1.00 $\pm$ 0.00 & 0.96 $\pm$ 0.02 \\
         &  & $\Delta_1$ (vs Hedge) & +213.33\% ($\downarrow$) & +1.69\% ($\uparrow$) & +1.57\% ($\downarrow$) & +5.05\% ($\uparrow$) & +1.95\% ($\uparrow$) & +4.99\% ($\uparrow$) & +1.06\% ($\uparrow$) & -1.13\% ($\downarrow$) & -0.82\% ($\downarrow$) & +4.53\% ($\uparrow$) \\
         &  & $\Delta_2$ (vs EXP3) & +219.83\% ($\downarrow$) & +1.69\% ($\uparrow$) & +1.57\% ($\downarrow$) & +5.05\% ($\uparrow$) & +2.01\% ($\uparrow$) & +5.02\% ($\uparrow$) & +1.06\% ($\uparrow$) & -1.11\% ($\downarrow$) & -0.82\% ($\downarrow$) & +4.53\% ($\uparrow$) \\
        \cmidrule(l){2-13}
         & \multirow{3}{*}{Full} & Baseline & 0.0005 $\pm$ 0.00 & 0.55 $\pm$ 0.06 & 3.19 $\pm$ 0.77 & 0.41 $\pm$ 0.12 & 0.78 $\pm$ 0.12 & 0.37 $\pm$ 0.11 & 0.22 $\pm$ 0.15 & 0.57 $\pm$ 0.21 & 1.00 $\pm$ 0.00 & 0.95 $\pm$ 0.04 \\
         &  & $\Delta_1$ (vs Hedge) & +153.15\% ($\downarrow$) & +1.69\% ($\uparrow$) & +1.57\% ($\downarrow$) & +5.05\% ($\uparrow$) & +2.48\% ($\uparrow$) & +5.28\% ($\uparrow$) & +1.06\% ($\uparrow$) & +0.03\% ($\uparrow$) & -3.94\% ($\downarrow$) & +5.00\% ($\uparrow$) \\
         &  & $\Delta_2$ (vs EXP3) & +134.54\% ($\downarrow$) & +1.69\% ($\uparrow$) & +1.57\% ($\downarrow$) & +5.05\% ($\uparrow$) & +2.39\% ($\uparrow$) & +5.26\% ($\uparrow$) & +1.06\% ($\uparrow$) & -0.04\% ($\downarrow$) & -3.94\% ($\downarrow$) & +5.00\% ($\uparrow$) \\
        \midrule
        \multirow{6}{*}{0.8} & \multirow{3}{*}{Clustered} & Baseline & 0.0002 $\pm$ 0.00 & 0.49 $\pm$ 0.04 & 2.30 $\pm$ 0.59 & 0.38 $\pm$ 0.10 & 0.82 $\pm$ 0.11 & 0.35 $\pm$ 0.09 & 0.19 $\pm$ 0.13 & 0.62 $\pm$ 0.23 & 1.00 $\pm$ 0.00 & 0.98 $\pm$ 0.02 \\
         &  & $\Delta_1$ (vs Hedge) & +294.20\% ($\downarrow$) & -0.01\% ($\downarrow$) & +2.61\% ($\downarrow$) & -0.61\% ($\downarrow$) & -0.43\% ($\downarrow$) & -1.06\% ($\downarrow$) & -5.85\% ($\downarrow$) & -5.06\% ($\downarrow$) & -0.59\% ($\downarrow$) & +2.12\% ($\uparrow$) \\
         &  & $\Delta_2$ (vs EXP3) & +287.40\% ($\downarrow$) & -0.01\% ($\downarrow$) & +2.61\% ($\downarrow$) & -0.61\% ($\downarrow$) & -0.50\% ($\downarrow$) & -1.07\% ($\downarrow$) & -5.85\% ($\downarrow$) & -5.18\% ($\downarrow$) & -0.59\% ($\downarrow$) & +2.12\% ($\uparrow$) \\
        \cmidrule(l){2-13}
         & \multirow{3}{*}{Full} & Baseline & 0.0003 $\pm$ 0.00 & 0.49 $\pm$ 0.04 & 2.30 $\pm$ 0.59 & 0.38 $\pm$ 0.10 & 0.82 $\pm$ 0.11 & 0.35 $\pm$ 0.10 & 0.19 $\pm$ 0.13 & 0.61 $\pm$ 0.24 & 1.00 $\pm$ 0.00 & 0.97 $\pm$ 0.02 \\
         &  & $\Delta_1$ (vs Hedge) & +345.37\% ($\downarrow$) & -0.01\% ($\downarrow$) & +2.61\% ($\downarrow$) & -0.61\% ($\downarrow$) & -0.14\% ($\downarrow$) & -0.95\% ($\downarrow$) & -5.85\% ($\downarrow$) & -4.32\% ($\downarrow$) & -0.64\% ($\downarrow$) & +2.76\% ($\uparrow$) \\
         &  & $\Delta_2$ (vs EXP3) & +321.35\% ($\downarrow$) & -0.01\% ($\downarrow$) & +2.61\% ($\downarrow$) & -0.61\% ($\downarrow$) & -0.14\% ($\downarrow$) & -0.95\% ($\downarrow$) & -5.85\% ($\downarrow$) & -4.32\% ($\downarrow$) & -0.64\% ($\downarrow$) & +2.76\% ($\uparrow$) \\
        \bottomrule
    \end{tabular}
    }
\end{table*}

\bibliographystyle{unsrturl}
\renewcommand{\bibfont}{\small}
\setlength{\bibsep}{4pt}
\bibliography{references}

\end{document}